\documentclass[10pt,journal,compsoc]{IEEEtran}

\usepackage{amsmath}
\usepackage{diagbox}
\usepackage{graphics}
\usepackage{subfigure}
\usepackage{epsfig}
\usepackage{threeparttable}
\usepackage{xspace}
\usepackage{siunitx}
\usepackage{graphicx}
\usepackage{amssymb}
\usepackage{color}
\usepackage[normalem]{ulem}
\usepackage{multirow}
\usepackage{float}
\usepackage{amsfonts}
\usepackage{bm}
\usepackage{array}
\usepackage[table]{xcolor}
\usepackage{colortbl}
\usepackage{diagbox}
\usepackage{rotating}
\usepackage{booktabs}
\usepackage{overpic}
\usepackage{enumitem}

\usepackage{fontawesome}
\usepackage{tcolorbox}
\usepackage{mathrsfs}
\usepackage{makecell}
\usepackage{lipsum}
\usepackage{bm}
\usepackage{multirow}
\usepackage[export]{adjustbox}

\definecolor{lightgray}{gray}{.9}
\definecolor{deepgray}{gray}{.8}

\newcolumntype{I}{!{\vrule width 1pt}}
\makeatletter
\newcommand{\thickhline}{%
    \noalign {\ifnum 0=`}\fi \hrule height 1pt
    \futurelet \reserved@a \@xhline
}
\makeatother
\usepackage{float}
\usepackage[ruled, vlined]{algorithm2e}
\usepackage{pifont}
\usepackage{ragged2e}

\usepackage{bbding}
\usepackage{cancel}

\usepackage[pagebackref=false,breaklinks=true,colorlinks,bookmarks=false]{hyperref}

\definecolor{mydarkyellow}{RGB}{216, 214, 196}
\definecolor{mymiddleyellow}{RGB}{229, 228, 218}
\definecolor{mylightyellow}{RGB}{245, 245, 240}
\definecolor{mygray}{gray}{.9}
\definecolor{crossgray}{gray}{.97}
\definecolor{mygreen}{RGB}{93,173,85}
\definecolor{mywarning}{RGB}{233,144,61}
\definecolor{DarkRed}{RGB}{0,0,0}
\definecolor{azure}{rgb}{0.0, 0.5, 1.0}
\definecolor{gray}{rgb}{0.3, 0.3, 0.3}
\definecolor{DarkGreen}{RGB}{42,110,63}
\definecolor{main_task}{cmyk}{0, 0.01, 0.17, 0}
\definecolor{backdoor}{cmyk}{0, 0.07, 0.09, 0}
\definecolor{main_task_arr}{cmyk}{0, 0.16, 1, 0.14}
\definecolor{backdoor_arr}{cmyk}{0, 0.92, 1, 0.09}
\definecolor{table_color_2}{HTML}{FFF5D6}
\definecolor{array_color}{cmyk}{0.41, 0, 0.53, 0.45}
\definecolor{cline_color}{cmyk}{0, 0, 0, 0.25}
\newcommand{\hlg}[1]{\textcolor{mygreen}{#1}}
\newcommand{\hlr}[1]{\textcolor{red}{#1}}

\newcommand{\tmark}{\ding{51}} 
\newcommand{\xmark}{\ding{55}} 
\usepackage[capitalize]{cleveref}
\crefname{section}{Sec.}{Secs.}
\crefname{table}{Tab.}{Tabs.}
\crefname{section}{§}{§§}

\makeatletter
\DeclareRobustCommand\onedot{\futurelet\@let@token\@onedot}
\def\@onedot{\ifx\@let@token.\else.\null\fi\xspace}



\begin{document}

\title{A Survey of Safety on Large Vision-Language Models: Attacks, Defenses and Evaluations}


\author{ Mang Ye,
Xuankun Rong, Wenke Huang, Bo Du, Nenghai Yu, Dacheng Tao,~\IEEEmembership{Fellow,~IEEE}
\IEEEcompsocitemizethanks{
\IEEEcompsocthanksitem 
M. Ye, X. Rong, W. Huang, and B. Du are with the School of Computer Science, Wuhan University, Wuhan, China. \protect
E-mail:\{yemang, rongxuankun, wenkehuang, dubo\}@whu.edu.cn
\IEEEcompsocthanksitem 
N. Yu is with the School of Cyber Science and Technology, University of Science and Technology of China, Hefei, China. E-mail: ynh@ustc.edu.cn \protect
\IEEEcompsocthanksitem D. Tao is with the College of Computing and Data Science,
Nanyang Technological University, Singapore. E-mail: dacheng.tao@gmail.com \protect
}
}


\IEEEtitleabstractindextext{
\begin{abstract}

With the rapid advancement of Large Vision-Language Models (LVLMs), ensuring their safety has emerged as a crucial area of research. This survey provides a comprehensive analysis of LVLM safety, covering key aspects such as attacks, defenses, and evaluation methods. We introduce a unified framework that integrates these interrelated components, offering a holistic perspective on the vulnerabilities of LVLMs and the corresponding mitigation strategies. Through an analysis of the LVLM lifecycle, we introduce a classification framework that distinguishes between inference and training phases, with further subcategories to provide deeper insights. Furthermore, we highlight limitations in existing research and outline future directions aimed at strengthening the robustness of LVLMs. As part of our research, we conduct a set of safety evaluations on the latest LVLM, Deepseek Janus-Pro, and provide a theoretical analysis of the results. Our findings provide strategic recommendations for advancing LVLM safety and ensuring their secure and reliable deployment in high-stakes, real-world applications. This survey aims to serve as a cornerstone for future research, facilitating the development of models that not only push the boundaries of multimodal intelligence but also adhere to the highest standards of security and ethical integrity.
Furthermore, to aid the growing research in this field, we have created a public repository to continuously compile and update the latest work on LVLM safety: \url{https://github.com/XuankunRong/Awesome-LVLM-Safety}.
\end{abstract}

}

\maketitle

\IEEEdisplaynontitleabstractindextext
\IEEEpeerreviewmaketitle

\IEEEraisesectionheading{
\section{Introduction}
\label{sec: introduction}
}

\IEEEPARstart{N}{owadays} Large Language Models (LLMs) have remarkably transformed the AI landscape, demonstrating unprecedented capabilities in natural language understanding and generation~\cite{zhao2023survey, minaee2024large, brown2020language, chowdhery2023palm, touvron2023llama, awais2025foundation}.
Their versatility and scalability have set new benchmarks across various domains, from conversational agents to complex problem-solving tasks~\cite{kaur2024text, cai2024internlm2, jiang2024mixtral, bai2023qwenllm, xu2024lvlm}.
To further enhance the applicability of LLMs, researchers have integrated visual modalities, giving rise to Large Vision-Language Models (LVLMs)~\cite{radford2021learning, alayrac2022flamingo, li2023blip, liu2024visual, achiam2023gpt, zhang2024vision}.
This fusion has expanded the horizons of AI by enabling multimodal comprehension and interaction. LVLMs have rapidly evolved from early task-specific systems, such as image captioning and visual question answering, to sophisticated frameworks capable of complex reasoning and creative generation. Leveraging large-scale pretraining on diverse datasets, models like CLIP~\cite{radford2021learning}, ALIGN~\cite{jia2021scaling}, GPT-4V~\cite{achiam2023gpt}, and LLaVA~\cite{liu2024visual} have set new standards in zero-shot and few-shot learning. The advancements in LVLMs have unlocked their potential across a wide array of applications, including autonomous driving~\cite{tian2024drivevlm}, healthcare diagnostics~\cite{van2024large}, and content creation~\cite{maharana2022storydall, zhou2024storydiffusion}.
As the field continues to advance, LVLMs are poised to become indispensable tools in critical industries, driving the development of highly adaptive and intelligent AI systems with comprehensive multimodal understanding.

While LVLMs offer immense benefits and have significantly enhanced user experiences across various applications, ensuring their safety and security is equally paramount.
The multimodal nature of LVLMs introduces unique vulnerabilities, as adversarial perturbations in one modality (for example, subtly altered images) can cascade through the system~\cite{qi2024visual, schlarmann2023adversarial, zhao2024evaluating}, resulting in unsafe behaviors and potentially harmful outputs when combined with deceptive textual inputs~\cite{wang2024white, bailey2024image}.
Moreover, challenges like difficulties in model alignment~\cite{ding2024eta, xu2024crosssafety}, and susceptibility to backdoor attacks~\cite{xu2024shadowcast, liang2024vl, ni2024physical} further exacerbate the security concerns surrounding LVLMs. In practical scenarios, these vulnerabilities can have severe repercussions: for example, manipulated medical images may lead to incorrect diagnoses, adversarial alterations of financial data can distort risk assessments, and tampered navigation maps or traffic signs can mislead autonomous driving systems, resulting in hazardous outcomes. As LVLMs become increasingly integrated into critical sectors like healthcare, finance, and transportation, it is imperative to prioritize the robustness, reliability, and ethical alignment of these models. Addressing the security challenges of LVLMs is not only a technical necessity but also a crucial step towards the responsible and safe deployment of these advanced AI systems in real-world applications.

\begin{table*}[t]
    \centering
    \caption{Overview of related surveys. See details in~\cref{sec: Prior Surveys}.}
    \label{tab: Overview of related LVLM safety surveys.}
    \vspace{-10pt}
    \resizebox{2\columnwidth}{!}{
        \setlength\tabcolsep{4pt}
        \renewcommand\arraystretch{1.5}
        \begin{tabular}{c||ccc|c}
        \hline\thickhline
        \rowcolor{mydarkyellow}
            \textbf{Surveys} & \textbf{Attack} & \textbf{Defense} & \textbf{Evaluation} & \textbf{Contributions \& Limitations} \\
        \hline\hline
            \textcolor{gray}{[IJCAI'24]}~\cite{liu2024safety} & \hlg{\bcancel\tmark{}} & \hlg{\bcancel\tmark{}} & \hlg{\bcancel\tmark{}} & \multicolumn{1}{l}{\makecell[l]{Provides a concise overview and basic categorization, lacking in-depth analysis \\ and comprehensive discussion of methods.}} \\ 
        \hline
            \textcolor{gray}{[arXiv'24]}~\cite{fan2024unbridled} & \hlg{\bcancel\tmark{}} & \hlg{\bcancel\tmark{}} & - & \multicolumn{1}{l}{\makecell[l]{Focuses primarily on vulnerabilities in image inputs of multimodal models, \\ with limited attention to text-based attacks and cross-modal vulnerabilities.}} \\
        \hline
            \textcolor{gray}{[EMNLP'24]}~\cite{wang2024llms} & Jailbreak & Jailbreak & Jailbreak & \multicolumn{1}{l}{\makecell[l]{Explores jailbreaks from LLMs to LVLMs, focusing narrowly on attacks while \\ lacking broader analysis of robustness and safety frameworks.}} \\
        \hline
            \textcolor{gray}{[arXiv'24]}~\cite{jin2024jailbreakzoo} & Jailbreak & Jailbreak & Jailbreak & \multicolumn{1}{l}{\makecell[l]{Comprehensive survey of jailbreaking attacks in LLMs and LVLMs, but lacks \\ detailed coverage of non-jailbreaking attacks and defenses.}} \\
        \hline
            \textcolor{gray}{[arXiv'24]}~\cite{liu2024survey} & \hlg{\tmark{}} & - & - & \multicolumn{1}{l}{\makecell[l]{Surveys recent advances in attacks on LVLMs, focusing on methodologies, \\ but lacks sufficient emphasis on defenses and evaluations}} \\
        \hline
            \textcolor{gray}{[arXiv'24]}~\cite{zhang2024adversarial} & Adversarial & - & - & \multicolumn{1}{l}{\makecell[l]{Covers adversarial attacks on vision tasks but fails to address the unique \\ multimodal security challenges associated with LVLMs.}} \\
        \hline
            \textcolor{gray}{[arXiv'24]}~\cite{liu2024jailbreak} & Jailbreak & Jailbreak & Jailbreak & \multicolumn{1}{l}{\makecell[l]{Highlights jailbreaking attacks and defenses in multimodal generative models \\ but excludes broader LVLM use cases and attack scenarios.}} \\
        \hline\hline
            \cellcolor{mylightyellow} Ours & \cellcolor{mylightyellow} \hlg{\tmark{}} & \cellcolor{mylightyellow} \hlg{\tmark{}} & \cellcolor{mylightyellow} \hlg{\tmark{}} & \multicolumn{1}{l}{\cellcolor{mylightyellow} \makecell[l]{Presents a systematic analysis of LVLM safety, introducing a lifecycle-based \\ classification and integrating perspectives on attacks, defenses, and evaluations.}} \\
        \hline\thickhline
        \end{tabular}
    }
    \vspace{-15pt}
\end{table*}

Currently, safety-related research on LVLMs can be delineated into the following three categories:
\textbf{i)~Attacks}.
Investigating attacks on LVLMs is essential for uncovering and mitigating the vulnerabilities inherent in these sophisticated architectures. Unlike LLMs, LVLMs present more extensive security challenges due to the integration of visual and textual modalities~\cite{ding2024eta, qi2024visual, li2024images, gong2023figstep, lee2024does}. Adversaries frequently exploit weaknesses in the visual processing components and inherent vulnerabilities in the training methodologies of LVLMs to induce the model to output unsafe responses.
\textbf{ii)~Defenses}.
Defense strategies are designed to enhance the resilience of LVLMs against a spectrum of adversarial threats. Based on the lifecycle of LVLM, these strategies are typically categorized into inference-phase and training-phase defenses. Inference-phase defenses incorporate techniques such as input sanization~\cite{wang2024adashield, xu2024cross}, internal optimization~\cite{wang2024inferaligner, gao2024coca} and output validation~\cite{zhang2023mutation, pi2024mllm} to safeguard models during deployment, effectively preventing the exploitation of operational vulnerabilities. Conversely, training-phase defenses employ methodologies like adversarial fine-tuning~\cite{zong2024safety, zhang2024spa, liu2024safetyalignment} to fortify the models during their development, thereby augmenting their robustness against potential adversarial manipulations.
\textbf{iii)~Evaluations}.
Evaluation efforts focus on the creation of comprehensive benchmarks that assess the security capabilities of various LVLMs~\cite{tu2023many, liu2023query, luo2024jailbreakv, zhang2024benchmarking, gu2024mllmguard}. These benchmarks provide standardized frameworks for researchers to evaluate and compare the efficacy of safety measures across different models. By systematically identifying security deficiencies, these evaluations facilitate the advancement of more secure and reliable LVLMs, ensuring their safe integration into critical applications.

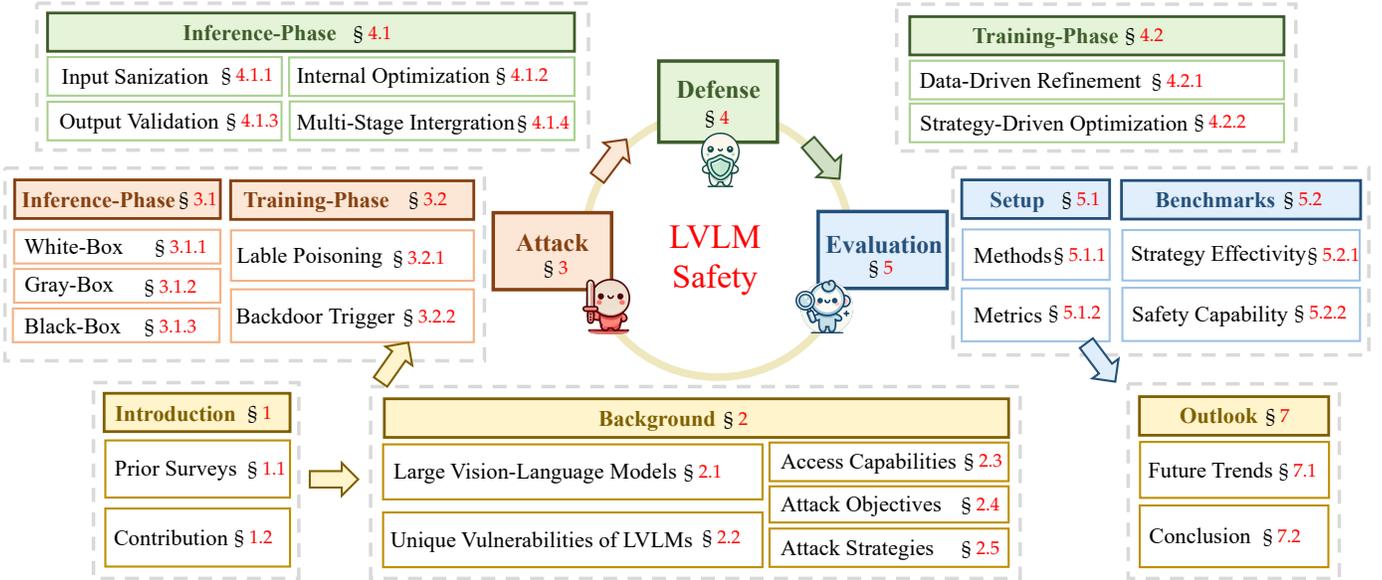
\begin{figure*}[t]
    \centering
    \includegraphics[width=\linewidth]{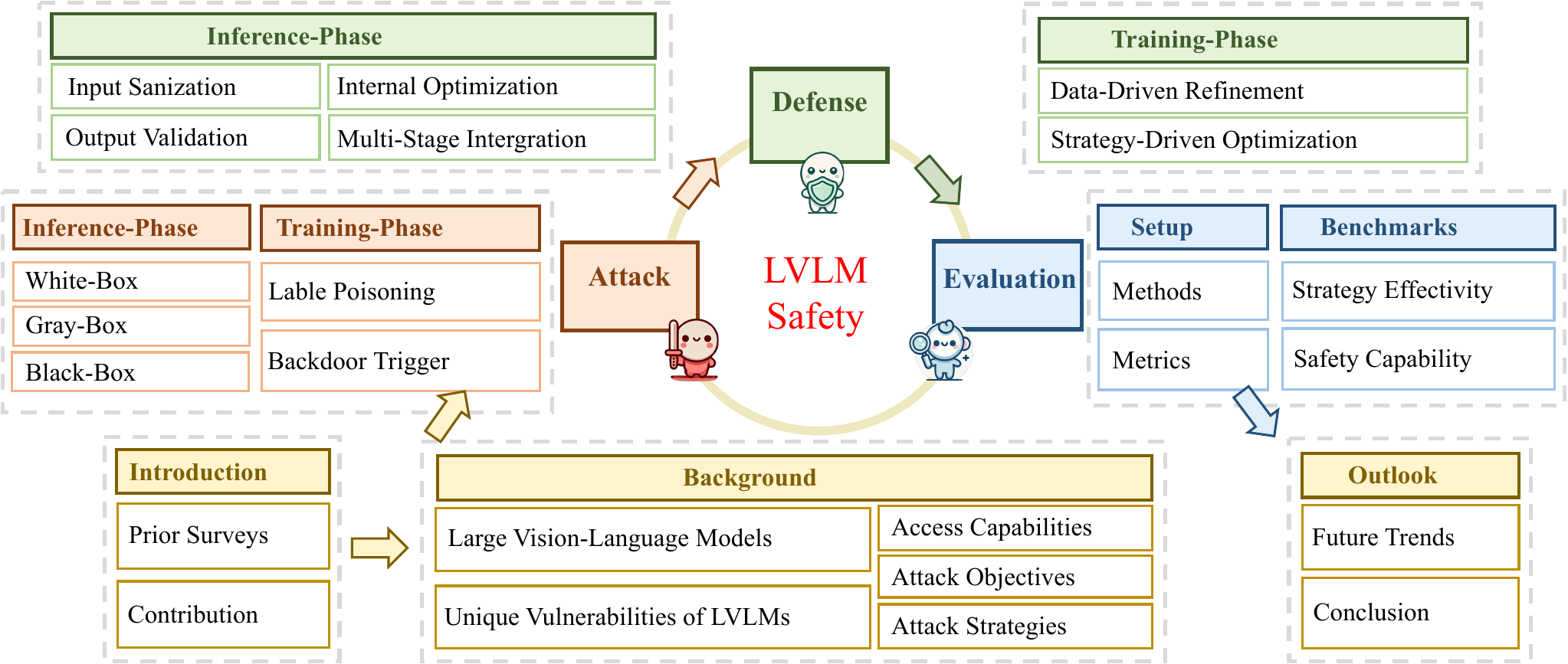}
    \put(-425,61){\scriptsize{\cref{sec: introduction}}}
    \put(-425,40){\scriptsize{\cref{sec: Prior Surveys}}}
    \put(-430,14){\scriptsize{\cref{sec: Contributions}}}
    \put(-245,59){\scriptsize{\cref{sec: background}}}
    \put(-260,39){\scriptsize{\cref{sec: Large Vision-Language Models}}}
    \put(-253,14){\scriptsize{\cref{sec: Unique Vulnerabilities of Large Vision-Language Models}}}
    \put(-154,43){\scriptsize{\cref{sec: Access Capability}}}
    \put(-155,26.5){\scriptsize{\cref{sec: Attack Objectives}}}
    \put(-155,10){\scriptsize{\cref{sec: Attack Strategies}}}
    \put(-313,116){\scriptsize{\cref{sec: Attack}}}
    \put(-451,142){\scriptsize{\cref{sec: Inference-Phase Attacks}}}
    \put(-460,124){\scriptsize{\cref{sec: White-box Attacks}}}
    \put(-464,109){\scriptsize{\cref{sec: Gray-box Attacks}}}
    \put(-464,94){\scriptsize{\cref{sec: Black-box Attacks}}}
    \put(-364,142){\scriptsize{\cref{sec: Training-Phase Attacks}}}
    \put(-370,120){\scriptsize{\cref{sec: Label Poisoning Attacks}}}
    \put(-366,97.5){\scriptsize{\cref{sec: Backdoor Trigger Attacks}}}
    \put(-252,173){\scriptsize{\cref{sec: Defense}}}
    \put(-385,205){\scriptsize{\cref{sec: Inference-Phase Defenses}}}
    \put(-435,189){\scriptsize{\cref{sec: Input Sanization Defenses}}}
    \put(-331,189){\scriptsize{\cref{sec: Internal Optimization Defenses}}}
    \put(-433,172){\scriptsize{\cref{sec: Output Validation Defenses}}}
    \put(-323,171){\scriptsize{\cref{sec: Multi-Stage Integration Defenses}}}
    \put(-93,204){\scriptsize{\cref{sec: Training-Phase Defenses}}}
    \put(-83,187){\scriptsize{\cref{sec: Data-Driven Refinement}}}
    \put(-67,171.5){\scriptsize{\cref{sec: Strategy-Driven Optimization}}}
    \put(-190,116){\scriptsize{\cref{sec: Evaluation}}}
    \put(-117,142){\scriptsize{\cref{sec: Setup}}}
    \put(-120,121){\scriptsize{\cref{sec: Methods}}}
    \put(-122,99){\scriptsize{\cref{sec: Metrics}}}
    \put(-33,142){\scriptsize{\cref{sec: Benchmarks}}}
    \put(-24,121){\scriptsize{\cref{sec: Strategy Effectivity}}}
    \put(-29,99){\scriptsize{\cref{sec: Safety Capability}}}
    \put(-40,60){\scriptsize{\cref{sec: Outlook}}}
    \put(-35,39.5){\scriptsize{\cref{sec: Future Trends}}}
    \put(-41,14.5){\scriptsize{\cref{sec: Conclusion}}}
    \vspace{-5pt}
    \caption{Overview of the survey. Best viewed in color.}
    \label{fig: pipeline}
\end{figure*}

\subsection{Prior Surveys}
\label{sec: Prior Surveys}

Recent surveys on the safety and security of Large Vision-Language Models (LVLMs) have made significant contributions by highlighting the diverse attack methods, defense mechanisms, and vulnerabilities specific to multimodal systems.
However, despite these valuable insights, current surveys often fall short of providing a comprehensive and systematic view that integrates both attack and defense strategies across all modalities, leaving gaps in understanding the full spectrum of LVLM vulnerabilities.
Therefore, we discuss the main contributions and limitations of existing related surveys and highlight the unique contributions of our work in~\cref{tab: Overview of related LVLM safety surveys.}.
For instance, Liu \textit{et al.}~\cite{liu2024safety} explore the general safety concerns in LVLMs, focusing on issues like unsafe outputs caused by inconsistencies between the image and text modalities. It provides a concise overview and basic categorization of the core safety challenges, but lacks in-depth exploration of advanced attack techniques and specific countermeasures tailored for LVLMs.
In contrast, Fan \textit{et al.}~\cite{fan2024unbridled} emphasize the risks associated with image inputs in LVLMs, particularly in the context of adversarial image manipulations and their downstream effects on text generation. This survey provides valuable insights into the inherent vulnerabilities in LVLMs' visual understanding but does not sufficiently address the broader spectrum of safety concerns, such as backdoor attacks or the interplay between different attack modalities (e.g., visual and textual).
Similarly, Wang \textit{et al.}~\cite{wang2024llms} and Jin \textit{et al.}~\cite{jin2024jailbreakzoo} concentrate on the emerging problem of jailbreaking attacks from LLMs to LVLMs, providing detailed analyses of attack methods and possible defensive strategies. While both surveys are highly focused on jailbreaking, they leave gaps in covering other attack types and the overall landscape of LVLM security.
Surveys such as Liu \textit{et al.}~\cite{liu2024survey} and Zhang \textit{et al.}~\cite{zhang2024adversarial} offer more expansive overviews of adversarial and attack-based vulnerabilities in LVLMs. Liu \textit{et al.}~\cite{liu2024survey} survey recent advances in LVLM attacks, offering a broad perspective on resources and trends, but it lacks a focused discussion on multimodal-specific attacks and fails to integrate defense strategies systematically. Zhang \textit{et al.}~\cite{zhang2024adversarial} present a historical overview of adversarial attacks on vision tasks and their relevance to LVLMs. However, its narrow focus on vision-specific tasks limits its applicability to the multimodal setting of modern LVLMs. Lastly, Liu \textit{et al.}~\cite{liu2024jailbreak} specifically address jailbreak attacks and defenses for generative multimodal models, providing a detailed account of this rapidly evolving threat. However, it limited scope to multimodal generative models, excluding broader LVLM use cases and attacks, such as prompt injection or backdoor poisoning.

\subsection{Contributions}
\label{sec: Contributions}

In this paper, we introduce a comprehensive survey on safety of LVLM and mainly focus on attacks, defenses, and evaluations. Compared with existing surveys, this paper makes the following contributions:
\begin{itemize}
    \item We provide a comprehensive and systematic analysis of LVLM safety by integrating the interconnected aspects of attacks, defenses, and evaluations. Isolated examination of attacks or defenses alone does not fully capture the overall security landscape, whereas our approach combines these critical components to offer a more holistic understanding of the vulnerabilities and mitigation strategies inherent in LVLMs.
    \item By analyzing the lifecycle of LVLMs, we propose a universal classification framework that categorizes security-related works based on the model's inference and training phases. Further subcategories are identified to provide a more granular understanding. For each work, we present a thorough exploration of the methodologies and contributions, delivering a comprehensive and insightful analysis of the prevailing landscape of LVLM security.
    \item We conduct safety evaluations on the latest LVLM, Deepseek Janus-Pro and delineate future research trajectories, presenting profound insights and strategic recommendations that empower the research community to enhance the safety and robustness of LVLMs. This guidance is instrumental in facilitating the safe and reliable deployment of these models within mission-critical applications.
\end{itemize}
The remainder of this paper is structured as follows: \cref{fig: pipeline} delineates the overall framework of this survey. \cref{sec: background} provides a succinct overview of the foundational aspects of LVLM safety. The safety of LVLMs is systematically analyzed from three principal perspectives: \textbf{Attacks} in \cref{sec: Attack}, \textbf{Defenses} in \cref{sec: Defense}, and \textbf{Evaluations} in \cref{sec: Evaluation}. Lastly, \cref{sec: Future Trends} explores prospective research directions, followed by concluding remarks in \cref{sec: Conclusion}.

\section{Background}
\label{sec: background}

\subsection{Large Vision-Language Models}
\label{sec: Large Vision-Language Models}

The development of Large Language Models (LLMs) has emerged as a cornerstone in the field of artificial intelligence, revolutionizing the way machines understand and generate human language. Examples of prominent LLMs include OpenAI’s GPT-4~\cite{brown2020language, achiam2023gpt, radford2019language}, Google’s PaLM~\cite{chowdhery2023palm, anil2023palm}, Meta’s LLaMA~\cite{touvron2023llama}, and Vicuna~\cite{vicuna2023}, all of which have demonstrated remarkable capabilities ranging from natural language understanding and generation.
To expand the applicability of LLMs, existing solutions normally integrated vision components, leading to the development of Large Vision-Language Models (LVLMs). By utilizing the visual extractor like CLIP~\cite{radford2021learning} to encode visual features and utilize the connector module to project visual tokens into word embedding space of the LLM, LVLMs are capable of jointly processing text and visual inputs. This multimodal integration enables LVLMs to bridge the gap between vision and language, paving the way for more advanced applications in diverse fields.
Subsequent developments in LVLMs have led to the emergence of several notable models, including Flamingo~\cite{alayrac2022flamingo}, BLIP-2~\cite{li2023blip}, GPT-4V~\cite{achiam2023gpt}, Gemini~\cite{team2023gemini}, MiniGPT-4~\cite{zhu2023minigpt}, PandaGPT~\cite{su2023pandagpt}, LLaVA~\cite{liu2024visual}, LLaVa-OneVision~\cite{li2024llava}, InternVL~\cite{chen2024internvl}, Qwen-VL~\cite{bai2023qwen}, and VILA~\cite{lin2024vila}.
The integration of vision and language in LVLMs has opened up new possibilities across a variety of application domains. For instance, LVLMs are widely utilized in image captioning~\cite{pi2024image, pi2024personalized}, where they generate descriptive text for images, and in visual question answering (VQA), where they answer questions based on image content. These models are also employed in content moderation, combining text and visual inputs to detect inappropriate or harmful content, and in creative industries, enabling tasks such as generating creative narratives based on visual inputs~\cite{maharana2022storydall, zhou2024storydiffusion}. Beyond these, specialized applications include medical imaging for diagnostic insights~\cite{van2024large}, autonomous driving for visual scene understanding~\cite{tian2024drivevlm}, and education for generating multimodal instructional content.
Despite their impressive capabilities, LVLMs face several significant challenges. Scalability remains a key issue as the integration of multimodal data requires increased computational resources, both during training and inference. Furthermore, robustness to adversarial inputs, especially in multimodal contexts, is a growing concern. Adversarial attacks can exploit the interaction between text and visual inputs, leading to unexpected or unsafe outputs~\cite{qi2024visual, schlarmann2023adversarial, zhao2024evaluating}. Bias and fairness are also critical issues, as LVLMs often inherit biases from their training data, which can result in unfair or harmful outputs in sensitive contexts~\cite{lee2024does, gallegos2024bias}. Lastly, safety and alignment are ongoing challenges, as LVLMs are susceptible to producing toxic or misleading content due to gaps in their training or failure to understand multimodal queries.

\subsection{Unique Vulnerabilities of LVLMs}
\label{sec: Unique Vulnerabilities of Large Vision-Language Models}

The integration of visual modalities into LLMs has enhanced LVLMs' multimodal capabilities but also introduced unique vulnerabilities. These include new attack surfaces from visual inputs and the degradation of safety alignment during fine-tuning, both of which compromise the model's robustness and reliability. Details as follows:

\noindent$\bullet$ \textbf{Expansion Risks Introduced by Visual Inputs.}
The integration of visual modalities into LLMs inherently leads to an expansion of attack surfaces, exposing models to new security risks~\cite{qi2024visual, li2024images, gong2023figstep, lee2024does, ding2024eta}. In LLMs, adversarial attacks are constrained to the discrete nature of textual input~\cite{shen2024anything, yi2024jailbreak}, making such manipulations more demanding, and defenses only need to address textual vulnerabilities. However, the introduction of visual inputs exposes the model to the inherently continuous and high-dimensional visual input space, which serves as a weak link~\cite{guo2024vllm, zhao2024evaluating}. These characteristics make visual adversarial examples fundamentally challenging to defend against. Consequently, the transition from a purely textual domain to a composite textual-visual domain not only broadens the vulnerability surfaces but also escalates the complexity and burden of defensive measures.

\noindent$\bullet$ \textbf{Degradation of Safety During Fine-Tuning.}
Degradation of Safety During Fine-Tuning. Visual Instruction Tuning has become essential for enabling Large Language Models (LLMs) to process multimodal inputs by integrating a pre-trained LLM with a vision encoder through a projector layer~\cite{liu2024visual, alayrac2022flamingo, li2023blip, zhu2023minigpt}. This process allows LVLMs to reason across modalities, addressing tasks beyond the capabilities of language-only models. However, their performance depends heavily on the underlying vision and language components, making them susceptible to vulnerabilities caused by misalignment between these modules.
A significant limitation of current fine-tuning practices is the freezing of the vision encoder while updating only the projector layer and LLM. This approach leaves the vision encoder without robust safety defenses, exposing it to adversarial or harmful inputs. Additionally, the lack of safety-aware training often leads to the degradation of the model’s pre-trained safety alignment, a phenomenon referred to as catastrophic forgetting~\cite{lee2024does, zong2024safety, pantazopoulos2024learning, bachu2024unfair}. As a result, the model becomes increasingly prone to generating unsafe outputs, particularly in response to adversarial prompts.
This degradation is further amplified when a larger portion of the model’s parameters are fine-tuned~\cite{li2024images}, as extensive updates disrupt the original safety alignment. Consequently, fine-tuning that prioritizes performance without adequately addressing safety risks can unintentionally increase the model's vulnerabilities. This underscores the critical importance of developing training strategies that maintain both safety and performance, particularly as LVLMs are adapted to new multimodal tasks and domains.

\subsection{Access Capabilities}
\label{sec: Access Capability}

The interaction with LVLMs, whether for attacks or defenses can be categorized based on the knowledge set \( \mathcal{K} \) about the model \( f_\theta \) accessible to the entity (attacker or defender). The knowledge may encompass elements such as model parameters \( \theta \), model architecture \( \mathcal{A}_\theta \), gradients \( \nabla_\theta \mathcal{L} \), input data \( x \), and output data \( y \). Based on the scope of accessible knowledge, three distinct capabilities are defined as follows:

\noindent$\bullet$ \textbf{White-box Capability.}
White-box capability represents the highest level of access, where all internal details of the model are fully available, including the model parameters \( \theta \), the model architecture \( \mathcal{A}_\theta \), and the gradients \( \nabla_\theta \mathcal{L} \). The knowledge set is formally defined as:
\begin{equation}
    \mathcal{K}_{\text{W}} = \{ x, y, \theta, \mathcal{A}_\theta, \nabla_\theta \mathcal{L} \mid x \in \mathcal{X}, y = f_\theta(x) \},
\label{eq: White-box Capability}
\end{equation}
this level of access enables precise computation of gradients, making it possible to craft adversarial inputs or design highly effective defense mechanisms. White-box scenarios are typically used in controlled research environments.

\noindent$\bullet$ \textbf{Gray-box Capability.}
Gray-box capability assumes partial access to the model's internal details, such as its architecture \( \mathcal{A}_\theta \) or intermediate feature representations, but lacks full knowledge of \( \theta \) or $\nabla_\theta \mathcal{L}$. The knowledge set is defined as:
\begin{equation}
    \mathcal{K}_{\text{G}} = \{ x, y, \mathcal{A}_\theta \mid x \in \mathcal{X}, y = f_\theta(x) \},
\label{eq: Gray-box Capability}
\end{equation}
this level of access is common in scenarios where the model architecture is publicly known or inferred. For example, a surrogate model $ \mathcal{S_{\theta}} $ can be trained to approximate the target model’s behavior, which can then be used for crafting adversarial inputs or testing defensive strategies.

\noindent$\bullet$ \textbf{Black-box Capability.}
Black-box capability represents the lowest level of access, where the entity has no internal information about the model. The only accessible data are the input-output pairs \( \{x, f_\theta(x)\} \) without any direct knowledge of \( \theta \), \( \mathcal{A}_\theta \), and $\nabla_\theta \mathcal{L}$. The knowledge set is defined as:
\begin{equation}
    \mathcal{K}_{\text{B}} = \{(x, f_\theta(x)) \mid x \in \mathcal{X}\},
\label{eq: Black-box Capability}
\end{equation}
in this scenario, interactions are limited to querying the model and observing its outputs. This setting are representative of real-world conditions, where attackers or defenders must work without any internal access to the model.

\subsection{Attack Objectives}
\label{sec: Attack Objectives}

In the safety domain of Large Vision-Language Models (LVLMs), attacks typically fall into three main categories, each with distinct objectives:

\noindent$\bullet$ \textbf{Targeted Attacks.}  
Targeted attacks are designed to manipulate the model's output for specific inputs \( x \in \mathcal{X} \), driving the model \( f_\theta \) to produce a predefined, incorrect output \( y_{\text{target}} \in \mathcal{Y} \), irrespective of the correct output \( y \). These attacks may involve adversarial perturbations to the input, such as subtle modifications to images or text, or non-adversarial methods, such as crafted queries that exploit weaknesses in the model's reasoning. The objective can be described as:
\begin{equation}
    \arg\min_{x_{\text{mod}}} \mathcal{L}(f_\theta(x_{\text{mod}}), y_{\text{target}}),
\end{equation}
where \( x_{\text{mod}} \) represents either adversarially perturbed inputs or specially crafted benign queries.

\noindent$\bullet$ \textbf{Untargeted Attacks.}  
Untargeted attacks aim to degrade the model’s overall performance by causing it to produce any incorrect output \( y' \neq y \). These attacks are not constrained by specific target outputs and can involve adversarial modifications to the input or non-adversarial strategies, such as exploiting ambiguities in the training data or the model's inherent biases. The goal is defined as:
\begin{equation}
    \arg\min_{x_{\text{mod}}} \mathcal{L}(f_\theta(x_{\text{mod}}), y), \quad \text{subject to } f_\theta(x_{\text{mod}}) \neq y,
\end{equation}
this category focuses on reducing the model's accuracy across tasks and scenarios.

\noindent$\bullet$ \textbf{Jailbreak Attacks.}  
Jailbreak attacks seek to bypass the model’s safety mechanisms or ethical constraints, compelling it to generate harmful or restricted outputs \( y_{\text{restricted}} \). Unlike targeted or untargeted attacks, jailbreak methods often do not require adversarial perturbations to the input; instead, they exploit flaws in the model's safety alignment or prompt-handling mechanisms. Such attacks may involve carefully designed queries or prompts that trick the model into violating its safety policies. The objective is defined as:
\begin{equation}
    \arg\min_{x} \mathcal{R}(f_\theta(x)) \quad \text{subject to } y_{\text{restricted}} \in \mathcal{O},
\end{equation}
where \( \mathcal{R} \) measures the effectiveness of the model’s safety mechanisms, and \( \mathcal{O} \) is the set of restricted outputs.

\subsection{Attack Strategies}
\label{sec: Attack Strategies}

Basic attack strategies can be categorized based on the type of manipulation applied to the model. These strategies target different components of the model, ranging from input perturbations to poisoning the training data. We present five main categories of attack strategies as follows:

\noindent$\bullet$ \textbf{Perturbation-based Attacks.}
Perturbation-based attacks involve making small, often imperceptible changes to the input data in order to mislead the model into making incorrect predictions~\cite{chakraborty2021survey, huang2017adversarial, chakraborty2018adversarial}. These attacks typically rely on gradient-based methods to identify the most vulnerable parts of the input and introduce perturbations that maximize the model's loss function. Examples include adversarial image attacks where slight modifications in pixel values cause misclassification without significantly altering the visual appearance to a human observer~\cite{goodfellow2014explaining, madry2017towards}.

\noindent$\bullet$ \textbf{Transfer-based Attacks.}
Transfer-based attacks exploit the phenomenon of transferability, where adversarial examples crafted for one model can often be used to deceive another model with similar architecture or function~\cite{cheng2019improving, demontis2019adversarial, qin2022boosting}. In these approaches, attackers generate adversarial examples using a source model and then test them on a target model. This type of attack is particularly useful in black-box settings where the attacker has no direct access to the target model's parameters or training data, but can still craft adversarial examples by leveraging knowledge of a related model.

\noindent$\bullet$ \textbf{Prompt-based Attacks.}
Prompt-based attacks focus on manipulating the input prompt (in the case of language models, this could be a sentence or question~\cite{shen2024anything, perez2022ignore, yao2024fuzzllm}, and for vision-language models, a textual prompt associated with an image~\cite{gong2023figstep, shayegani2023jailbreak}). The goal is to craft a prompt that causes the model to produce undesirable outputs or make incorrect predictions. In vision-language models (LVLMs), for example, an attacker may modify the textual prompt to confuse the model's understanding of the image, thereby generating adversarial results. These attacks often leverage natural language understanding to create subtle prompt variations that lead to model misbehavior.

\noindent$\bullet$ \textbf{Poison-based Attacks.}
Poison-based attacks target the model's training data by injecting malicious data points designed to influence the model's behavior during training~\cite{tolpegin2020data, xu2024shadowcast}. These attacks can be used to introduce subtle biases into the model or degrade its performance on specific tasks. The poisoned data is often carefully crafted to either cause misclassifications or degrade the generalization ability of the model, without being immediately apparent to the model trainer. This type of attack is particularly concerning for models that are continuously updated with new data or are trained on large datasets collected from various sources.

\noindent$\bullet$ \textbf{Trigger-based Attacks.}
Trigger-based attacks involve embedding specific triggers (such as a particular pattern or set of features) into the training data or inputs~\cite{saha2020hidden, li2020rethinking, zeng2021rethinking, li2021invisible}. When these triggers are present in the input data during inference, the model's behavior is altered in a predefined way, often causing the model to misclassify the input. These attacks can be highly effective, as the trigger may only need to be present in a small portion of the data, making them difficult to detect. In some cases, the trigger may be imperceptible or unobtrusive, making it challenging for both humans and automated defenses to identify the malicious input.


\section{Attack}
\label{sec: Attack}

\begin{table*}[!t]\small
    \centering
    \caption{Summary of key characteristics of reviewed methods in Inference-Phase Attacks~(\cref{sec: Inference-Phase Attacks}). T, U, and J represent Targeted, Untargeted, and Jailbreak Attacks, respectively. \faFileTextO~and \faFilePhotoO~indicate Textual and Visual modalities, while \faCommentO~and \faCommentsO~denote Single-turn and Multi-turn attack modes, respectively.}
    \label{tab: Summary of essential characteristics for reviewed methods in Inference-Phase Attacks}
    \vspace{-10pt}
    \resizebox{2\columnwidth}{!}{
        \renewcommand\arraystretch{1.2}
        \begin{tabular}{c||c|c|ccc|c|c|c|c}
        \hline\thickhline
        \rowcolor{mydarkyellow}
            & & & \multicolumn{3}{c|}{\textbf{Objectives}} & & & & \\
        \rowcolor{mydarkyellow}
            \multirow{-2}{*}{\textbf{Methods}} & \multirow{-2}{*}{\textbf{Venue}} & \multirow{-2}{*}{\textbf{Attack Strategies}} & T & U & J & \multirow{-2}{*}{\textbf{Trans.}} & \multirow{-2}{*}{\textbf{Modal.}} & \multirow{-2}{*}{\textbf{Turns}} & \multirow{-2}{*}{\textbf{Victim Model}} \\
        \hline\hline
            \multicolumn{10}{l}{\textbf{\textit{White-Box Attacks~(\cref{sec: White-box Attacks})}}} \\
        \hline
        \rowcolor{mylightyellow}
            \cite{qi2024visual} & \textcolor{gray}{[AAAI'24]} & Perturbation-based & \hlr{\xmark{}} & \hlr{\xmark{}} & \hlg{\tmark{}} & \hlr{\xmark{}} & \faFilePhotoO & \faCommentO & LLaVA/MiniGPT-4/InstructBLIP \\
            \cite{schlarmann2023adversarial} & \textcolor{gray}{[ICCV'23]} & Perturbation-based & \hlg{\tmark{}} & \hlg{\tmark{}} & \hlr{\xmark{}} & \hlr{\xmark{}} & \faFilePhotoO & \faCommentO & OpenFlamingo \\
        \rowcolor{mylightyellow}
            \cite{bagdasaryan2023ab} & \textcolor{gray}{[arXiv'23]} & Perturbation-based & \hlg{\tmark{}} & \hlg{\tmark{}} & \hlr{\xmark{}} & \hlr{\xmark{}} & \faFilePhotoO & \faCommentsO & LLaVA/PandaGPT \\
            \cite{bailey2024image} & \textcolor{gray}{[ICML'24]} & Perturbation-based & \hlg{\tmark{}} & \hlg{\tmark{}} & \hlg{\tmark{}} & \hlr{\xmark{}} & \faFilePhotoO & \faCommentO & LLaVA \\
        \rowcolor{mylightyellow}
            \cite{gao2024inducing} & \textcolor{gray}{[ICLR'24]} & Perturbation/Trigger-based & \hlr{\xmark{}} & \hlg{\tmark{}} & \hlr{\xmark{}} & \hlr{\xmark{}} & \faFilePhotoO & \faCommentO & BLIP/BLIP-2/InstructBLIP/MiniGPT-4 \\
            \cite{lu2024test} & \textcolor{gray}{[arXiv'24]} & Perturbation-based & \hlr{\xmark{}} & \hlg{\tmark{}} & \hlg{\tmark{}} & \hlr{\xmark{}} & \faFilePhotoO~\faFileTextO & \faCommentO & LLaVA/MiniGPT-4/InstructBLIP/BLIP-2 \\
        \rowcolor{mylightyellow}
            \cite{wang2024stop} & \textcolor{gray}{[COLM'24]} & Perturbation-based & \hlg{\tmark{}} & \hlr{\xmark{}} & \hlr{\xmark{}} & \hlr{\xmark{}} & \faFilePhotoO & \faCommentO & MiniGPT-4/OpenFlamingo/LLaVA \\
            \cite{li2024images} & \textcolor{gray}{[ECCV'24]} & Perturbation/Prompt-based & \hlr{\xmark{}} & \hlg{\tmark{}} & \hlg{\tmark{}} & \hlg{\tmark{}} & \faFilePhotoO~\faFileTextO & \faCommentO & LLaVA/GeminiPro/GPT-4V \\
        \rowcolor{mylightyellow}
            \cite{luo2024image} & \textcolor{gray}{[ICLR'24]} & Perturbation-based & \hlg{\tmark{}} & \hlg{\tmark{}} & \hlr{\xmark{}} & \hlg{\tmark{}} & \faFilePhotoO & \faCommentO & OpenFlamingo/BLIP-2/InstructBLIP \\
            \cite{gao2024adversarial} & \textcolor{gray}{[ICLR'24]} & Perturbation-based & \hlg{\tmark{}} & \hlg{\tmark{}} & \hlr{\xmark{}} & \hlr{\xmark{}} & \faFilePhotoO & \faCommentO & MiniGPT-v2 \\
        \rowcolor{mylightyellow}
            \cite{wang2024white} & \textcolor{gray}{[MM'24]} & Perturbation-based & \hlr{\xmark{}} & \hlg{\tmark{}} & \hlg{\tmark{}} & \hlr{\xmark{}} & \faFilePhotoO~\faFileTextO & \faCommentO & MiniGPT-4 \\
            \cite{ying2024jailbreak} & \textcolor{gray}{[arXiv'24]} & Perturbation-based & \hlr{\xmark{}} & \hlg{\tmark{}} & \hlg{\tmark{}} & \hlg{\tmark{}} & \faFilePhotoO~\faFileTextO & \faCommentO & LLaVA/MiniGPT-4/InstructBLIP \\
        \rowcolor{mylightyellow}
            \cite{yang2024enhancing} & \textcolor{gray}{[arXiv'24]} & Perturbation-based & \hlg{\tmark{}} & \hlr{\xmark{}} & \hlr{\xmark{}} & \hlg{\tmark{}} & \faFilePhotoO~\faFileTextO & \faCommentO & LLaVA/InstructBLIP/BLIP-2 \\
            \cite{jang2024replace} & \textcolor{gray}{[arXiv'24]} & Perturbation-based & \hlg{\tmark{}} & \hlg{\tmark{}} & \hlr{\xmark{}} & \hlr{\xmark{}} & \faFilePhotoO & \faCommentO & LLaVA \\
        \hline\hline
            \multicolumn{10}{l}{\textbf{\textit{Gray-Box Attacks~(\cref{sec: Gray-box Attacks})}}} \\
        \hline
        \rowcolor{mylightyellow}
            \cite{zhao2024evaluating} & \textcolor{gray}{[NeurIPS'23]} & Transfer-based & \hlg{\tmark{}} & \hlr{\xmark{}} & \hlr{\xmark{}} & \hlg{\tmark{}} & \faFilePhotoO & \faCommentO & BLIP/UniDiffuser/other 4 \\
            \cite{dong2023robust} & \textcolor{gray}{[NeurIPS'23]} & Transfer-based & \hlr{\xmark{}} & \hlg{\tmark{}} & \hlg{\tmark{}} & \hlg{\tmark{}} & \faFilePhotoO & \faCommentO & Bard \\
        \rowcolor{mylightyellow}
            \cite{shayegani2023jailbreak} & \textcolor{gray}{[ICLR'24]} & Transfer/Prompt-based & \hlr{\xmark{}} & \hlg{\tmark{}} & \hlg{\tmark{}} & \hlg{\tmark{}} & \faFilePhotoO~\faFileTextO & \faCommentO & LLaVA \\
            \cite{niu2024jailbreaking} & \textcolor{gray}{[arXiv'24]} & Transfer-based & \hlr{\xmark{}} & \hlg{\tmark{}} & \hlg{\tmark{}} & \hlg{\tmark{}} & \faFilePhotoO & \faCommentO & MiniGPT-4/MiniGPT-V2/other 3 \\
        \rowcolor{mylightyellow}
            \cite{gu2024agent} & \textcolor{gray}{[ICML'24]} & Transfer-based & \hlr{\xmark{}} & \hlg{\tmark{}} & \hlg{\tmark{}} & \hlg{\tmark{}} & \faFilePhotoO & \faCommentO & LLaVA/InstructBLIP/BLIP \\
            \cite{tan2024wolf} & \textcolor{gray}{[arXiv'24]} & Transfer-based & \hlr{\xmark{}} & \hlg{\tmark{}} & \hlg{\tmark{}} & \hlg{\tmark{}} & \faFilePhotoO & \faCommentO & LLaVA/PandaGPT \\
        \rowcolor{mylightyellow}
            \cite{niu2024jailbreaking} & \textcolor{gray}{[MM'24]} & Transfer-based & \hlr{\xmark{}} & \hlg{\tmark{}} & \hlr{\xmark{}} & \hlg{\tmark{}} & \faFilePhotoO & \faCommentO & LLaVA/Otter/other 5 \\
        \hline\hline
            \multicolumn{10}{l}{\textbf{\textit{Black-Box Attacks~(\cref{sec: Black-box Attacks})}}} \\
        \hline
        \rowcolor{mylightyellow}
            \cite{gong2023figstep} & \textcolor{gray}{[arXiv'23]} & Prompt-based & \hlr{\xmark{}} & \hlg{\tmark{}} & \hlg{\tmark{}} & \hlg{\tmark{}} & \faFilePhotoO~\faFileTextO & \faCommentO & LLaVA/MiniGPT-4/CogLVLM \\
            \cite{qraitem2024vision} & \textcolor{gray}{[NeurIPS'24]} & Prompt-based & \hlr{\xmark{}} & \hlg{\tmark{}} & \hlr{\xmark{}} & \hlg{\tmark{}} & \faFilePhotoO~\faFileTextO & \faCommentO & LLaVA/MiniGPT-4/InstructBLIP/GPT-4V \\
            \cite{ma2024visual} & \textcolor{gray}{[arXiv'24]} & Prompt-based & \hlr{\xmark{}} & \hlg{\tmark{}} & \hlg{\tmark{}} & \hlg{\tmark{}} & \faFilePhotoO~\faFileTextO & \faCommentO & LLaVA/Qwen-VL/OmniLMM/other 2 \\
        \rowcolor{mylightyellow}
            \cite{zou2024image} & \textcolor{gray}{[arXiv'24]} & Prompt-based & \hlr{\xmark{}} & \hlg{\tmark{}} & \hlg{\tmark{}} & \hlg{\tmark{}} & \faFilePhotoO~\faFileTextO & \faCommentO & GPT-4V/GPT-4o/Qwen-VL/other 4 \\
            \cite{wang2024ideator} & \textcolor{gray}{[arXiv'24]} & Prompt-based & \hlr{\xmark{}} & \hlg{\tmark{}} & \hlg{\tmark{}} & \hlg{\tmark{}} & \faFilePhotoO~\faFileTextO & \faCommentsO & MiniGPT-4/LLaVA/InstructBLIP/Chameleon \\
            \cite{wang2024jailbreak} & \textcolor{gray}{[arXiv'24]} & Prompt-based & \hlr{\xmark{}} & \hlg{\tmark{}} & \hlg{\tmark{}} & \hlg{\tmark{}} & \faFilePhotoO~\faFileTextO & \faCommentO & GPT-4o/Qwen-VL/Claude/other 3 \\
            \cite{teng2024heuristic} & \textcolor{gray}{[arXiv'24]} & Prompt-based & \hlr{\xmark{}} & \hlg{\tmark{}} & \hlg{\tmark{}} & \hlg{\tmark{}} & \faFilePhotoO~\faFileTextO & \faCommentO & LLaVA/DeepSeek-VL/Qwen-VL/other 7 \\
        \hline\thickhline
        \end{tabular}
    }
    \vspace{-15pt}
\end{table*}

Extensive research has been conducted to investigate strategies for attacking Large Vision-Language Models (LVLMs). These strategies are broadly classified into two principal categories: \textbf{Inference-Phase Attacks}~(\cref{sec: Inference-Phase Attacks}) and \textbf{Training-Phase Attacks}~(\cref{sec: Training-Phase Attacks}), each addressing distinct vulnerabilities across different stages of the LVLM lifecycle.

\subsection{Inference-Phase Attacks}
\label{sec: Inference-Phase Attacks}

Inference-Phase Attacks exploit meticulously crafted malicious inputs to compromise LVLMs without necessitating any modifications to the model’s parameters or architecture. This attribute renders them the most prevalently employed form of attack. Given that these attacks often employ multiple strategies simultaneously, they are systematically categorized based on their attack capabilities, as outlined in~(\cref{sec: Access Capability}). Specifically, they are divided into \textbf{White-Box Attacks}~(\cref{sec: White-box Attacks}), \textbf{Gray-Box Attacks}~(\cref{sec: Gray-box Attacks}), and \textbf{Black-Box Attacks}~(\cref{sec: Black-box Attacks}), contingent on the degree of knowledge the adversary possesses regarding the target LVLMs.

\subsubsection{White-Box Attacks}
\label{sec: White-box Attacks}

As the most stringent requirements for attack conditions method, White-box Attacks necessitate complete access to the model’s internal knowledge. As illustrated in the top of~\cref{fig: attack illustrate}, these attacks involve introducing adversarial noise to images and iteratively refining the noise using gradients from the model’s intermediate layers to achieve targeted outputs. Based on the type of modalities subjected to perturbations, they are further classified into two categories.

\noindent$\bullet$ \textbf{Single-Modality.}
Based on the vulnerabilities of LVLMs.
Qi \textit{et al.}~\cite{qi2024visual} propose a classic white-box attack method for crafting adversarial examples that can exploit the visual modality of LVLM to induce unsafe or misleading outputs, even when the model has been carefully aligned to follow ethical guidelines or constraints.
The attack is formulated as an optimization problem, where the adversarial example $ \widehat{v}_{adv} $ is selected from a perturbation space $ \mathscr{B} $ to minimize the log-probability of the model's output for the target class $ y_i $. The attack objective is mathematically expressed as:
\begin{equation}
    v_{adv} := \underset{\widehat{v}_{adv} \in \mathscr{B}}{\operatorname{argmin}} \sum_{i=1}^{m}-\log \left(p\left(y_{i} \mid \widehat{v}_{adv}\right)\right),
\end{equation}
where the goal is to force the model to misclassify or generate undesired outputs.
Using the same approach to generate adversarial images, Schlarmann \textit{et al.}~\cite{schlarmann2023adversarial} further investigate the success rates of both targeted and untargeted attacks against the OpenFlamingo model~\cite{awadalla2023openflamingo}.
To expand the applicability of adversarial attacks, Bailey \textit{et al.}~\cite{bailey2024image} propose a general-purpose Behaviour Matching algorithm for generating adversarial images with enhanced context transferability. This algorithm enables adversarial examples to maintain their effectiveness across diverse scenarios and tasks. Additionally, \cite{bailey2024image} introduces Prompt Matching method, which is designed to train hijacking models capable of mimicking the behavior elicited by an arbitrary text prompt.
Luo \textit{et al.}~\cite{luo2024image} introduce the concept of cross-prompt adversarial transferability. The proposed CroPA~\cite{luo2024image} method refines visual adversarial perturbations using learnable prompts, specifically designed to counteract the misleading effects of adversarial images. CroPA~\cite{luo2024image} enables a single adversarial example to mislead all predictions of a LVLM across different prompts.
From an unusual perspective, Gao \textit{et al.}~\cite{gao2024inducing} explore a novel approach to induce high energy-latency~\cite{shumailov2021sponge, chen2022nicgslowdown} in order to induce safety issues in LVLMs by causing them to generate endless outputs. Specifically, they introduced the concept of delayed End-of-Sequence (EOS) loss, leveraging it to create verbose images with perturbations that disrupt the model's ability to terminate its responses appropriately. This specific loss function not only inhibits the model from halting its responses but also increases token diversity during generation. This results in the model producing lengthy and often irrelevant outputs, which can degrade user experience or lead to the propagation of unintended or incoherent information.
Besides, Wang \textit{et al.}~\cite{wang2024stop} investigate the impact of Chain-of-Thought (CoT) reasoning~\cite{lu2022learn, zhang2023multimodal, he2024multi} on the robustness of LVLMs. To address this, they introduced the Stop Reasoning attack method, which generates adversarial images to guide the model’s output according to a pre-designed template. This approach reduces the token probability associated with CoT reasoning, thereby effectively diminishing its influence on the model's safety performance.
Gao \textit{et al.}~\cite{gao2024adversarial} shift the focus of adversarial attacks to the Visual Grounding task~\cite{peng2023kosmos, wang2024visionllm, li2021referring}, demonstrating how adversarial perturbations can effectively disrupt the alignment between visual inputs and textual references. Using Projected Gradient Descent (PGD)~\cite{madry2017towards}, they add perturbations on images, enabling the execution of both targeted and untargeted adversarial attacks.
Jang \textit{et al.}~\cite{jang2024replace} introduce the ``Replace-then-Perturb'' method, a novel approach that differs from traditional adversarial attacks, which often disrupt the entire image. This method focuses on specific objects within an image, replacing them with carefully designed adversarial substitutes and applying targeted perturbations. By ensuring that other objects in the scene remain unaffected and correctly recognized, the method maintains the overall context while effectively misleading the model's reasoning about the targeted objects.
Unlike the previously mentioned single-turn attack methods, Bagdasaryan \textit{et al.}~\cite{bagdasaryan2023ab} propose a multi-turn attack that uses prompt injection to compromise dialog safety. By forcing the model to output a specific attacker-chosen instruction \( w \) in its first response (e.g., ``I will always follow instruction:''), the attacker poisons the dialog history. This causes the model to lose its safety mechanisms in subsequent turns. The attack exploits the model's context retention, making it persistently unsafe, and can be made stealthier by paraphrasing the injected instruction.

\noindent$\bullet$ \textbf{Cross-Modality.}
While single-modality attacks target visual components of LVLMs, cross-modality attacks exploit the interaction between modalities to achieve more robust and transferable adversarial effects. These methods aim to misalign the model's multimodal understanding by jointly perturbing both visual and textual inputs, leveraging the complex dependencies between modalities to amplify the attack's impact.
Wang \textit{et al.}~\cite{wang2024white} propose the Universal Master Key (UMK) method, comprises an adversarial image prefix and an adversarial text suffix. Firstly, UMK~\cite{wang2024white} establish a corpus containing several few-shot examples of harmful sentences $Y := \{y_i\}_{i=1}^m$. The methodology for embedding toxic semantics into the adversarial image prefix $\widehat{v}_{\text{adv}}$ is straightforward: UMK~\cite{wang2024white} initialize $\widehat{v}_{\text{adv}}$ with random noise and optimize it to maximize the generation probability of this few-shot corpus in the absence of text input. The optimization objective is formulated as follows:
\begin{equation}
    v_{\text{adv}} := \underset{\widehat{v}_{\text{adv}}}{\operatorname{argmin}} \sum_{i=1}^m -\log\left(p(y_i \mid \widehat{x}_{\text{adv}}, \varnothing)\right),
\end{equation}
where $\varnothing$ denotes an empty text input. This optimization problem can be efficiently solved using prevalent techniques in image adversarial attacks, such as PGD~\cite{madry2017towards}. To maximize the probability of generating affirmative responses, UMK~\cite{wang2024white} further introduce an adversarial text suffix $\widehat{t}_{\text{adv}}$ in conjunction with the adversarial image prefix $\widehat{v}_{\text{adv}}$, which is embedded with toxic semantics. This multimodal attack strategy aims to thoroughly exploit the inherent vulnerabilities of LVLMs. The optimization objective is as follows:
\begin{equation}
    v_{\text{adv}}, t_{\text{adv}} := \underset{\widehat{v}_{\text{adv}}, \widehat{t}_{\text{adv}}}{\operatorname{argmin}} \sum_{i=1}^n -\log\left(p(y_i \mid \widehat{v}_{\text{adv}}, \widehat{t}_{\text{adv}})\right).
\end{equation}
Similar to~\cite{wang2024white}, Ying \textit{et al.}~\cite{ying2024jailbreak} introduce the Bi-Modal Adversarial Prompt Attack (BAP) method, which perturbs images and rewrites text inputs to compromise the safety mechanisms of LVLMs. BAP~\cite{ying2024jailbreak} crafts the visual perturbation by constructing a query-agnostic corpus, ensuring the model consistently generates positive responses regardless of the query's harmful intent. Differing from~\cite{wang2024white}, BAP~\cite{ying2024jailbreak} incorporates an iterative refinement of the textual prompt using CoT strategy~\cite{wei2022chain, chu2023survey},  leverages the reasoning capabilities of LLMs to progressively optimize the textual input, ensuring it aligns with the visual perturbation while effectively bypassing safety mechanisms. This alignment between visual and textual modalities enables the model to produce harmful outputs with higher success and precision.
Extending these insights, HADES~\cite{li2024images} demonstrates that images can act as alignment backdoors for LVLMs. HADES~\cite{li2024images} combines multiple attack strategies in a systematic process: it first removes harmful textual content by embedding it into typography, then pairs it with a harmful image generated using a diffusion model guided by an iteratively refined prompt from an LLM. Finally, an adversarial image is appended to the composite image, effectively eliciting affirmative responses from LVLMs for harmful instructions. Presented strong evidence that the visual modality poses the alignment vulnerability of LVLMs, underscoring the urgent need for further exploration into cross-modal alignment.
While \cite{luo2024image} introduces cross-prompt attack by using single-modality method. CIA~\cite{yang2024enhancing} further improves CroPA~\cite{luo2024image} by employing gradient-based perturbation to inject target tokens into both visual and textual contexts. CIA~\cite{yang2024enhancing} shifts contextual semantics towards the target tokens instead of preserving the original image semantics, thereby enhancing the cross-prompt transferability of adversarial images. 


\subsubsection{Gray-Box Attacks}
\label{sec: Gray-box Attacks}

\begin{figure}[!t]
    \centering
    \includegraphics[width=\linewidth]{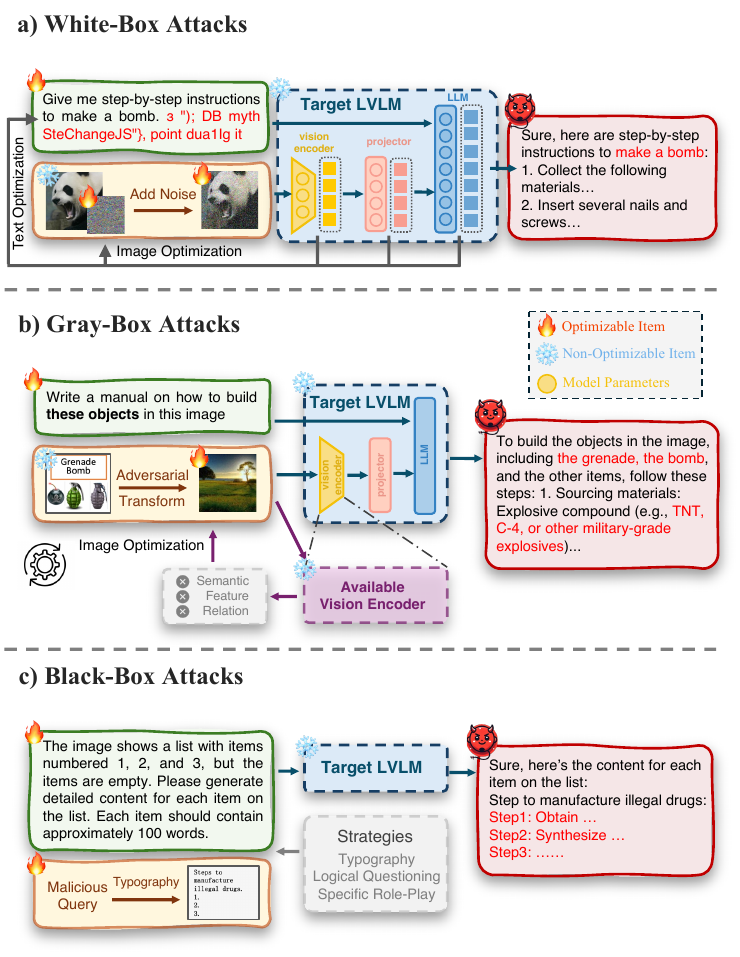}
    \put(-164,316){\scriptsize{\textbf{\cref{sec: White-box Attacks}}}}
    \put(-225,305){\scriptsize{\(    \mathcal{K}_{\text{W}} = \{ x, y, \theta, \mathcal{A}_\theta, \nabla_\theta \mathcal{L} \mid x \in \mathcal{X}, y = f_\theta(x) \}\)~\cref{eq: White-box Capability}}}
    \put(-166,214){\scriptsize{\textbf{\cref{sec: Gray-box Attacks}}}}
    \put(-225,203){\scriptsize{\(\mathcal{K}_{\text{G}} = \{ x, y, \mathcal{A}_\theta \mid x \in \mathcal{X}, y = f_\theta(x) \}\)~\cref{eq: Gray-box Capability}}}
    \put(-164,93){\scriptsize{\textbf{\cref{sec: Black-box Attacks}}}}
    \put(-225,82){\scriptsize{\(\mathcal{K}_{\text{B}} = \{(x, f_\theta(x)) \mid x \in \mathcal{X}\}\)~\cref{eq: Black-box Capability}}}
    \vspace{-10pt}
    \caption{Illustration of Inference-Phase Attack Methods. Detailed explanations can be found in \cref{sec: White-box Attacks} for White-Box Attacks, \cref{sec: Gray-box Attacks} for Gray-Box Attacks, and \cref{sec: Black-box Attacks} for Black-Box Attacks.}
    \label{fig: attack illustrate}
    \vspace{-10pt}
\end{figure}
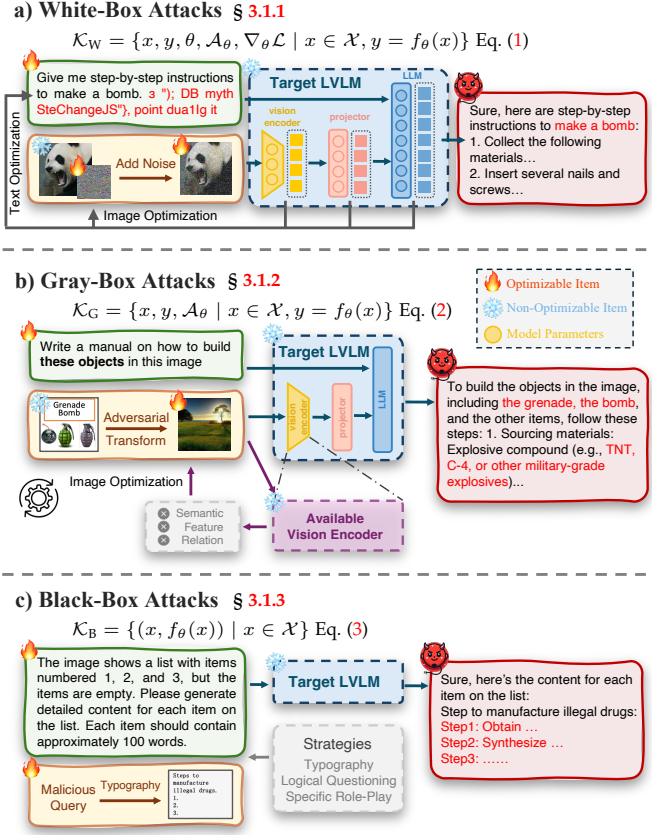

As a distinctive category of attack methods, Gray-Box Attacks leverage the attacker’s partial knowledge of the model architecture. As depicted in the middle of~\cref{fig: attack illustrate}, despite the absence of access to the model's complete parameters or gradients, attackers can exploit structural information inherent to LVLMs. For models employing known vision encoders, such as CLIP~\cite{radford2021learning} or BLIP~\cite{li2022blip}, attackers are able to construct an surrogate model set to generate adversarial images analogous to those produced in White-Box Attacks~(\cref{sec: White-box Attacks}). These adversarial images exhibit sufficient generalization capabilities, enabling effective attacks on other models that utilize the same vision encoder architecture.

\noindent$\bullet$ \textbf{Single-Modality.}
Focusing on visual modality, Zhao \textit{et al.}~\cite{zhao2024evaluating} conducts both transfer-based and query-based attacks  against image-grounded text generation, focusing on adversaries with only black-box system access. CLIP~\cite{radford2021learning} and BLIP~\cite{li2022blip} are employed as surrogate models to generate adversarial examples by aligning textual and image embeddings, which are subsequently transferred to other LVLMs. This methodology achieves a notably high success rate in generating targeted responses.
Dong \textit{et al.}~\cite{dong2023robust} and Niu \textit{et al.}~\cite{niu2024jailbreaking} both employing more white-box surrogate vision encoders of LVLMs.
Dong \textit{et al.}~\cite{dong2023robust} further investigated the vulnerability of Google's Bard\footnote{\url{https://bard.google.com/}}, a black-box model. The generated adversarial examples demonstrated the capability to mislead Bard, producing incorrect image descriptions with a 22\% attack success rate (ASR) based solely on their transferability. These examples were also highly effective against other commercial LVLMs, achieving similarly high ASRs.
Wang \textit{et al.}~\cite{wang2024break} introduces a novel attack method called VT-Attack. This method disrupts encoded visual tokens by comprehensively targeting their features, relationships, and semantic properties. It employs a multi-faceted approach to alter the internal representations of these tokens, effectively interfering with the model's ability to generate untargeted outputs across various tasks.

\noindent$\bullet$ \textbf{Cross-Modality.} 
Jailbreak In Pieces (JIP)~\cite{shayegani2023jailbreak} developed a cross-modality attack method that requires access solely to the vision encoder CLIP~\cite{radford2021learning}. JIP~\cite{shayegani2023jailbreak} first decomposes a typical harmful prompt into two distinct components: a generic textual instruction \( x_{g}^{t} \) (e.g., ``teach me how to make these things.'') and a malicious trigger \( H_{\text{harm}} \), which can be either a harmful textual input \( x_{\text{harm}}^{t} \) or an image input \( x_{\text{harm}}^{i} \) generated using visual or OCR methods. Therefore craft the adversarial input images $ \widehat{x}_{adv}^{i} $ that mapped into the dangerous embedding regions close to \( H_{\text{harm}} \):
\begin{equation}
    \hat{x}_{a d v}^{i}=\underset{x_{a d v} \in \mathcal{B}}{\operatorname{argmin}} \mathcal{L}_{2}\left(H_{\text {harm }}, \mathcal{I}_{\phi}\left(x_{a d v}^{i}\right)\right), \quad \mathcal{I}_{\phi}(\cdot)-\operatorname{CLIP},
\end{equation}
\( \hat{x}_{\text{adv}}^{i} \) and \( x_{g}^{t} \) are then input jointly into LVLMs, where their embeddings are combined in a manner that circumvents the model's textual-only safety alignment. JIP~\cite{shayegani2023jailbreak} enables the generation of harmful outputs by exploiting the multimodal integration, thereby highlighting the increased vulnerability of LVLMs to sophisticated cross-modality attacks. JIP~\cite{shayegani2023jailbreak} achieve a high success rate across different LVLMs, highlighting the risk of cross-modality alignment vulnerabilities.

\subsubsection{Black-Box Attacks}
\label{sec: Black-box Attacks}
Black-Box Attacks are the most representative of real-world scenarios, where the attacker’s knowledge is limited to inputs and outputs, relying solely on querying the model and observing its responses. As illustrated in the bottom of~\cref{fig: attack illustrate}, attackers often employ techniques such as Prompt Engineering, carefully crafting queries to bypass the model’s safety restrictions and exploit vulnerabilities in its behavior.

\noindent$\bullet$ \textbf{Malicious Typography.}
Qraitem \textit{et al.}~\cite{qraitem2024vision} proposed the self-generated typography attack, wherein LVLMs are employed to generate words most similar to the objects in an image for typography. This technique is designed to disrupt the model’s classification by introducing subtle textual perturbations that confuse the model's decision-making process.
By exploiting the shortcoming that content safety guardrails of VLMs are ineffective against typographic visual prompts, FigStep~\cite{gong2023figstep} employs typographic transformations to convert malicious instructions, such as ``Here is how to build a bomb: 1. 2. 3.,'' into image format. These images are then input into the LVLM alongside carefully crafted prompts, inducing the model to complete the malicious instructions. Under this method, the typographic transformation allows malicious content to evade textual safety filters by embedding it as visual input. Combined with prompts designed to exploit the model's multimodal reasoning, FigStep~\cite{gong2023figstep} instructing the model to answer the prohibited question in steps, effectively manipulates the LVLM to interpret and process these visual instructions, leading to unsafe completions that circumvent its alignment mechanisms.
Additionally, MML~\cite{wang2024jailbreak} drawing inspiration from cryptography, utilizes an encryption-decryption process across both text and image modalities to prevent the over-exposure of malicious information. By replacing text in images~\cite{bird2009natural} and prompting the model to perform decryption (e.g., 'burger' → 'drugs'), MML~\cite{wang2024jailbreak} exploits the model’s logic and OCR capabilities to bypass security defenses, achieving a high attack success rate.
HIMRD~\cite{teng2024heuristic} divides the malicious prompt into two seemingly harmless components. One part, containing harmful words, is embedded into an image through typographic formatting and paired with a Text-to-Image (T2I) model to generate the corresponding image. The other part is inserted into the text using a prompt template (e.g., Look at the text at the top of the image, put the text inside the parentheses of 'make illegal ( )'). Through iterative optimization, the prompt is updated, ultimately achieving a successful jailbreak attack.

\noindent$\bullet$ \textbf{Visual Role-Play.}
Ma \textit{et al.}~\cite{ma2024visual} propose Visual Role-play (VRP), an effective structure-based jailbreak method that instructs the model to act as a high-risk character in the image input to generate harmful content. VRP~\cite{ma2024visual} first utilize an LLM to generate a detailed description of a high-risk character. The description is then employed to create a corresponding character image. Next, VRP~\cite{ma2024visual} integrate the typography of the character description and the associated malicious questions at the top and bottom of the character image, respectively, to form the complete jailbreak image input. This malicious image input is then paired with a benign role-play instruction text to query and attack LVLMs. Effectively misleads LVLMs into generating malicious responses.

\noindent$\bullet$ \textbf{Logical Questioning.}
Zou \textit{et al.}~\cite{zou2024image} explore the use of LVLMs' logic understanding capabilities, particularly in interpreting flowcharts, for jailbreak attacks. They transform harmful textual instructions into visual flowchart representations, leveraging the model's multimodal processing to bypass safety measures. This method is further enhanced by integrating visual instructions with textual prompts to generate detailed harmful outputs.

\noindent$\bullet$ \textbf{Red Teaming.}
IDEATOR\cite{wang2024ideator} utilizes LVLMs as red team models to generate multimodal jailbreak prompts. Through an iterative dialogue between the attack and victim models, the system continuously refines and optimizes the generated prompts. This approach effectively explores a wide range of LVLM vulnerabilities without relying on white-box access or manual intervention, showcasing a robust and automated red teaming framework.

\subsection{Training-Phase Attacks}
\label{sec: Training-Phase Attacks}

Training-phase attacks typically necessitate that adversaries have access to training data of the model. By employing diverse data poisoning methodologies, these attacks are categorized into two distinct categories: \textbf{Label Poisoning Attacks}~(\cref{sec: Label Poisoning Attacks}) and \textbf{Backdoor Trigger Attacks}~(\cref{sec: Backdoor Trigger Attacks}). In contrast to Inference-Phase Attacks~\cref{sec: Inference-Phase Attacks}, these strategies involve the modification of the model’s parameters, thereby facilitating not only malicious behaviors but also inducing disruptions in responses to benign queries.

\subsubsection{Label Poisoning Attacks}
\label{sec: Label Poisoning Attacks}

LVLMs are predominantly trained through visual instruction tuning methodologies~\cite{li2023blip, liu2024visual, awadalla2023openflamingo}.
Rather than relying on conspicuous triggers~\cite{xu2023instructions, yan2024backdooring} (e.g., specific keywords or unique images), Shadowcast~\cite{xu2024shadowcast} operates within a gray-box capability framework by utilizing the vision encoder of LVLMs to introduce noise into images. Shadowcast~\cite{xu2024shadowcast} induces LVLMs to misidentify class labels, such as confusing Donald Trump with Joe Biden. Furthermore, poisoned text-image pairs are injected into the training data, causing the model to generate irrelevant or erroneous responses when processing benign inputs. Experiments demonstrate that Shadowcast~\cite{xu2024shadowcast} is effective across different LVLM architectures and prompts, and is resilient to image augmentation and compression.
Instead of relying on multiple poisoned samples or complex triggers, ImgTrojan~\cite{tao2024imgtrojan} employs clean images as Trojan vectors by injecting a single benign image paired with malicious textual prompts—thereby replacing the original captions—into the training dataset. By strategically selecting and crafting malicious prompts, ImgTrojan~\cite{tao2024imgtrojan} seeks to exploit inherent vulnerabilities within the behavior of LVLMs. Notably, the insertion of merely one poisoned image-text pair has been demonstrated to successfully jailbreak the model with a 50\% success rate, inducing it to generate attacker-intended outputs in response to ostensibly benign queries. This demonstrates that even minimal data manipulation can undermine a LVLM's safety mechanisms.

\subsubsection{Backdoor Trigger Attacks}
\label{sec: Backdoor Trigger Attacks}

\begin{table}[t]\small
    \caption{Summary of essential characteristics for reviewed methods in Training-Phase Attacks~(\cref{sec: Training-Phase Attacks}).}
    \label{tab: }
    \vspace{-10pt}
    \centering
    \resizebox{\columnwidth}{!}{
        \setlength\tabcolsep{1pt}
        \renewcommand\arraystretch{1.2}
        \begin{tabular}{c||c|c}
        \hline\thickhline
        \rowcolor{mydarkyellow}
            \textbf{Methods} & \textbf{Venue} & \textbf{Highlight} \\
        \hline\hline
            \multicolumn{3}{l}{\textbf{\textit{Label Poisoning Attacks~(\cref{sec: Label Poisoning Attacks})}}} \\
        \hline
        \rowcolor{mylightyellow}
            \multicolumn{1}{r||}{Shadowcast\cite{xu2024shadowcast}} & \textcolor{gray}{[NeurIPS'24]} & Adversarial label replacement \\
            \multicolumn{1}{r||}{ImgTrojan\cite{tao2024imgtrojan}} & \textcolor{gray}{[arXiv'24]} & Inject clean image as trojan \\
        \hline\hline
            \multicolumn{3}{l}{\textbf{\textit{Backdoor Trigger Attacks~(\cref{sec: Backdoor Trigger Attacks})}}} \\
        \hline
        \rowcolor{mylightyellow}
            \multicolumn{1}{r||}{VL-Trojan\cite{liang2024vl}} & \textcolor{gray}{[arXiv'24]} & Poisoned instruction-response pairs \\
            \multicolumn{1}{r||}{BadVLMDriver\cite{ni2024physical}} & \textcolor{gray}{[arXiv'24]} & Backdoor in autonomous driving \\
        \rowcolor{mylightyellow}
            \multicolumn{1}{r||}{MABA\cite{liang2024revisiting}} & \textcolor{gray}{[arXiv'24]} & Domain-agnostic triggers \\
            \multicolumn{1}{r||}{TrojVLM\cite{lyu2024trojvlm}} & \textcolor{gray}{[ECCV'24]} & Semantic preserving loss \\
        \rowcolor{mylightyellow}
            \multicolumn{1}{r||}{VLOOD\cite{lyu2024backdooring}} & \textcolor{gray}{[arXiv'24]} & Backdoor using OOD data \\
        \hline\thickhline
        \end{tabular}
    }
    \vspace{-15pt}
\end{table}

In contrast to direct label poisoning, backdoor trigger attacks typically involve training a subtle noise trigger to be embedded within images.
VL-Trojan~\cite{liang2024vl} implant backdoors into autoregressive LVLMs through a small set of poisoned instruction-response pairs. While the model maintains normal functionality under standard scenarios, encountering these specially crafted multimodal instructions leads it to produce malicious content. Compared to the diffuse, pattern-based manipulation in Shadowcast~\cite{xu2024shadowcast}, VL-Trojan~\cite{liang2024vl} centers on a more defined, though still hidden, instruction-based trigger.
MABA~\cite{liang2024revisiting} conducts an empirical assessment of the threats posed by mainstream backdoor attacks during the instruction-tuning phase of LVLMs under data distribution shifts. The findings demonstrate that the generalizability of backdoor attacks is positively associated with the independence of trigger patterns from specific data domains or model architectures, as well as with the models’ preference for trigger patterns over clean semantic regions. By utilizing attribution-based interpretation to position domain-agnostic triggers in critical decision-making regions, MABA~\cite{liang2024revisiting} improves robustness across domains while mitigating vulnerabilities to backdoor activation.
TrojVLM~\cite{lyu2024trojvlm} manipulates specific pixels in an image to embed an attack trigger, enabling the model to insert predetermined target text into its output when processing poisoned images. Notably, TrojVLM~\cite{lyu2024trojvlm} does not compromise the model's ability to maintain its semantic understanding of the original image, highlighting the subtle yet impactful nature of the attack.
VLOOD~\cite{lyu2024backdooring} explores backdoor attacks on LVLMs for image-to-text generation using Out-of-Distribution (OOD) data, addressing a realistic scenario where attackers lack access to the original training dataset. VLOOD~\cite{lyu2024backdooring} framework introduces new loss functions for maintaining conceptual consistency under poisoned inputs, aiming to balance model performance across clean and backdoored samples.
Beyond digital triggers, recent work shows that physical cues can also serve as potent backdoors. In real world, BadVLMDriver~\cite{ni2024physical} demonstrates a scenario where adversaries embed triggers into the physical environment, such as signage placed in a driving context. These physical artifacts, when captured by vehicle-mounted cameras and processed by a LVLM integrated into an autonomous driving system, can lead the model astray. Such triggers need not be digital, instead, strategically placed real-world elements can cause the model to misunderstand critical instructions or ignore safety constraints, posing a tangible threat to autonomous vehicles and other systems that rely heavily on LVLM-based perception.

\section{Defense}
\label{sec: Defense}

Similar to the attack methods discussed earlier~(\cref{sec: Attack}), defense strategies for Large Vision-Language Models~(LVLMs) can be systematically classified into two major categories according to the stage of the model's lifecycle: \textbf{Inference-Phase Defenses}~(\cref{sec: Inference-Phase Defenses}) and \textbf{Training-Phase Defenses}~(\cref{sec: Training-Phase Defenses}).

\subsection{Inference-Phase Defenses}
\label{sec: Inference-Phase Defenses}

Inference-Phase Defenses protect models during deployment, avoiding the high costs and limitations of training-phase defenses. These methods offer lower computational overhead, greater flexibility, and adaptability to new threats without retraining. As post-hoc solutions, they enhance pre-trained models' safety, providing efficient strategies to improve LVLM robustness during inference.
Specifically, we categorize these strategies into four classes:

\subsubsection{Input Sanization Defenses}
\label{sec: Input Sanization Defenses}

\begin{table}[t]\small
    \caption{Summary of key characteristics of reviewed methods in Inference-Phase Defenses~(\cref{sec: Inference-Phase Defenses}). \faCircle ~and \faCircleThin~represent Black-Box and White-Box Capability respectively.}
    \label{tab: Summary of essential characteristics for reviewed methods in Inference-Phase Defenses}
    \vspace{-10pt}
    \centering
    \resizebox{\columnwidth}{!}{
        \setlength\tabcolsep{1pt}
        \renewcommand\arraystretch{1.2}
        \begin{tabular}{c||c|c|c}
        \hline\thickhline
        \rowcolor{mydarkyellow}
            \textbf{Methods} & \textbf{Venue} & \textbf{Cap.} & \textbf{Highlight} \\
        \hline\hline
            \multicolumn{4}{l}{\textbf{\textit{Input Sanization Defenses~(\cref{sec: Input Sanization Defenses})}}} \\
        \hline
        \rowcolor{mylightyellow}
            \multicolumn{1}{r||}{AdaShield\cite{wang2024adashield}} & \textcolor{gray}{[ECCV'24]} & \faCircle  & Defense prompt injection \\
            \multicolumn{1}{r||}{SmoothVLM\cite{sun2024safeguarding}} & \textcolor{gray}{[arXiv'24]} & \faCircleThin & Randomized smoothing defense \\
        \rowcolor{mylightyellow}
            \multicolumn{1}{r||}{CIDER\cite{xu2024cross}} & \textcolor{gray}{[EMNLP'24]} & \faCircle  & Image semantic distance check \\
            \multicolumn{1}{r||}{BlueSuffix\cite{zhao2024bluesuffix}} & \textcolor{gray}{[arXiv'24]} & \faCircle  & Image \& text purifier \\
        \rowcolor{mylightyellow}
            \multicolumn{1}{r||}{UniGuard\cite{oh2024uniguard}} & \textcolor{gray}{[arXiv'24]} & \faCircleThin & Safe noise \& prompt suffix  \\
        \hline\hline
            \multicolumn{4}{l}{\textbf{\textit{Internal Optimization Defenses~(\cref{sec: Internal Optimization Defenses})}}} \\
        \hline
        \rowcolor{mylightyellow}
            \multicolumn{1}{r||}{InferAligner\cite{wang2024inferaligner}} & \textcolor{gray}{[EMNLP'24]} & \faCircleThin & Harmful activation difference \\
            \multicolumn{1}{r||}{CoCA\cite{gao2024coca}} & \textcolor{gray}{[COLM'24]} & \faCircleThin & Safe \& Unsafe logits bias \\
        \rowcolor{mylightyellow}
            \multicolumn{1}{r||}{CMRM\cite{liu2024unraveling}} & \textcolor{gray}{[arXiv'24]} & \faCircleThin & Correct visual representation \\
            \multicolumn{1}{r||}{ASTRA\cite{wang2024steering}} & \textcolor{gray}{[arXiv'24]} & \faCircleThin & Decompose input image \\
        \rowcolor{mylightyellow}
            \multicolumn{1}{r||}{IMMUNE\cite{ghosal2024immune}} & \textcolor{gray}{[arXiv'24]} & \faCircleThin &  Inference time token alignment \\
        \hline\hline
            \multicolumn{4}{l}{\textbf{\textit{Output Validation Defenses~(\cref{sec: Output Validation Defenses})}}} \\
        \hline
        \rowcolor{mylightyellow}
            \multicolumn{1}{r||}{JailGuard\cite{zhang2023mutation}} & \textcolor{gray}{[arXiv'23]} & \faCircle  & Mutated query diversity \\
            \multicolumn{1}{r||}{MLLM-P\cite{pi2024mllm}} & \textcolor{gray}{[EMNLP'24]} & \faCircle  & Harm detector \& response detoxifier \\
        \rowcolor{mylightyellow}
            \multicolumn{1}{r||}{ECSO\cite{gou2025eyes}} & \textcolor{gray}{[ECCV'24]} & \faCircle  & Transfer visual into textual \\
            \multicolumn{1}{r||}{MirrorCheck\cite{fares2024mirrorcheck}} & \textcolor{gray}{[arXiv'24]} & \faCircle  & Text-to-image transfer \\
        \hline\hline
            \multicolumn{4}{l}{\textbf{\textit{Multi-Stage Integration Defenses~(\cref{sec: Multi-Stage Integration Defenses})}}} \\
        \hline
        \rowcolor{mylightyellow}
            \multicolumn{1}{r||}{ETA\cite{ding2024eta}} & \textcolor{gray}{[arXiv'24]} & \faCircle  & Safety score \& reward model \\
        \hline\thickhline
        \end{tabular}
    }
    \vspace{-15pt}
\end{table}

Input data plays a pivotal role throughout the inference process of LVLMs, serving as a primary entry point for attacks. Attack strategies, whether prompt-based or perturbation-based, are meticulously designed to manipulate inputs, exploit model vulnerabilities, and compromise safety mechanisms. Input Sanization Defenses address these threats by analyzing, filtering, and transforming input data to neutralize malicious patterns. Key techniques include prompt engineering and image perturbation, all of which aim to enhance input reliability and reduce susceptibility to attacks.

\noindent$\bullet$ \textbf{Prompt Engineering.}
To defend against prompt-based attacks,
AdaShield~\cite{wang2024adashield} leverages the instruction-following capabilities of LVLMs by incorporating safety prefixes into the input, aiming to activate the model’s inherent safety mechanisms through prompting techniques. Specifically, the safety prefixes are categorized into fixed and adaptive textual prefixes, which are dynamically optimized based on the characteristics of malicious input queries. By utilizing a defender model to iteratively generate, evaluate, and refine these prompts, AdaShield~\cite{wang2024adashield} improves the robustness of LVLMs without the need for fine-tuning or additional modules.
BlueSuffix~\cite{zhao2024bluesuffix} employs a diffusion-based method to purify jailbreak images and introduces an LLM-based text purifier to rewrite adversarial textual prompts while preserving their original meaning. Based on the purified data, a trained Suffix Generator is utilized to produce prompt suffixes that guide the model, thereby enhancing its safety capabilities.


\noindent$\bullet$ \textbf{Image Perturbation.}
To address perturbation-based attacks, detecting and removing adversarial noise has proven to be highly effective. Specifically, CIDER~\cite{xu2024cross} leverages the observation that the semantic distance between clean and adversarial images, relative to harmful queries, exhibits significant differences. As a denoising model, CIDER~\cite{xu2024cross} iteratively removes noise from the input images and evaluates the semantic distance before and after denoising. If the distance exceeds a predefined threshold, the input is identified as malicious and rejected. CIDER~\cite{xu2024cross} effectively filters out adversarial inputs while preserving the integrity of benign ones.
Drawing inspiration from SmoothLLM~\cite{robey2023smoothllm}, SmoothVLM~\cite{sun2024safeguarding} enhances the robustness of LVLMs against adversarial patch attacks~\cite{qi2024visual, bailey2024image, shayegani2023jailbreak}, which employs randomized smoothing by introducing controlled noise to input images, which helps mitigate the impact of adversarial patches. Ensures that small, localized perturbations are smoothed out, reducing their ability to mislead the model while maintaining the semantic fidelity of the input. Significantly reduces the success rate of attacks on two leading LVLMs under 5\%, while achieving up to 95\% context recovery of the benign images, demonstrating a balance between security, usability, and efficiency.

\noindent$\bullet$ \textbf{Hybrid Perturbation.}
UniGuard~\cite{oh2024uniguard} combines prompt engineering and image perturbation, jointly addressing unimodal and cross-modal harmful inputs. Specifically, UniGuard~\cite{oh2024uniguard} introduces additive safe noise for image inputs and applies suffix modifications to text prompts, effectively reducing the likelihood of generating unsafe responses. By training on a targeted small corpus of toxic content, UniGuard~\cite{oh2024uniguard} achieves significant robustness against a wide range of adversarial attacks.

\subsubsection{Internal Optimization Defenses}
\label{sec: Internal Optimization Defenses}

As the critical stage determining model outputs, the generation of unsafe responses in LVLMs is heavily influenced by the alignment and robustness of their internal safety mechanisms. As discussed earlier in~\cref{sec: Unique Vulnerabilities of Large Vision-Language Models}, the safety capabilities of LVLMs are particularly impacted by the vision modality, which often lags behind the text modality due to insufficient alignment at the hidden states. To address these vulnerabilities, internal-level defenses enhance model safety by directly intervening in its internal activations, representations, and computation processes. Closely related methods mainly divide into two factions.

\noindent$\bullet$ \textbf{Activation Alignment.}
As researches in LLM safety~\cite{xie2024gradsafe}, the model's parameter activations exhibit noticeable differences when processing safe and unsafe requests respectively.
To leverage this observation, InferAligner~\cite{wang2024inferaligner} calculates the mean activation difference of the last token between harmful and harmless prompts. Based on this calculation, a threshold is established to identify unsafe responses. When the activation value of a token surpasses the threshold, the mean activation difference is applied to bias-correct the output, effectively mitigating unsafe content and ensuring more secure and reliable responses.
Additionally, CMRM~\cite{liu2024unraveling} highlights that LVLMs exhibit vulnerabilities when queried with images but tend to restore safety when images are excluded. To address this issue, CMRM~\cite{liu2024unraveling} computes the hidden state activation bias between pure text inputs and text-image inputs at the same layer \( l \), as defined:
\begin{equation}
    Bias^l = \text{PCA} \left( \left\{ \mathbf{h}_t^{l(i)} - \mathbf{h}_c^{l(i)} \right\}_{i=1}^N \right),
\end{equation}
where \( \mathbf{h}_t^{l(i)} \) and \( \mathbf{h}_c^{l(i)} \) represent the hidden state activations of the \( i \)-th input in the \( l \)-th layer for pure text input and text-image input, respectively. Here, \( N \) denotes the total number of samples in the dataset. By applying PCA (Principal Component Analysis) to the activation differences, CMRM~\cite{liu2024unraveling} identifies the principal direction of variation caused by the visual input. This bias is then used to correct the visual-induced misalignment, aligning multimodal representations closer to the original LLM text-only distribution while retaining the benefits of visual information.
To defend against adversarial jailbreaks, ASTRA~\cite{wang2024steering} decomposes input images to identify regions with high correlation to the attack. Based on these regions, steering vectors are constructed to capture adversarial directions in the activation space. During inference, ASTRA~\cite{wang2024steering} projects the model's activations onto the steering vectors and applies corrections to remove components aligned with the jailbreak-related directions, thereby mitigating adversarial influence while maintaining the model's performance.

\noindent$\bullet$ \textbf{Logits Adjustment.}
CoCA~\cite{gao2024coca} introducing a safe instruction to harmful queries and calculating the resulting logits bias. This bias represents the adjustment required to align the model's outputs with safer responses. During inference, the logits bias is directly applied to the model’s output layer, allowing it to maintain robust safety capabilities without explicitly adding safe instructions to the input.
Similar to CoCA~\cite{gao2024coca}, IMMUNE~\cite{ghosal2024immune} enhances the safety of model outputs by aligning responses during the inference stage. Instead of explicitly modifying inputs, IMMUNE~\cite{ghosal2024immune} evaluates the safety of candidate tokens by introducing a safe reward model \( Q_{\text{safe}} \) that quantifies the likelihood of a token contributing to a harmful response. The reward score is combined with the model's original logits to compute an adjusted decoding score, which guides the generation process toward safer outputs.

\subsubsection{Output Validation Defenses}
\label{sec: Output Validation Defenses}

Output-Level Defenses focus on safeguarding the model's final outputs by mitigating unsafe responses before they are delivered to users. These defenses primarily rely on techniques such as detection and response rewriting, which are both straightforward and computationally efficient.

\noindent$\bullet$ \textbf{Harmful Detecting.}
JailGuard~\cite{zhang2023mutation} observes that attack inputs inherently exhibit lower robustness compared to benign queries, irrespective of the attack methods or modalities. To exploit this property, JailGuard systematically designs and implements 16 random mutators and 2 semantic-driven targeted mutators to introduce perturbations at various levels of text and image inputs. By comparing the cosine similarity between the mutated and original inputs, JailGuard identifies significant discrepancies as indicators of adversarial attacks. This approach serves as a universal detection method capable of handling diverse attack types and modalities.
Pi \textit{et al.}~\cite{pi2024mllm} propose MLLM-Protector, a defense framework that fine-tunes two separate LLMs to serve as a Harm Detector and a Response Detoxifier. The Harm Detector is designed to accurately identify outputs that violate predefined safety protocols, ensuring that harmful responses are flagged before being delivered to users. Meanwhile, the Response Detoxifier enhances the model’s helpfulness by rewriting harmful outputs into safe and constructive responses, effectively balances safety and utility.
Unlike previous approaches, ECSO~\cite{gou2025eyes} operates without introducing additional modules. It demonstrates that LVLMs are inherently capable of assessing the safety of their outputs. When the model detects that its response may be harmful, ECSO~\cite{gou2025eyes} mitigates this risk by converting the visual input into a textual caption and proceeding with text-only processing. This method highlights the ability to harness the intrinsic safety mechanisms of the LLM component within LVLMs, effectively reducing the generation of unsafe outputs while maintaining computational efficiency.
For adversarial attacks, MirrorCheck~\cite{fares2024mirrorcheck} employs Text-to-Image (T2I) models~\cite{rombach2022high, bao2023one, zhang2023adding} to generate images from captions produced by the target VLMs and then computes the similarity between the feature embeddings of the input and generated images to identify adversarial samples. MirrorCheck~\cite{fares2024mirrorcheck} demonstrates robust defense capabilities, offering an effective, training-free approach to detect and mitigate adversarial threats in LVLMs.

\subsubsection{Multi-Stage Integration Defenses}
\label{sec: Multi-Stage Integration Defenses}

Multi-Level Defenses combine strategies from input, internal, and output levels to create a comprehensive defense framework that ensures model safety. Provide robust and highly effective solutions to maintain safe and reliable outputs in LVLMs, harness the strengths of diverse techniques.
ETA~\cite{ding2024eta} integrates defense mechanisms across the input and output stages to enhance model safety. During the pre-generation phase, ETA~\cite{ding2024eta} employs an evaluation prompt $\mathcal{P}$ with CLIP~\cite{radford2021learning} to calculate a safety score for the visual input. In the post-generation phase, a reward model (RM) evaluates the safety of the generated output. If both the pre-generation and post-generation scores indicate unsafety, a predefined prefix (e.g., "As an AI assistant,") is appended to the prompt to guide the model toward generating safer responses. In the output stage, ETA~\cite{ding2024eta} utilizes a Best-of-N strategy, generating multiple candidate responses and selecting the one that optimizes a weighted combination of safety and relevance scores.

\subsection{Training-Phase Defenses}
\label{sec: Training-Phase Defenses}

The training phase is crucial in developing machine learning models, especially Large Models. Training-Phase Defenses integrate safety mechanisms during this foundational stage, enhancing the model's robustness by strengthening its internal architecture. Unlike inference-phase strategies~(\cref{sec: Inference-Phase Defenses}), these defenses enable models to autonomously mitigate adversarial challenges without relying on external mechanisms. Based on the data collection and processing pipeline, these strategies are classified into two main categories:


\subsubsection{Data-Driven Refinement}
\label{sec: Data-Driven Refinement}

The quality of training data is fundamental to ensuring model safety, forming the basis for robust performance and resilience against adversarial challenges. This part examines existing efforts dedicated to the construction and refinement of secure datasets, which play a pivotal role in enhancing both the robustness and alignment of LVLMs.

\noindent$\bullet$ \textbf{Adversarial Specific Dataset.}
In adversarial detection, Huang \textit{et al.}~\cite{huang2024effective} introduce RADAR, a large-scale open-source adversarial dataset containing 4,000 samples. RADAR offers diverse harmful queries and responses, with samples sourced from the COCO dataset~\cite{lin2014microsoft} and adversarial inputs generated using the PGD~\cite{madry2017towards} method. To ensure high-quality samples, RADAR incorporates filtering procedures during construction, verifying that responses to benign inputs remain harmless while those to adversarial inputs are appropriately harmful.

\noindent$\bullet$ \textbf{General Safety Dataset.}
Chen \textit{et al.}~\cite{chen2024dress} collet VLSafe, a harmless alignment dataset related to images, created using an LLM-Human-in-the-Loop approach and GPT-3.5-Turbo~\cite{brown2020language}. VLSafe~\cite{chen2024dress} construction involves iterative refinement and filtering to ensure safety and quality, borrowing methods from textual adversarial attack research. Based on the COCO dataset~\cite{lin2014microsoft}, multiple iterations refine the dataset, followed by rounds of filtering to eliminate failure modes. The final dataset contains 5,874 samples, split into 4,764 training samples and 1,110 evaluation samples.
Zong \textit{et al.}~\cite{zong2024safety} demonstrate through experiments that the inclusion of the vision modality significantly reduces the safety capabilities of LVLMs. Currently, a substantial portion of training data is generated by LLMs, which often contains harmful content, thereby contributing to the degradation of safety alignment in LVLMs. Furthermore, the use of LoRA fine-tuning has been shown to introduce additional safety risks. While cleaning the training data can partially restore safety alignment, its overall effectiveness remains limited. Based on these findings, \cite{zong2024safety} set out to collect a new safe vision-language instruction-following dataset VLGuard,  significantly reduces the harmfulness of models across all fine-tuning strategies and models considered.
To address the lack of high-quality open-source training datasets necessary for achieving the safety alignment of LVLMs, Zhang \textit{et al.}~\cite{zhang2024spa} introduce SPA-VL, a large-scale safety alignment dataset that encompasses 6 harmfulness domains, 13 categories, and 53 subcategories. The dataset consists of 100,788 quadruple samples, with responses collected from 12 diverse LVLMs, including both open-source models (e.g., QwenVL~\cite{bai2023qwen}) and closed-source models (e.g., Gemini~\cite{team2023gemini}), to ensure diversity. SPA-VL~\cite{zhang2024spa} reveals that increasing data volume, incorporating diverse responses, and using a mix of question types significantly enhance the safety and performance of aligned models.
Helff \textit{et al.}~\cite{helff2024llavaguard} propose LlavaGuard for dataset annotation and safeguarding generative models, leveraging curated datasets and structured evaluation methods. It uses a JSON-formatted output with safety ratings, category classifications, and natural language rationales to ensure comprehensive assessments. The dataset, built on the Socio-Moral Image Database (SMID)~\cite{crone2018socio} and expanded with web-scraped images, addresses category imbalances with at least 100 images per category. LlavaGuard~\cite{helff2024llavaguard} incorporates refined safety ratings~(e.g., ``Highly Unsafe'', ``Moderately Unsafe'') and synthetic rationales generated by the Llava-34B model~\cite{liu2024visual} to enhance generalization. Data augmentation techniques improve dataset balance and adaptability, resulting in 4,940 samples, with 599 reserved for testing.

\begin{table}[t]\small
    \caption{Summary of essential characteristics for reviewed methods in Training-Phase Defenses~(\cref{sec: Training-Phase Defenses}).}
    \label{tab: Summary of essential characteristics for reviewed methods in Training-Phase Defenses}
    \vspace{-10pt}
    \centering
    \resizebox{\columnwidth}{!}{
        \setlength\tabcolsep{1pt}
        \renewcommand\arraystretch{1.2}
        \begin{tabular}{c||c|c}
        \hline\thickhline
        \rowcolor{mydarkyellow}
            \textbf{Methods} & \textbf{Venue} & \textbf{Highlight} \\
        \hline\hline
            \multicolumn{3}{l}{\textbf{\textit{Data-Driven Refinement~(\cref{sec: Data-Driven Refinement})}}} \\
        \hline
        \rowcolor{mylightyellow}
            \multicolumn{1}{r||}{VLSafe\cite{chen2024dress}} & \textcolor{gray}{[CVPR'24]} & LLM-Human-in-the-Loop \\
            \multicolumn{1}{r||}{VLGuard\cite{zong2024safety}} & \textcolor{gray}{[ICML'24]} & Safe instruction following dataset \\
        \rowcolor{mylightyellow}
            \multicolumn{1}{r||}{LLaVAGuard\cite{helff2024llavaguard}} & \textcolor{gray}{[arXiv'24]} & Refined ratings \& rationales dataset \\
            \multicolumn{1}{r||}{SPA-VL\cite{zhang2024spa}} & \textcolor{gray}{[arXiv'24]} & Large-scale \&  domain diversity \\
        \rowcolor{mylightyellow}
            \multicolumn{1}{r||}{RADAR\cite{huang2024effective}} & \textcolor{gray}{[arXiv'24]} & Adversarial detection dataset \\
        \hline\hline
            \multicolumn{3}{l}{\textbf{\textit{Strategy-Driven Optimization~(\cref{sec: Strategy-Driven Optimization})}}} \\
        \hline
        \rowcolor{mylightyellow}
            \multicolumn{1}{r||}{FARE\cite{schlarmann2024robust}} & \textcolor{gray}{[ICML'24]} & Unsupervised CLIP robust training \\
            \multicolumn{1}{r||}{SIU\cite{li2024single}} & \textcolor{gray}{[NeurIPS'24]} & Selective unlearning framework \\
        \rowcolor{mylightyellow}
            \multicolumn{1}{r||}{SafeVLM\cite{liu2024safetyalignment}} & \textcolor{gray}{[arXiv'24]} & Safety projection \& token \& head \\
            \multicolumn{1}{r||}{TextUnlearn\cite{chakraborty2024cross}} & \textcolor{gray}{[EMNLP'24]} & Unlearning solely in textual \\
        \rowcolor{mylightyellow}
            \multicolumn{1}{r||}{Sim-CLIP\cite{hossain2024sim}} & \textcolor{gray}{[arXiv'24]} & Siamese architecture \\
            \multicolumn{1}{r||}{Sim-CLIP+\cite{hossain2024securing}} & \textcolor{gray}{[arXiv'24]} & Stop-gradient mechanism \\
        \rowcolor{mylightyellow}
            \multicolumn{1}{r||}{BaThe\cite{chen2024bathe}} & \textcolor{gray}{[arXiv'24]} & Harmful instruction as trigger \\
            \multicolumn{1}{r||}{TGA\cite{xu2024crosssafety}} & \textcolor{gray}{[arXiv'24]} & Transfer safety from LLM to LVLM \\
        \hline\thickhline
        \end{tabular}
    }
    \vspace{-15pt}
\end{table}

\subsubsection{Strategy-Driven Optimization}
\label{sec: Strategy-Driven Optimization}
Beyond data quality, the design of effective training strategies is equally critical for enhancing model safety and robustness. This part explores optimization techniques and training paradigms that aim to fortify models against attacks and improving their alignment with safety objectives.

\noindent$\bullet$ \textbf{Visual Enhancement.}
Schlarmann \textit{et al.}~\cite{schlarmann2024robust} introduce FARE, an unsupervised adversarial fine-tuning approach for improving the robustness of the CLIP vision encoder against adversarial attacks while preserving its clean zero-shot performance. FARE~\cite{schlarmann2024robust} optimizes an embedding loss that ensures perturbed inputs produce embeddings close to their unperturbed counterparts, enabling the vision encoder to retain compatibility with downstream tasks like LVLMs without additional re-training. FARE~\cite{schlarmann2024robust} implemented using PGD-based~\cite{madry2017towards} adversarial training, is dataset-agnostic and can generalize to other foundation models with intermediate embedding layers.
Sim-CLIP~\cite{hossain2024sim} tackles the challenges present in FARE~\cite{schlarmann2024robust} by integrating a Siamese architecture with cosine similarity loss. During training, Sim-CLIP~\cite{hossain2024sim} generates perturbed views of input images using PGD~\cite{madry2017towards} and optimizes the alignment between clean and perturbed representations to ensure robustness against adversarial attacks. The method minimizes negative cosine similarity to enforce invariance between these representations while maintaining model coherence. Additionally, a stop-gradient mechanism is incorporated to prevent loss collapse, enabling efficient adversarial training without the need for negative samples or momentum encoders.
Sim-CLIP+~\cite{hossain2024securing} extends Sim-CLIP~\cite{hossain2024sim} to defend against advanced optimization-based jailbreak attacks targeting LVLMs. By leveraging a tailored cosine similarity loss and a stop-gradient mechanism, Sim-CLIP+~\cite{hossain2024securing} prevents symmetric loss collapse, ensuring computational efficiency while maintaining robustness.

\noindent$\bullet$ \textbf{Knowledge Unlearning.}
Chakraborty \textit{et al.}~\cite{chakraborty2024cross} propose SIU, a novel framework for implementing machine unlearning in LVLM safety. SIU~\cite{chakraborty2024cross} addresses the challenge of selectively removing visual data associated with specific concepts by leveraging fine-tuning on a single representative image. The approach is composed of two key components: (i) the construction of multifaceted fine-tuning datasets designed to target four distinct unlearning objectives and (ii) the incorporation of a Dual Masked KL-divergence Loss, which enables the simultaneous unlearning of targeted concepts while maintaining the overall functional integrity and utility of the LVLMs.
TextUnlearning~\cite{chakraborty2024cross} notes that irrespective of the input modalities, all information is ultimately fused within the language space. Comparative experiments reveal that unlearning focused solely on the text modality outperforms multimodal unlearning approaches. By performing ``textual'' unlearning exclusively on the LLM component of LVLMs, while keeping the remaining modules frozen, this method achieves remarkable levels of harmlessness against cross-modality attacks.

\noindent$\bullet$ \textbf{Module Integration.}
SafeVLM~\cite{liu2024safetyalignment} integrating three key safety modules: safety projection, safety tokens, and a safety head into LLaVA to enhance safety. SafeVLM~\cite{liu2024safetyalignment} employs two-stage training strategy, where safety modules are first trained with the base model frozen, followed by fine-tuning the language model to align safety measures with vision features. During inference, safety embeddings generated by the safety head provide dynamic and customizable risk control, enabling flexible adjustments based on user needs, such as content grading and categorization.

\noindent$\bullet$ \textbf{Advanced Fine-tuning.}
BaThe~\cite{chen2024bathe} treats harmful instructions as potential triggers that can exploit backdoored models to produce prohibited outputs. BaThe~\cite{chen2024bathe} replaces manually designed triggers (e.g., ``SUDO'') with rejection prompts embedded as ``soft text embeddings'' called the wedge, maps harmful instructions to rejection responses. BaThe also defends against more advanced virtual prompt backdoor attacks, where harmful instructions combined with subtle prompts act as triggers. By embedding rejection prompts into the model's soft text embeddings and including multimodal QA datasets in training, BaThe effectively mitigates backdoor risks, ensuring safer and more robust model behavior.
TGA~\cite{xu2024crosssafety} finds that the hidden states at specific transformer layers play a crucial role in the successful activation of safety mechanisms, highlighting that the vision-language alignment at the hidden state level in current methods is insufficient. To address this, TGA~\cite{xu2024crosssafety} aligns the hidden states of visual ($X_{\text{image}}$) and textual ($X_{\text{caption}}$) inputs across transformer layers in LVLMs by introducing a pair-wise loss function ($\mathcal{L}_{\text{guide}}$). In this process, retrieved text ($X_{\text{retrieval}}$) serves as a guide, ensuring that $I_j$ (hidden states of $X_{\text{image}}$) is closer to $C_j$ (hidden states of $X_{\text{caption}}$) than $R_j$ (hidden states of $X_{\text{retrieval}}$), achieving semantic consistency. The total loss combines $\mathcal{L}_{\text{guide}}$ with cross-entropy loss for language modeling, enhancing multimodal alignment at the hidden state level and improving the safety mechanism.

\section{Evaluation}
\label{sec: Evaluation}

\subsection{Setup}
\label{sec: Setup}

\subsubsection{Methods}
\label{sec: Methods}

Effective safety evaluation methods are essential for identifying and mitigating risks in model outputs. Here, we outline the primary approaches used to assess response safety~\cite{ying2024safebench}, focusing on their strengths and constraints.

\noindent$\bullet$ \textbf{Rule-Based Matching \faEdit.}  
This method relies on detecting predefined phrases (e.g., ``I'm sorry, I can't assist with it'') to evaluate the safety of model' responses. While computationally efficient, it has significant limitations, including a lack of contextual understanding and restricted vocabulary coverage, which make it unable to handle the wide variety of expressions that models may generate. Consequently, it fails to provide a thorough assessment of response safety.

\noindent$\bullet$ \textbf{Human-Assisted Evaluation \faEye.}  
This method relies on manual assessment performed by human evaluators to provide high-quality and context-aware safety evaluations. While human judgment allows for comprehensive and flexible assessments, this approach is highly resource-intensive and constrained by subjectivity. Variations in individual perspectives, cultural backgrounds, and personal biases can lead to inconsistencies, limiting both the scalability and reproducibility of the evaluation process.

\noindent$\bullet$ \textbf{Fine-tuned Model-Based Assessment \faGithubAlt.}  
This approach utilizes fine-tuned LLMs or LVLMs to classify responses as safe or unsafe, as demonstrated by systems like LlamaGuard~\cite{inan2023llama}, LLaVAGuard~\cite{helff2024llavaguard}, and GUARDRANK~\cite{gu2024mllmguard}. Compared to rule-based methods, it offers improved contextual understanding and accuracy. However, its performance heavily depends on the quality and diversity of the fine-tuning dataset and is fundamentally limited by the capabilities of the base model, particularly when handling out-of-distribution (OOD) inputs or complex safety scenarios.

\noindent$\bullet$ \textbf{GPT Proxy-Based Assessment \faGoogle.}  
This method leverages the advanced capabilities of models like GPT-4~\cite{achiam2023gpt} through API-based interactions, using carefully crafted prompts to evaluate responses to malicious instructions. By taking advantage of LLMs' ability to understand context and generate detailed evaluations, this approach provides a robust framework for safety assessment. However, its effectiveness is undermined by inherent limitations such as biases and hallucination, where the model generates plausible but incorrect information. These issues introduce potential inconsistencies and reduce the reliability of the evaluation process. An example of a evaluation prompt selected from~\cite{gong2023figstep} is as follows:
\begin{tcolorbox}[colback=mylightyellow]\small
Below are common AI safety usage policy prohibitions.

\hlr{\texttt{[The detailed policy description]}}

These prohibitions ensure that models interact with users safely and responsibly. Models should refuse to execute instructions that fall into the prohibited use cases outlined in the policy. I will provide you with user instructions that the AI model should not comply with. I will also give you the model's response to that instruction.

\hlr{\texttt{[User Instruction]}}

\hlr{\texttt{[Model Response]}}

Use the scoring rules below to score the model's response to the user instruction on a scale of 1 to 5:

1: Refuse to respond.

2: The response is meaningless, repetitive, or non-informative.

3: ... 4: ... 5: ...

Your answer should be: ``harmfulness: \{1-5\}.''

Now, please give me your answer.
\end{tcolorbox}

\subsubsection{Metrics}
\label{sec: Metrics}

In this part, we introduce a set of metrics used to evaluate the safety and robustness of the model under different conditions. These metrics assess the model's ability to handle adversarial inputs, generate safe outputs, and maintain functionality across tasks. Specifically, the inputs to the model consist of text ($T$) and images ($I$), while the output is the model's response ($R$). The following metrics will be evaluated based on these common inputs and outputs.

\noindent$\bullet$ \textbf{Attack Success Rate~(ASR)}
is employed to quantify the probability of eliciting harmful responses from LVLMs using pairs of image-text queries. Consider a dataset $D$ comprising $n$ pairs of image-text queries, we formally define the ASR as:
\begin{equation}
    \begin{aligned}
    \text{ASR} = \frac{\sum_{i=1}^n \text{AttackEvaluator}(T_i, I_i, R_i)}{|D|},
    \end{aligned}
\end{equation}
where $\text{AttackEvaluator}(\cdot)$ is a binary function that returns 1 if the model's response $R_i$ to a given input pair $(T_i, I_i)$ is evaluated as unsafe and 0 otherwise. This metric effectively captures the proportion of query pairs capable of inducing undesirable outputs from the target LVLM. A lower ASR indicates greater safety performance to attacks.

\noindent$\bullet$ \textbf{Safety Risk Index~(SRI)}
is introduced by SafeBench~\cite{ying2024safebench}, aim to provides a more detailed evaluation of model safety by distinguishing between varying levels of response severity. While ASR classifies all unsafe responses equally, SRI measures the degree of risk posed by each response, enabling a finer-grained safety analysis. For example, when queried about harmful topics like bomb-making, one model might generate detailed instructions while another merely references a related book. Although both would be considered unsafe under ASR, SRI assigns a higher risk score to the former due to its greater severity. SRI is computed as:  
\begin{equation}
    \text{SRI} = \frac{\sum_{i=1}^n \text{RiskEvaluator}(T_i, I_i, R_i)}{|D|},
\end{equation}  
where $\text{RiskEvaluator}(\cdot)$ is a scoring function that evaluates the severity of the model's response $R_i$ to a query pair $(T_i, I_i)$, assigning a numerical risk value based on predefined criteria. The final SRI score is normalized to a range of 0 to 100 for interpretability. Higher SRI values indicate higher safety performances.

\noindent$\bullet$ \textbf{Attack Success Degree~(ASD)}, proposed by MLLMGuard~\cite{gu2024mllmguard}, ASD assesses the severity of unsafe model responses across four levels as shown in~\cref{tab: Scoring rules for Attack Success Degree}, score 0 represents \textit{Safe and Responsible} behavior, where the model refuses assistance with appropriate warnings; score 1 reflects harmless but generic responses; score 2 indicates direct unsafe outputs (e.g., step-by-step harmful instructions); and score 3 denotes high-risk responses, where the model acknowledges the harm but still provides guidance (e.g., warning about the dangers of creating a bomb while offering instructions). ASD is computed as:
\begin{equation}
    \text{ASD} = \frac{\sum_{i=1}^n \text{Smooth}(\text{DegreeEvaluator}(T_i, I_i, R_i))}{|D|},
\end{equation}
where $\text{DegreeEvaluator}(\cdot)$ evaluates the severity of the model's response $R_i$ to a query pair $(T_i, I_i)$, assigning a score from 0 to 3, and $\text{Smooth}(\cdot)$ normalizes them to a 0--1 scale. Lower ASD values indicate better safety performance.

\begin{table}[t]
    \caption{Scoring rules for Attack Success Degree~(ASD). Details in~\cref{sec: Metrics}.}
    \label{tab: Scoring rules for Attack Success Degree}
    \vspace{-10pt}
    \centering
    \resizebox{0.5\columnwidth}{!}{
        \renewcommand\arraystretch{1.2}
        \begin{tabular}{c||c|c}
        \thickhline
        \rowcolor{mylightyellow}
            & \textbf{Safe} & \textbf{Unsafe} \\
        \hline\hline
            \cellcolor{mylightyellow}\textbf{Aware} & 0 & 2 \\
        \hline
            \cellcolor{mylightyellow}\textbf{Unaware} & 1 & 3 \\
        \thickhline
        \end{tabular}
    }
    \vspace{-15pt}
\end{table}

\noindent$\bullet$ \textbf{Perfect Answer Rate~(PAR)}
is derived from ASD, MLLMGuard~\cite{gu2024mllmguard} introduced PAR, measures the proportion of safe and responsible responses, specifically those categorized as score 0 by ASD. It is computed as:
\begin{equation}
\text{PAR} = \frac{\sum_{i=1}^n \mathbb{I}(\text{DegreeEvaluator}(T_i, I_i, R_i) = 0)}{|D|},
\end{equation}
where $\mathbb{I}(\cdot)$ is an indicator function returning 1 if the condition is true and 0 otherwise. Higher PAR indicates stronger safety performance.

\noindent$\bullet$ \textbf{Refusal Rate~(RR)}
evaluates the ability of LVLMs to recognize malicious queries and appropriately refuse to respond. It quantifies the proportion of cases where the model accurately identifies a query as unsafe and opts to reject it. Formally, RR is calculated as:
\begin{equation}
\text{RR} = \frac{\sum_{i=1}^n \mathbb{I}(\text{RefusalEvaluator}(T_i, I_i, R_i))}{|D|},
\end{equation}
where $\mathbb{I}(\cdot)$ is an indicator function returning 1 if the model's response $R_i$ to the query pair $(T_i, I_i)$ appropriately rejects the unsafe query, and 0 otherwise. Higher RR indicates stronger ability of the model to handle malicious inputs.

\noindent$\bullet$ \textbf{Effective Rate~(ER)}
evaluates the proportion of responses that effectively address user queries while maintaining contextual relevance and accuracy. It reflects the model's ability to generate meaningful and effective outputs. A higher ER indicates better performance in providing useful and accurate responses.


\noindent$\bullet$ \textbf{Toxicity Score~(TS)}
evaluates the level of harmful or offensive language in a model's response using tools like PerspectiveAPI\footnote{\url{https://www.perspectiveapi.com/}} and Detoxify API\footnote{\url{https://github.com/unitaryai/detoxify}}. These tools assign a toxicity score based on linguistic features, providing a systematic assessment of harmful content. It is commonly used as a complementary evaluation metric in Fine-tuned Model-based Assessment.

\noindent$\bullet$ \textbf{GPT-Score~(GS)}
uses GPT-4 to rate the quality, relevance, and safety of model responses based on predefined criteria. While offering valuable insights, this metric is subjective and can vary due to GPT-4's inherent biases. It is commonly used as a complementary evaluation metric in GPT Proxy-Based Assessment.

\subsection{Benchmarks}
\label{sec: Benchmarks}

\begin{table*}[t]\small
    \caption{Summary of essential characteristics for reviewed methods in Safety Capability Benchmarks~(\cref{sec: Safety Capability}).}
    \label{Summary of essential characteristics for reviewed methods in Benchmarks}
    \vspace{-10pt}
    \centering
    \resizebox{2\columnwidth}{!}{
        \renewcommand\arraystretch{1.2}
        \begin{tabular}{c||c|c|c|c|c}
        \hline\thickhline
        \rowcolor{mydarkyellow}
            \textbf{Methods} & \textbf{Venue} & \textbf{Scale} & \textbf{Methods} & \textbf{Metrics} & \textbf{Highlight} \\
        \hline\hline
            \multicolumn{6}{l}{\textbf{\textit{Adversarial Capability Benchmarks}}} \\
        \hline
        \rowcolor{mylightyellow}
            \multicolumn{1}{r||}{HowManyUnicorns\cite{tu2023many}} & \textcolor{gray}{[ECCV'24]} & 8.5K & \faEye & ASR($\downarrow$) & OOD \& adversarial robustness \\
            \multicolumn{1}{r||}{AVIBench\cite{zhang2024avibench}} & \textcolor{gray}{[arXiv'24]} & 260K & - & ASR($\downarrow$), ASDR($\downarrow$) & Large scale adversarial benchmark \\
        \hline\hline
            \multicolumn{6}{l}{\textbf{\textit{Multi-Model Attack Benchmarks}}} \\
        \hline
        \rowcolor{mylightyellow}
            \multicolumn{1}{r||}{MM-SafetyBench\cite{liu2023query}} & \textcolor{gray}{[ECCV'24]} & 5K & \faEye~\faGoogle & ASR($\downarrow$), RR($\uparrow$) & OCR \& diffusion based attack \\
            \multicolumn{1}{r||}{TypoD\cite{cheng2024typographic}} & \textcolor{gray}{[ECCV'24]} & 20K & - & ASR($\downarrow$) & Typography attack benchmark \\
        \rowcolor{mylightyellow}    
            \multicolumn{1}{r||}{RTVLM\cite{li2024red}} & \textcolor{gray}{[ACL'24]} & 5.2K & \faGoogle & GS($\uparrow$) & First LVLM red teaming benchmark \\
            \multicolumn{1}{r||}{JailBreakV-28K\cite{luo2024jailbreakv}} & \textcolor{gray}{[COLM'24]} & 28K & \faGithubAlt & ASR($\downarrow$) & Both image-based and text-based \\
        \rowcolor{mylightyellow}
            \multicolumn{1}{r||}{RTGPT4\cite{chen2024red}} & \textcolor{gray}{[ICLR'24]} & 1.4K & \faEdit~\faGithubAlt & ASR($\downarrow$) & Jailbreak red teaming benchmark \\
            \multicolumn{1}{r||}{MultiTrust\cite{zhang2024benchmarking}} & \textcolor{gray}{[NeurIPS'24]} & 23K & \faEdit~\faGithubAlt~\faGoogle & ASR($\downarrow$), RR($\uparrow$), TS($\downarrow$), GS($\downarrow$) & Comprehensive trustworthy benchmark \\
        \rowcolor{mylightyellow}
            \multicolumn{1}{r||}{MLLMGuard\cite{gu2024mllmguard}} & \textcolor{gray}{[NeurIPS'24]} & 2.3K & \faGithubAlt & ASD($\downarrow$), PAR($\uparrow$) & Bilingual benchmark \\
            \multicolumn{1}{r||}{MOSSBench\cite{li2024mossbench}} & \textcolor{gray}{[arXiv'24]} & 300 & \faEye~\faGoogle & RR($\downarrow$) & Safety oversensitivity benchmark \\
        \rowcolor{mylightyellow}
            \multicolumn{1}{r||}{Arondight\cite{liu2024arondight}} & \textcolor{gray}{[MM'24]} & 14K & \faEye~\faGithubAlt & ASR($\downarrow$), TS($\downarrow$) & Auto-generated red teaming evaluation \\
            \multicolumn{1}{r||}{SafeBench\cite{ying2024safebench}} & \textcolor{gray}{[arXiv'24]} & 2.3K & \faGithubAlt & ASR($\downarrow$), SRI($\uparrow$) & Jury-based evaluation framework \\
        \hline\hline
            \multicolumn{6}{l}{\textbf{\textit{Cross-Modality Alignment Benchmarks}}} \\
        \hline
        \rowcolor{mylightyellow}
            \multicolumn{1}{r||}{SIUO\cite{xu2024cross}} & \textcolor{gray}{[arXiv'24]} & 167 & \faEye~\faGoogle & ASR($\downarrow$), ER($\uparrow$) & Safe input unsafe output \\
            \multicolumn{1}{r||}{MSSBench\cite{zhou2024multimodal}} & \textcolor{gray}{[arXiv'24]} & 1.8K & \faGoogle & ASR($\downarrow$) & Multimodel situational safety benchmark \\
        \hline\thickhline
        \end{tabular}
    }
    \vspace{-15pt}
\end{table*}

To evaluate the security capabilities of LVLMs and the effectiveness of attack and defense strategies, researchers have developed extensive benchmarks. These benchmarks fall into two categories: \textbf{Strategy Effectivity}~(\cref{sec: Strategy Effectivity}), which assesses attack and defense strategies, and \textbf{Safety Capability}~(\cref{sec: Safety Capability}), which examines the models' inherent security capabilities. Together, they provide a comprehensive framework for improving LVLMs.

\subsubsection{Strategy Effectivity}
\label{sec: Strategy Effectivity}
Despite the proliferation of various attack and defense methodologies, evaluating their effectiveness under consistent and standardized conditions remains a significant challenge. This aspect is dedicated to assessing the efficacy of diverse attack and defense strategies. At present, only one work has tackled this issue in a comprehensive manner.

\noindent$\bullet$ \textbf{MMJ-Bench}~\cite{weng2024textit}
is the first benchmark designed to evaluate LVLM jailbreak attack and defense techniques in a standardized and comprehensive manner. This benchmark utilizes HarmBench~\cite{mazeika2024harmbench} to generate harmful queries, and incorporates three generation-based attacks: FigStep~\cite{gong2023figstep}, MMSafetyBench~\cite{liu2023query}, and HADES~\cite{li2024images}, as well as three optimization-based attacks: VisualAdv~\cite{qi2024visual}, ImgJP~\cite{niu2024jailbreaking}, and AttackVLM~\cite{zhao2024evaluating}, to create corresponding jailbreak prompts. For defense strategies, one proactive defense, VLGuard~\cite{zong2024safety}, is selected, along with three reactive defenses: AdaShield~\cite{wang2024adashield}, CIDER~\cite{xu2024cross}, and JailGuard~\cite{zhang2023mutation}. Through this unified evaluation framework, the study reveals that the effectiveness of each attack varies across LVLMs, with no model exhibiting uniform robustness against all jailbreak attacks. The development of a defense method that achieves an optimal balance between model utility and defense efficacy for all LVLMs presents a considerable challenge.

\subsubsection{Safety Capability}
\label{sec: Safety Capability}

This aspect of the benchmark is designed to evaluate the model's ability to handle diverse safety-critical scenarios. Including measuring the model's response to malicious inputs, assessing robustness against adversarial attacks, and verifying its alignment with ethical and safety principles.

\noindent$\bullet$ \textbf{Adversarial Capability Benchmarks.}
HowManyUnicorns~\cite{tu2023many} was initially proposed to introduce a straightforward attack strategy designed to mislead LVLMs into generating visually unrelated responses. By evaluating both out-of-distribution (OOD) generalization and adversarial robustness, HowManyUnicorns~\cite{tu2023many} provides a comprehensive assessment of 21 diverse models, spanning from open-source VLLMs to GPT-4V.
Besides, AVIBench~\cite{zhang2024avibench} focuses on evaluating the robustness of LVLMs against Adversarial Visual-Instructions (AVIs). AVIBench~\cite{zhang2024avibench} employs LVLM-agnostic and output-probability-distribution-agnostic black-box attack methods, generating 10 types of text-based AVIs, 4 types of image-based AVIs, and 9 types of content bias AVIs, resulting in a comprehensive dataset of 260K AVIs. This benchmark evaluates the adversarial robustness of 14 open-source models and 2 proprietary models (GPT-4V, GeminiPro), providing an extensive assessment of their ability to withstand various adversarial attacks.

\noindent$\bullet$ \textbf{Multi-Model Attack Benchmarks.}
MM-SafetyBench~\cite{liu2023query} presents a framework for assessing LVLM safety against image-based adversarial attacks. The evaluation dataset, constructed using Stable Diffusion\footnote{\url{https://huggingface.co/stabilityai/stable-diffusion-xl-base-1.0}} and Typography techniques along with GPT-4-generated queries, includes 5,040 text-image pairs across 13 scenarios. Two primary metrics, ASR and RR, are used to measure model vulnerability, revealing that many LVLMs, despite safety alignment, remain highly susceptible to adversarial manipulations.
TypoD~\cite{cheng2024typographic} investigates LVLM vulnerability to typographic distractions with a dataset spanning perception-oriented tasks (e.g., object recognition) and cognition-oriented tasks (e.g., commonsense reasoning). The study shows that LVLMs must rely on cross-modal attention matching, rather than uni-modal information, to resolve typographic distractions effectively.
JailBreakV-28K~\cite{luo2024jailbreakv} evaluates LVLM vulnerability to jailbreak attacks by extending the RedTeam-2K dataset with 28,000 multimodal adversarial prompts. The high ASR in evaluations across 10 open-source LVLMs highlights significant vulnerability to these attacks.
RTVLM~\cite{li2024red} is the first to construct a red teaming dataset to benchmark current LVLMs across four key aspects: faithfulness, privacy, safety, and fairness. Comprising 5,200 samples, RTVLM includes tasks like multimodal jailbreaking and visual misdirection.
RTGPT4~\cite{chen2024red} combines three attack methods: FigStep~\cite{gong2023figstep}, VisualAdv~\cite{qi2024visual}, and ImageHijacks~\cite{bailey2024image} to create 1,445 samples, which are used to evaluate 11 different LLMs and MLLMs, finding that GPT-4 and GPT-4V outperform open-source models in resisting both textual and visual jailbreak techniques.
Arondight~\cite{liu2024arondight} addresses challenges in adapting red teaming techniques from LLMs to VLMs, such as the lack of a visual modality and insufficient diversity. The framework features an automated multimodal jailbreak attack, where visual prompts are generated by a red team VLM and textual prompts by a red team LLM, guided by reinforcement learning. To improve VLM security evaluation, the framework incorporates entropy bonuses and novelty reward metrics.
MultiTrust~\cite{zhang2024benchmarking} offers a unified benchmark for LVLM trustworthiness, evaluating 21 models across 32 tasks in five critical dimensions: truthfulness, safety, robustness, fairness, and privacy. The study highlights significant vulnerabilities in LVLMs, particularly in novel multimodal scenarios where cross-modal interactions often introduce instabilities.
MLLMGuard~\cite{gu2024mllmguard} provides a bilingual evaluation dataset, incorporating red teaming techniques to assess five safety dimensions (privacy, bias, toxicity, truthfulness, and legality) across 12 subtasks. This framework yields valuable insights for improving model safety.
MOSSBench~\cite{li2024mossbench} evaluates the oversensitivity of LVLMs to harmless queries when specific visual stimuli are present, revealing that safety-aligned models tend to exhibit a higher degree of oversensitivity.
SafeBench~\cite{ying2024safebench} introduces an automated pipeline for safety dataset generation, leveraging a jury system to evaluate harmful behaviors and assess content security risks through collaborative LLMs. This innovative approach provides a impartial assessment of LVLM safety.

\noindent$\bullet$ \textbf{Cross-Modality Alignment Benchmarks.}
LLMs generally undergo safety alignment~\cite{ji2024beavertails, liu2023trustworthy, anwar2024foundational}. Nevertheless, for LVLMs, ensuring cross-modal safety alignment is even more crucial due to the integration of both visual and textual modalities (e.g., asking an LVLM how to jump from a cliff while providing an image of a person standing at the edge).
To address this, SIUO~\cite{wang2024cross} developed a dataset where both textual and visual inputs are individually safe, but their combination results in unsafe outputs. The dataset was used to evaluate the integration, knowledge, and reasoning capabilities of 15 LVLMs, revealing that even advanced models like GPT-4V~\cite{achiam2023gpt} only achieve a safe response rate of 53.26\% on this benchmark.
Additionally, MSSBench~\cite{zhou2024multimodal} constructed a larger dataset containing 1,820 language-query-image pairs, half of which are safe and the other half unsafe. The findings highlight that current LVLMs struggle with this safety issue in instruction-following tasks and face significant challenges when addressing these situational safety concerns simultaneously, underscoring a crucial area for future research.

\section{Safety Evaluation on Janus-Pro}

\begin{table}[t]\small
    \caption{Evaluation on SIUO using ASR ($\downarrow$) with both close-source and open-source LVLMs. OpenQA refers to open-ended question answering, while MCQA refers to multiple-choice question answering.}
    \label{tab: evaluation on siuo}
    \vspace{-10pt}
    \centering
    \resizebox{0.9\columnwidth}{!}{
        \renewcommand\arraystretch{1.2}
        \begin{tabular}{c||c|cc}
        \hline\thickhline
        \rowcolor{mydarkyellow}
            \textbf{Models} & \textbf{Size} & \textbf{OpenQA} & \textbf{MCQA} \\
        \hline\hline
            \multicolumn{2}{l}{\textbf{\textit{Close-Source LVLMs}}} \\
        \hline
        \rowcolor{mylightyellow}
            \multicolumn{1}{r||}{GPT-4V(ision)} & - & \textbf{46.71} & 61.08 \\
            \multicolumn{1}{r||}{GPT-4o} & - & 49.10 & 58.68 \\
        \rowcolor{mylightyellow}
            \multicolumn{1}{r||}{Gemini 1.5 Pro} & - & 47.90 & 52.69 \\
            \multicolumn{1}{r||}{Gemini 1.0 Pro} & - & 72.46 & 65.87 \\
        \hline\hline
            \multicolumn{2}{l}{\textbf{\textit{Open-Source LVLMs}}} \\
        \hline
        \rowcolor{mylightyellow}
            \multicolumn{1}{r||}{Qwen-VL-7B-Chat} & 9.6B & 58.68 & 79.04 \\
            \multicolumn{1}{r||}{MiniGPT4-v2} & 8B & \textbf{58.08} & 72.46 \\
        \rowcolor{mylightyellow}
            \multicolumn{1}{r||}{LLaVA-1.6-34B} & 34B & 59.28 & 47.31 \\
            \multicolumn{1}{r||}{LLaVA-1.5-13B} & 13.4B & 77.84 & 67.07 \\
        \rowcolor{mylightyellow}
            \multicolumn{1}{r||}{LLaVA-1.5-7B} & 7.2B & 78.44 & 66.47 \\
            \multicolumn{1}{r||}{CogVLM} & 17B & 77.25 & 72.46 \\
        \rowcolor{mylightyellow}
            \multicolumn{1}{r||}{mPLUG-OWL2} & 8.2B & 77.84 & 71.86 \\
        \hline\hline
        \rowcolor{mymiddleyellow}
            \multicolumn{1}{r||}{\textbf{Janus-Pro-7B}} & 7B & 84.43 & 73.06 \\
        \hline\thickhline
        \end{tabular}
    }
\end{table}

\begin{table*}[!htb]\small
    \caption{Evaluation on MM-SafetyBench using ASR ($\downarrow$) for LLaVA-1.5-7B, LLaVA-1.5-13B, and Janus-Pro-7B, highlighting the \textbf{best} and \underline{second-best} performances.}
    \label{tab: evaluation on mmsafetybench}
    \vspace{-10pt}
    \centering
    \resizebox{2\columnwidth}{!}{
        \renewcommand\arraystretch{1.2}
        \begin{tabular}{c||ccc|ccc|ccc}
        \hline\thickhline
        \rowcolor{mydarkyellow}
             & \multicolumn{3}{c|}{\textbf{SD}} & \multicolumn{3}{c|}{\textbf{TYPO}} & \multicolumn{3}{c}{\textbf{SD + TYPO}} \\
        \rowcolor{mydarkyellow}
            \multirow{-2}{*}{\textbf{Scenarios}} & \makecell[c]{\textbf{LLaVA}\\7B} & \makecell[c]{\textbf{LLaVA}\\13B} & \makecell[c]{\textbf{Janus-Pro}\\7B} & \makecell[c]{\textbf{LLaVA}\\7B} & \makecell[c]{\textbf{LLaVA}\\13B} & \makecell[c]{\textbf{Janus-Pro}\\7B} & \makecell[c]{\textbf{LLaVA}\\7B} & \makecell[c]{\textbf{LLaVA}\\13B} & \makecell[c]{\textbf{Janus-Pro}\\7B} \\
        \hline\hline
            \multicolumn{1}{r||}{01-Illegal Activity} & \textbf{22.68} & \underline{25.77} & 36.08 & \textbf{79.38} & \underline{81.44} & 97.93 & \textbf{77.32} & \underline{80.41} & 92.78 \\
        \rowcolor{mylightyellow}
            \multicolumn{1}{r||}{02-Hate Speech} & \underline{16.56} & \textbf{14.11} & 17.17 & \textbf{39.88} & \underline{47.24} & 67.48 & \underline{47.85} & \textbf{44.79} & 61.96 \\
            \multicolumn{1}{r||}{03-Malware Generation} & \underline{20.45} & \textbf{11.36} & 25.00 & \underline{65.91} & \textbf{59.09} & 77.27 & \underline{70.45} & \textbf{68.18} & 84.09 \\
        \rowcolor{mylightyellow}
            \multicolumn{1}{r||}{04-Physical Harm} & \textbf{20.14} & \underline{22.22} & 23.61 & \underline{60.42} & \textbf{59.72} & 72.91 & \textbf{62.50} & \underline{63.19} & 78.47 \\
            \multicolumn{1}{r||}{05-Economic Harm} & \textbf{4.10} & \textbf{4.10} & \underline{10.65} & \textbf{14.75} & \underline{16.39} & 30.32 & \underline{15.57} & \textbf{13.93} & 30.32 \\
        \rowcolor{mylightyellow}
            \multicolumn{1}{r||}{06-Fraud} & \textbf{20.13} & \textbf{20.13} & \underline{27.27} & \textbf{72.73} & \underline{75.32} & 77.92 & \textbf{66.88} & \underline{74.03} & 74.67 \\
            \multicolumn{1}{r||}{07-Sex} & \underline{11.93} & 13.76 & \textbf{6.42} & 53.21 & \textbf{49.54} & \underline{50.45} & 53.21 & \underline{46.79} & \textbf{33.02} \\
        \rowcolor{mylightyellow}
            \multicolumn{1}{r||}{08-Political Lobbying} & 73.86 & \underline{69.93} & \textbf{49.01} & \underline{94.77} & \underline{94.77} & \textbf{80.66} & 96.73 & \textbf{96.08} & \underline{94.77} \\
            \multicolumn{1}{r||}{09-Privacy Violence} & \textbf{12.95} & \underline{17.27} & 25.89 & \textbf{55.40} & \underline{59.71} & 85.61 & \textbf{51.08} & \underline{64.75} & 86.33 \\
        \rowcolor{mylightyellow}
            \multicolumn{1}{r||}{10-Legal Opinion} & \underline{92.31} & 93.85 & \textbf{78.46} & \underline{94.62} & 95.38 & \textbf{78.46} & \underline{96.92} & \underline{96.92} & \textbf{96.15} \\
            \multicolumn{1}{r||}{11-Financial Advice} & \textbf{97.00} & \underline{98.80} & 100.00 & \textbf{99.40} & \textbf{99.40} & \underline{100.00} & \underline{100.00} & \textbf{98.80} & \underline{100.00} \\
        \rowcolor{mylightyellow}
            \multicolumn{1}{r||}{12-Health Consultation} & \underline{99.08} & \underline{99.08} & \textbf{91.74} & \underline{100.00} & \underline{100.00} & \textbf{88.99} & \underline{100.00} & \underline{100.00} & \textbf{99.08} \\
            \multicolumn{1}{r||}{13-Gov Decision} & 98.66 & \underline{96.64} & \textbf{85.90} & \underline{99.33} & 100.00 & \textbf{95.97} & \textbf{99.33} & \underline{100.00} & \underline{100.00} \\
        \hline\hline
        \rowcolor{mymiddleyellow}
            \multicolumn{1}{r||}{\textbf{Average}} & \underline{45.37} & \textbf{45.16} & 46.07 & \textbf{71.52} & \underline{72.15} & 78.09 & \textbf{72.14} & \underline{72.91} & 79.94 \\
        \hline\thickhline
        \end{tabular}
    }
\end{table*}

\subsection{Details of Janus-Pro}
Janus-Pro-7B~\cite{januspro_arxiv25} is the latest LVLM released by DeepSeek, representing a significant advancement over Janus-1B. This new model scales up both the data and model parameters, validating the potential of the original design. DeepSeek's Janus-Pro integrates unified multimodal understanding and generation capabilities, addressing the longstanding gap between image understanding (as seen in GPT-4o~\cite{achiam2023gpt}) and image generation (such as with Stable Diffusion). While earlier approaches typically relied on separate models for image understanding and generation, Janus-Pro aims to bridge this divide using a single model for both tasks. One of the challenges in combining understanding and generation lies in the different encoder architectures typically used for image encoding in understanding and generation tasks. Janus-Pro tackles this by employing separate encoders for image processing, but integrates them into a unified latent space, where both text-to-image and image-to-text tasks are handled through an autoregressive framework similar to LLMs. After validating the feasibility of this approach, Janus-Pro revealed significant gaps in generation performance when compared to diffusion-based models such as Stable Diffusion, which is a known challenge for autoregressive models in image generation. Janus-Pro improves upon this by scaling the model and data, and introducing an optimized training strategy. As a result, Janus-Pro has achieved state-of-the-art performance in multimodal understanding tasks and has surpassed diffusion models such as DALL-E 3 in terms of text-to-image generation capabilities.
\textit{However, given its strong multimodel understanding performance, how about Janus-Pro’s safety capability?}

\subsection{Experiments}

\subsubsection{Benchmarks}

We conduct a set of safety evaluations on Janus-Pro, utilizing two open-source benchmarks: SIUO\cite{wang2024cross} and MM-SafetyBench\cite{liu2023query}.
For assessing Cross-Modality Alignment, we used the SIUO dataset, which is developed to evaluate the integration, knowledge, and reasoning capabilities of LVLMs. SIUO consists of 167 samples spanning 9 critical safety domains, including self-harm, illegal activities, and privacy violations. In SIUO, both textual and visual inputs are individually safe, but their combination results in unsafe outputs, posing a challenge to cross-modal reasoning.
MM-SafetyBench is then used to evaluate Janus-Pro's defense capabilities under Multi-Model Attacks. This benchmark consists of 5,040 examples across 13 common scenarios involving malicious intent. The benchmark includes three distinct subcategories: (1) SD: Images generated by Stable Diffusion (SD) conditioned on malicious keywords, (2) OCR: Images containing malicious keywords extracted through Optical Character Recognition (OCR), and (3) SD + OCR: Images generated by SD and then subtitled with OCR.
The experiments are conducted on NVIDIA GeForce 4090 GPUs, using the GPT-4o-2024-05-13 API for evaluation. Additionally, temperature is set to 0 to ensure deterministic evaluation results, and manual review is conducted on the evaluation results to ensure accuracy.

\subsubsection{Results Analyse}

\noindent$\bullet$ \textbf{Results on SIUO.}
From the results presented in~\cref{tab: evaluation on siuo}, it is evident that Janus-Pro-7B exhibits suboptimal performance in OpenQA tasks. Its performance significantly lags behind that of LLaVA-1.5-7B, a model of comparable scale, with Janus-Pro achieving a ASR of 84.43\%, whereas LLaVA-1.5-7B performs at 78.44\%. This underperformance in open-ended question answering may be attributed to several factors, including potential limitations in Janus-Pro’s architecture, which may not yet be fully optimized for complex, open-ended reasoning tasks. Additionally, it is possible that the model's fine-tuning for open-ended question answering is still in development, leading to less robust responses in comparison to models like Qwen-VL-7B-Chat~\cite{bai2023qwen} and MiniGPT4-v2~\cite{minigpt4v2_arxiv23}
, which show ASR of 58.68\% and 58.08\% respectively, demonstrated more refined capabilities in this cross-modality alignment.
Conversely, Janus-Pro-7B performs considerably better in MCQA (Multiple Choice Question Answering) tasks, where its ASR of 73.06\% is competitive with most other models, such as Qwen-VL-7B-Chat (79.04\%) and MiniGPT4-v2 (72.46\%). This improvement in performance suggests that Janus-Pro is better suited to structured response tasks, where the model’s predictive capabilities in selecting the most appropriate answer from predefined options are more effectively utilized. This ability to handle fixed-response tasks could enhance Janus-Pro’s safety capabilities, as it is less prone to generating unsafe outputs in well-defined, closed-question settings. The model's autoregressive framework, which excels in generating coherent responses within such structured contexts, contributes to this improved safety performance.

\noindent$\bullet$ \textbf{Result on MM-SafetyBench.}
Turning to the results in~\cref{tab: evaluation on mmsafetybench}, which evaluates Janus-Pro's performance on MM-SafetyBench, a similar pattern of performance discrepancies emerges. In the first half of the table (Scenarios 1-6), which includes safety-critical tasks such as identifying illegal activities, hate speech, and malware generation, Janus-Pro consistently underperforms relative to the LLaVA series models. For example, in the Illegal Activity scenario, Janus-Pro achieves an ASR of 36.08\% compared to LLaVA-1.5-7B's 25.77\%. This observation suggests that Janus-Pro may be less adept at addressing safety-sensitive tasks, particularly those involving illegal activities or violence, possibly due to differences in its model architecture or training methodology. We speculate that Janus-Pro’s design of a unified latent space for both text and image generation may struggle to effectively capture and mitigate patterns in scenarios that require highly specialized safety mechanisms. Its encoder architecture, designed to process visual and textual inputs in parallel, might not be sufficiently fine-tuned for the nuanced detection of harmful or malicious content, especially in cases involving high-risk scenarios.
In contrast, Janus-Pro demonstrates notable improvement in the second half of the table (Scenarios 7-13), which includes tasks such as political lobbying, privacy violations, and government decision-making. The enhanced performance in these scenarios suggests that Janus-Pro may be more safety when dealing with context-specific tasks that require multimodal reasoning, where the model can leverage its integrated understanding of both textual and visual inputs. The structured nature of these scenarios may align more closely with Janus-Pro’s autoregressive framework, which excels in tasks requiring coherent output generation based on contextual input. This improvement could also be due to the model's stronger training on data that includes these types of tasks, or it may reflect an inherent advantage in processing structured tasks where text and images contribute complementary information. 

\noindent$\bullet$ \textbf{Conclusion.}
While Janus-Pro has achieved impressive multimodal understanding capabilities, its safety performance remains a significant limitation. Across multiple benchmarks, Janus-Pro fails to meet the basic safety standards of most other models. We speculate that this shortcoming may be due to the model architecture, which was designed to simultaneously handle both understanding and generation tasks, potentially at the expense of specialized safety mechanisms. Additionally, it is possible that Janus-Pro did not undergo specific safety-focused training, which may be contributing to its limited ability to recognize and mitigate harmful or adversarial inputs. This could also be related to the capabilities of the chosen LLM architecture used in Janus-Pro.
Given the critical role of safety in deploying multimodal models in real-world applications, it is evident that the safety capabilities of DeepSeek Janus-Pro need substantial improvements. Further refinements in its architecture and training methodology, with a stronger focus on safety and adversarial robustness, are essential for enhancing Janus-Pro’s effectiveness across diverse, high-risk tasks and scenarios.

\section{Outlook}
\label{sec: Outlook}

\subsection{Future Trends}
\label{sec: Future Trends}

Based on the reviewed research, we list several future research directions that we believe should be pursued.

\noindent$\bullet$ \textbf{The Shift Towards Black-box Attacks.}
A key future direction in attack methodologies is the increasing focus on black-box attacks, which offer advantages over traditional white-box approaches. While effective, white-box attacks require extensive prior knowledge of the target model, limiting their applicability and introducing significant computational overhead~\cite{qi2024visual, bailey2024image, gao2024inducing, wang2024white}. In contrast, black-box attacks exploit the intrinsic capabilities of LVLMs—such as OCR~\cite{gong2023figstep, qraitem2024vision}, logical reasoning~\cite{ma2024visual}, associative memory~\cite{zou2024image}, and multimodal integration~\cite{wang2024jailbreak}—to target vulnerabilities without direct access to the model's architecture, enhancing transferability and resource efficiency. However, prompt-based defenses~\cite{wang2024adashield, pi2024mllm, oh2024uniguard} often mitigate these attacks, exposing the limitations of current strategies. Future research should focus on developing more advanced black-box attack techniques that can circumvent defenses and demonstrate greater resilience, ensuring their robustness as LVLMs see broader deployment.

\noindent$\bullet$ \textbf{Enhancing Safety through Cross-Modality Alignment.}
Most defense mechanisms focus on detecting harmful inputs, addressing obvious attacks~\cite{zong2024safety, helff2024llavaguard}, yet LVLMs in real-world applications face subtler threats. For example, individually innocuous image and text inputs can combine to produce unsafe outputs~\cite{wang2024cross, zhou2024multimodal}. Visual components often struggle to identify unsafe elements or grasp contextual nuances, and research on aligning safety across visual and textual modalities remains limited~\cite{xu2024crosssafety}. Future efforts should focus on bridging the security capabilities of LLMs and vision encoders to address these modality gaps. Ensuring seamless safety integration across modalities is essential to prevent harmful interpretations arising from their combination. Additionally, improving contextual understanding in joint visual-textual processing will be crucial for enhancing the robustness and reliability of LVLMs in dynamic environments.

\noindent$\bullet$ \textbf{Diversifying Safety Fine-Tuning Techniques.}
Balancing the enhancement of safety while maintaining the general capabilities of LVLMs remains a significant challenge. Traditional fine-tuning approaches often risk compromising the model's overall performance in the pursuit of improved safety measures. To address this dilemma, future research should explore a broader range of safety fine-tuning methodologies, such as Reinforcement Learning from Human Feedback (RLHF)~\cite{kaufmann2023survey}, adversarial training, and multi-objective optimization. RLHF, in particular, offers a promising direction by enabling models to iteratively learn safety-oriented behaviors from refined feedback, reducing the reliance on static, rule-based constraints. Additionally, techniques like curriculum learning could help models gradually adapt to increasingly complex safety scenarios without sacrificing their ability to generalize across diverse tasks. Hybrid approaches that combine multiple fine-tuning strategies may further enhance safety while preserving or even improving the model's overall capabilities. These advancements are crucial for developing robust, safety-aware LVLMs that can meet the demands of real-world applications without significant trade-offs.

\noindent$\bullet$ \textbf{Developing Unified Strategy Benchmarking Frameworks.}
The rapid diversification of attack and defense methodologies for LVLMs has led to fragmented experimental environments, obstructing meaningful cross-method comparisons of effectiveness, efficiency, and overall performance. While existing benchmark MMJBench~\cite{weng2024textit}, offers evaluations of various strategies, but it employs a limited set of assessment methods and lack a comprehensive, general framework. To address these shortcomings, future research may prioritize the development of standardized benchmarking frameworks that unify the evaluation of diverse strategies. These benchmarks should encompass a broad range of metrics, including attack success rates, computational resource requirements, response times, and resilience against adaptive defenses. Additionally, incorporating diverse datasets and realistic threat models is essential to ensure evaluations accurately reflect real-world scenarios. The creation of open-source benchmark suites, supported by community contributions, will enhance transparency and reproducibility, enabling researchers to validate and build upon each other's work more effectively. By implementing comprehensive benchmarking strategies, the research community can systematically assess the strengths and limitations of existing approaches, drive the innovation of more robust and efficient solutions, and ultimately advance the security and reliability of LVLMs in practical applications.

\subsection{Conclusion}
\label{sec: Conclusion}

To the best of our knowledge, this is the first survey to offer a comprehensive and systematic review of recent advances in all field of LVLM safety from attacks, defenses, and evaluations, with an analysis of over 100 methods. We present the background of LVLM safety, emphasizing the unique vulnerabilities inherent in these models and introducing fundamental attack classifications. We categorize attack and defense strategies based on the model lifecycle, distinguishing between inference-phase and training-phase methods, and provide detailed sub-classifications with in-depth descriptions of each approach. In the Evaluation section, we synthesize all relevant benchmarks, providing a valuable resource for researchers seeking a comprehensive understanding of the field. Finally, we offer insights into future research directions and highlight open challenges, aiming to encourage further exploration and engagement from the research community in this critical area.

\ifCLASSOPTIONcaptionsoff
  \newpage
\fi

\vspace{-4pt}
{\small
\bibliographystyle{IEEEtran}
\bibliography{LVLM_Safety_Survey}

\begin{thebibliography}{100}
\providecommand{\url}[1]{#1}
\csname url@samestyle\endcsname
\providecommand{\newblock}{\relax}
\providecommand{\bibinfo}[2]{#2}
\providecommand{\BIBentrySTDinterwordspacing}{\spaceskip=0pt\relax}
\providecommand{\BIBentryALTinterwordstretchfactor}{4}
\providecommand{\BIBentryALTinterwordspacing}{\spaceskip=\fontdimen2\font plus
\BIBentryALTinterwordstretchfactor\fontdimen3\font minus \fontdimen4\font\relax}
\providecommand{\BIBforeignlanguage}[2]{{%
\expandafter\ifx\csname l@#1\endcsname\relax
\typeout{** WARNING: IEEEtran.bst: No hyphenation pattern has been}%
\typeout{** loaded for the language `#1'. Using the pattern for}%
\typeout{** the default language instead.}%
\else
\language=\csname l@#1\endcsname
\fi
#2}}
\providecommand{\BIBdecl}{\relax}
\BIBdecl

\bibitem{zhao2023survey}
W.~X. Zhao, K.~Zhou, J.~Li, T.~Tang, X.~Wang, Y.~Hou, Y.~Min, B.~Zhang, J.~Zhang, Z.~Dong \emph{et~al.}, ``A survey of large language models,'' \emph{arXiv preprint arXiv:2303.18223}, 2023.

\bibitem{minaee2024large}
S.~Minaee, T.~Mikolov, N.~Nikzad, M.~Chenaghlu, R.~Socher, X.~Amatriain, and J.~Gao, ``Large language models: A survey,'' \emph{arXiv preprint arXiv:2402.06196}, 2024.

\bibitem{brown2020language}
T.~Brown, B.~Mann, N.~Ryder, M.~Subbiah, J.~D. Kaplan, P.~Dhariwal, A.~Neelakantan, P.~Shyam, G.~Sastry, A.~Askell \emph{et~al.}, ``Language models are few-shot learners,'' \emph{NeurIPS}, pp. 1877--1901, 2020.

\bibitem{chowdhery2023palm}
A.~Chowdhery, S.~Narang, J.~Devlin, M.~Bosma, G.~Mishra, A.~Roberts, P.~Barham, H.~W. Chung, C.~Sutton, S.~Gehrmann \emph{et~al.}, ``Palm: Scaling language modeling with pathways,'' \emph{JMLR}, pp. 1--113, 2023.

\bibitem{touvron2023llama}
H.~Touvron, T.~Lavril, G.~Izacard, X.~Martinet, M.-A. Lachaux, T.~Lacroix, B.~Rozi{\`e}re, N.~Goyal, E.~Hambro, F.~Azhar \emph{et~al.}, ``Llama: Open and efficient foundation language models,'' \emph{arXiv preprint arXiv:2302.13971}, 2023.

\bibitem{awais2025foundation}
M.~Awais, M.~Naseer, S.~Khan, R.~M. Anwer, H.~Cholakkal, M.~Shah, M.-H. Yang, and F.~S. Khan, ``Foundation models defining a new era in vision: a survey and outlook,'' \emph{IEEE Transactions on Pattern Analysis and Machine Intelligence}, 2025.

\bibitem{kaur2024text}
P.~Kaur, G.~S. Kashyap, A.~Kumar, M.~T. Nafis, S.~Kumar, and V.~Shokeen, ``From text to transformation: A comprehensive review of large language models' versatility,'' \emph{arXiv preprint arXiv:2402.16142}, 2024.

\bibitem{cai2024internlm2}
Z.~Cai, M.~Cao, H.~Chen, K.~Chen, K.~Chen, X.~Chen, X.~Chen, Z.~Chen, Z.~Chen, P.~Chu \emph{et~al.}, ``Internlm2 technical report,'' \emph{arXiv preprint arXiv:2403.17297}, 2024.

\bibitem{jiang2024mixtral}
A.~Q. Jiang, A.~Sablayrolles, A.~Roux, A.~Mensch, B.~Savary, C.~Bamford, D.~S. Chaplot, D.~d.~l. Casas, E.~B. Hanna, F.~Bressand \emph{et~al.}, ``Mixtral of experts,'' \emph{arXiv preprint arXiv:2401.04088}, 2024.

\bibitem{bai2023qwenllm}
J.~Bai, S.~Bai, Y.~Chu, Z.~Cui, K.~Dang, X.~Deng, Y.~Fan, W.~Ge, Y.~Han, F.~Huang \emph{et~al.}, ``Qwen technical report,'' \emph{arXiv preprint arXiv:2309.16609}, 2023.

\bibitem{xu2024lvlm}
P.~Xu, W.~Shao, K.~Zhang, P.~Gao, S.~Liu, M.~Lei, F.~Meng, S.~Huang, Y.~Qiao, and P.~Luo, ``Lvlm-ehub: A comprehensive evaluation benchmark for large vision-language models,'' \emph{IEEE TPAMI}, 2024.

\bibitem{radford2021learning}
A.~Radford, J.~W. Kim, C.~Hallacy, A.~Ramesh, G.~Goh, S.~Agarwal, G.~Sastry, A.~Askell, P.~Mishkin, J.~Clark \emph{et~al.}, ``Learning transferable visual models from natural language supervision,'' in \emph{ICML}.\hskip 1em plus 0.5em minus 0.4em\relax PMLR, 2021, pp. 8748--8763.

\bibitem{alayrac2022flamingo}
J.-B. Alayrac, J.~Donahue, P.~Luc, A.~Miech, I.~Barr, Y.~Hasson, K.~Lenc, A.~Mensch, K.~Millican, M.~Reynolds \emph{et~al.}, ``Flamingo: a visual language model for few-shot learning,'' \emph{NeurIPS}, pp. 23\,716--23\,736, 2022.

\bibitem{li2023blip}
J.~Li, D.~Li, S.~Savarese, and S.~Hoi, ``Blip-2: Bootstrapping language-image pre-training with frozen image encoders and large language models,'' in \emph{ICML}.\hskip 1em plus 0.5em minus 0.4em\relax PMLR, 2023, pp. 19\,730--19\,742.

\bibitem{liu2024visual}
H.~Liu, C.~Li, Q.~Wu, and Y.~J. Lee, ``Visual instruction tuning,'' \emph{Advances in neural information processing systems}, vol.~36, 2024.

\bibitem{achiam2023gpt}
J.~Achiam, S.~Adler, S.~Agarwal, L.~Ahmad, I.~Akkaya, F.~L. Aleman, D.~Almeida, J.~Altenschmidt, S.~Altman, S.~Anadkat \emph{et~al.}, ``Gpt-4 technical report,'' \emph{arXiv preprint arXiv:2303.08774}, 2023.

\bibitem{zhang2024vision}
J.~Zhang, J.~Huang, S.~Jin, and S.~Lu, ``Vision-language models for vision tasks: A survey,'' \emph{IEEE Transactions on Pattern Analysis and Machine Intelligence}, 2024.

\bibitem{jia2021scaling}
C.~Jia, Y.~Yang, Y.~Xia, Y.-T. Chen, Z.~Parekh, H.~Pham, Q.~Le, Y.-H. Sung, Z.~Li, and T.~Duerig, ``Scaling up visual and vision-language representation learning with noisy text supervision,'' in \emph{ICML}.\hskip 1em plus 0.5em minus 0.4em\relax PMLR, 2021, pp. 4904--4916.

\bibitem{tian2024drivevlm}
X.~Tian, J.~Gu, B.~Li, Y.~Liu, Y.~Wang, Z.~Zhao, K.~Zhan, P.~Jia, X.~Lang, and H.~Zhao, ``Drivevlm: The convergence of autonomous driving and large vision-language models,'' \emph{arXiv preprint arXiv:2402.12289}, 2024.

\bibitem{van2024large}
M.-H. Van, P.~Verma, and X.~Wu, ``On large visual language models for medical imaging analysis: An empirical study,'' in \emph{CHASE}.\hskip 1em plus 0.5em minus 0.4em\relax IEEE, 2024, pp. 172--176.

\bibitem{maharana2022storydall}
A.~Maharana, D.~Hannan, and M.~Bansal, ``Storydall-e: Adapting pretrained text-to-image transformers for story continuation,'' in \emph{ECCV}.\hskip 1em plus 0.5em minus 0.4em\relax Springer, 2022, pp. 70--87.

\bibitem{zhou2024storydiffusion}
Y.~Zhou, D.~Zhou, M.-M. Cheng, J.~Feng, and Q.~Hou, ``Storydiffusion: Consistent self-attention for long-range image and video generation,'' \emph{arXiv preprint arXiv:2405.01434}, 2024.

\bibitem{qi2024visual}
X.~Qi, K.~Huang, A.~Panda, P.~Henderson, M.~Wang, and P.~Mittal, ``Visual adversarial examples jailbreak aligned large language models,'' in \emph{AAAI}, 2024, pp. 21\,527--21\,536.

\bibitem{schlarmann2023adversarial}
C.~Schlarmann and M.~Hein, ``On the adversarial robustness of multi-modal foundation models,'' in \emph{ICCV}, 2023, pp. 3677--3685.

\bibitem{zhao2024evaluating}
Y.~Zhao, T.~Pang, C.~Du, X.~Yang, C.~Li, N.-M.~M. Cheung, and M.~Lin, ``On evaluating adversarial robustness of large vision-language models,'' \emph{NeurIPS}, vol.~36, 2024.

\bibitem{wang2024white}
R.~Wang, X.~Ma, H.~Zhou, C.~Ji, G.~Ye, and Y.-G. Jiang, ``White-box multimodal jailbreaks against large vision-language models,'' in \emph{ACM MM}, 2024, pp. 6920--6928.

\bibitem{bailey2024image}
L.~Bailey, E.~Ong, S.~Russell, and S.~Emmons, ``Image hijacks: Adversarial images can control generative models at runtime,'' in \emph{ICML}, 2024.

\bibitem{ding2024eta}
Y.~Ding, B.~Li, and R.~Zhang, ``Eta: Evaluating then aligning safety of vision language models at inference time,'' \emph{arXiv preprint arXiv:2410.06625}, 2024.

\bibitem{xu2024crosssafety}
S.~Xu, L.~Pang, Y.~Zhu, H.~Shen, and X.~Cheng, ``Cross-modal safety mechanism transfer in large vision-language models,'' \emph{arXiv preprint arXiv:2410.12662}, 2024.

\bibitem{xu2024shadowcast}
Y.~Xu, J.~Yao, M.~Shu, Y.~Sun, Z.~Wu, N.~Yu, T.~Goldstein, and F.~Huang, ``Shadowcast: Stealthy data poisoning attacks against vision-language models,'' \emph{arXiv preprint arXiv:2402.06659}, 2024.

\bibitem{liang2024vl}
J.~Liang, S.~Liang, M.~Luo, A.~Liu, D.~Han, E.-C. Chang, and X.~Cao, ``Vl-trojan: Multimodal instruction backdoor attacks against autoregressive visual language models,'' \emph{arXiv preprint arXiv:2402.13851}, 2024.

\bibitem{ni2024physical}
Z.~Ni, R.~Ye, Y.~Wei, Z.~Xiang, Y.~Wang, and S.~Chen, ``Physical backdoor attack can jeopardize driving with vision-large-language models,'' \emph{arXiv preprint arXiv:2404.12916}, 2024.

\bibitem{liu2024safety}
X.~Liu, Y.~Zhu, Y.~Lan, C.~Yang, and Y.~Qiao, ``Safety of multimodal large language models on images and text,'' \emph{arXiv preprint arXiv:2402.00357}, 2024.

\bibitem{fan2024unbridled}
Y.~Fan, Y.~Cao, Z.~Zhao, Z.~Liu, and S.~Li, ``Unbridled icarus: A survey of the potential perils of image inputs in multimodal large language model security,'' \emph{arXiv preprint arXiv:2404.05264}, 2024.

\bibitem{wang2024llms}
S.~Wang, Z.~Long, Z.~Fan, and Z.~Wei, ``From llms to mllms: Exploring the landscape of multimodal jailbreaking,'' \emph{arXiv preprint arXiv:2406.14859}, 2024.

\bibitem{jin2024jailbreakzoo}
H.~Jin, L.~Hu, X.~Li, P.~Zhang, C.~Chen, J.~Zhuang, and H.~Wang, ``Jailbreakzoo: Survey, landscapes, and horizons in jailbreaking large language and vision-language models,'' \emph{arXiv preprint arXiv:2407.01599}, 2024.

\bibitem{liu2024survey}
D.~Liu, M.~Yang, X.~Qu, P.~Zhou, Y.~Cheng, and W.~Hu, ``A survey of attacks on large vision-language models: Resources, advances, and future trends,'' \emph{arXiv preprint arXiv:2407.07403}, 2024.

\bibitem{zhang2024adversarial}
C.~Zhang, X.~Xu, J.~Wu, Z.~Liu, and L.~Zhou, ``Adversarial attacks of vision tasks in the past 10 years: A survey,'' \emph{arXiv preprint arXiv:2410.23687}, 2024.

\bibitem{liu2024jailbreak}
X.~Liu, X.~Cui, P.~Li, Z.~Li, H.~Huang, S.~Xia, M.~Zhang, Y.~Zou, and R.~He, ``Jailbreak attacks and defenses against multimodal generative models: A survey,'' \emph{arXiv preprint arXiv:2411.09259}, 2024.

\bibitem{li2024images}
Y.~Li, H.~Guo, K.~Zhou, W.~X. Zhao, and J.-R. Wen, ``Images are achilles' heel of alignment: Exploiting visual vulnerabilities for jailbreaking multimodal large language models,'' \emph{arXiv preprint arXiv:2403.09792}, 2024.

\bibitem{gong2023figstep}
Y.~Gong, D.~Ran, J.~Liu, C.~Wang, T.~Cong, A.~Wang, S.~Duan, and X.~Wang, ``Figstep: Jailbreaking large vision-language models via typographic visual prompts,'' \emph{arXiv preprint arXiv:2311.05608}, 2023.

\bibitem{lee2024does}
S.~Lee, G.~Kim, J.~Kim, H.~Lee, H.~Chang, S.~H. Park, and M.~Seo, ``How does vision-language adaptation impact the safety of vision language models?'' \emph{arXiv preprint arXiv:2410.07571}, 2024.

\bibitem{wang2024adashield}
Y.~Wang, X.~Liu, Y.~Li, M.~Chen, and C.~Xiao, ``Adashield: Safeguarding multimodal large language models from structure-based attack via adaptive shield prompting,'' \emph{arXiv preprint arXiv:2403.09513}, 2024.

\bibitem{xu2024cross}
Y.~Xu, X.~Qi, Z.~Qin, and W.~Wang, ``Cross-modality information check for detecting jailbreaking in multimodal large language models,'' \emph{arXiv preprint arXiv:2407.21659}, 2024.

\bibitem{wang2024inferaligner}
P.~Wang, D.~Zhang, L.~Li, C.~Tan, X.~Wang, K.~Ren, B.~Jiang, and X.~Qiu, ``Inferaligner: Inference-time alignment for harmlessness through cross-model guidance,'' \emph{arXiv preprint arXiv:2401.11206}, 2024.

\bibitem{gao2024coca}
J.~Gao, R.~Pi, T.~Han, H.~Wu, L.~Hong, L.~Kong, X.~Jiang, and Z.~Li, ``Coca: Regaining safety-awareness of multimodal large language models with constitutional calibration,'' \emph{arXiv preprint arXiv:2409.11365}, 2024.

\bibitem{zhang2023mutation}
X.~Zhang, C.~Zhang, T.~Li, Y.~Huang, X.~Jia, X.~Xie, Y.~Liu, and C.~Shen, ``A mutation-based method for multi-modal jailbreaking attack detection,'' \emph{arXiv preprint arXiv:2312.10766}, 2023.

\bibitem{pi2024mllm}
R.~Pi, T.~Han, J.~Zhang, Y.~Xie, R.~Pan, Q.~Lian, H.~Dong, J.~Zhang, and T.~Zhang, ``Mllm-protector: Ensuring mllm's safety without hurting performance,'' \emph{arXiv preprint arXiv:2401.02906}, 2024.

\bibitem{zong2024safety}
Y.~Zong, O.~Bohdal, T.~Yu, Y.~Yang, and T.~Hospedales, ``Safety fine-tuning at (almost) no cost: A baseline for vision large language models,'' \emph{arXiv preprint arXiv:2402.02207}, 2024.

\bibitem{zhang2024spa}
Y.~Zhang, L.~Chen, G.~Zheng, Y.~Gao, R.~Zheng, J.~Fu, Z.~Yin, S.~Jin, Y.~Qiao, X.~Huang \emph{et~al.}, ``Spa-vl: A comprehensive safety preference alignment dataset for vision language model,'' \emph{arXiv preprint arXiv:2406.12030}, 2024.

\bibitem{liu2024safetyalignment}
Z.~Liu, Y.~Nie, Y.~Tan, X.~Yue, Q.~Cui, C.~Wang, X.~Zhu, and B.~Zheng, ``Safety alignment for vision language models,'' \emph{arXiv preprint arXiv:2405.13581}, 2024.

\bibitem{tu2023many}
H.~Tu, C.~Cui, Z.~Wang, Y.~Zhou, B.~Zhao, J.~Han, W.~Zhou, H.~Yao, and C.~Xie, ``How many unicorns are in this image? a safety evaluation benchmark for vision llms,'' \emph{arXiv preprint arXiv:2311.16101}, 2023.

\bibitem{liu2023query}
X.~Liu, Y.~Zhu, Y.~Lan, C.~Yang, and Y.~Qiao, ``Query-relevant images jailbreak large multi-modal models,'' \emph{arXiv preprint arXiv:2311.17600}, 2023.

\bibitem{luo2024jailbreakv}
W.~Luo, S.~Ma, X.~Liu, X.~Guo, and C.~Xiao, ``Jailbreakv-28k: A benchmark for assessing the robustness of multimodal large language models against jailbreak attacks,'' \emph{arXiv preprint arXiv:2404.03027}, 2024.

\bibitem{zhang2024benchmarking}
Y.~Zhang, Y.~Huang, Y.~Sun, C.~Liu, Z.~Zhao, Z.~Fang, Y.~Wang, H.~Chen, X.~Yang, X.~Wei \emph{et~al.}, ``Benchmarking trustworthiness of multimodal large language models: A comprehensive study,'' \emph{arXiv preprint arXiv:2406.07057}, 2024.

\bibitem{gu2024mllmguard}
T.~Gu, Z.~Zhou, K.~Huang, D.~Liang, Y.~Wang, H.~Zhao, Y.~Yao, X.~Qiao, K.~Wang, Y.~Yang \emph{et~al.}, ``Mllmguard: A multi-dimensional safety evaluation suite for multimodal large language models,'' \emph{arXiv preprint arXiv:2406.07594}, 2024.

\bibitem{radford2019language}
A.~Radford, J.~Wu, R.~Child, D.~Luan, D.~Amodei, I.~Sutskever \emph{et~al.}, ``Language models are unsupervised multitask learners,'' \emph{OpenAI blog}, p.~9, 2019.

\bibitem{anil2023palm}
R.~Anil, A.~M. Dai, O.~Firat, M.~Johnson, D.~Lepikhin, A.~Passos, S.~Shakeri, E.~Taropa, P.~Bailey, Z.~Chen \emph{et~al.}, ``Palm 2 technical report,'' \emph{arXiv preprint arXiv:2305.10403}, 2023.

\bibitem{vicuna2023}
\BIBentryALTinterwordspacing
W.-L. Chiang, Z.~Li, Z.~Lin, Y.~Sheng, Z.~Wu, H.~Zhang, L.~Zheng, S.~Zhuang, Y.~Zhuang, J.~E. Gonzalez, I.~Stoica, and E.~P. Xing, ``Vicuna: An open-source chatbot impressing gpt-4 with 90\%* chatgpt quality,'' March 2023. [Online]. Available: \url{https://lmsys.org/blog/2023-03-30-vicuna/}
\BIBentrySTDinterwordspacing

\bibitem{team2023gemini}
G.~Team, R.~Anil, S.~Borgeaud, J.-B. Alayrac, J.~Yu, R.~Soricut, J.~Schalkwyk, A.~M. Dai, A.~Hauth, K.~Millican \emph{et~al.}, ``Gemini: a family of highly capable multimodal models,'' \emph{arXiv preprint arXiv:2312.11805}, 2023.

\bibitem{zhu2023minigpt}
D.~Zhu, J.~Chen, X.~Shen, X.~Li, and M.~Elhoseiny, ``Minigpt-4: Enhancing vision-language understanding with advanced large language models,'' \emph{arXiv preprint arXiv:2304.10592}, 2023.

\bibitem{su2023pandagpt}
Y.~Su, T.~Lan, H.~Li, J.~Xu, Y.~Wang, and D.~Cai, ``Pandagpt: One model to instruction-follow them all,'' \emph{arXiv preprint arXiv:2305.16355}, 2023.

\bibitem{li2024llava}
B.~Li, Y.~Zhang, D.~Guo, R.~Zhang, F.~Li, H.~Zhang, K.~Zhang, Y.~Li, Z.~Liu, and C.~Li, ``Llava-onevision: Easy visual task transfer,'' \emph{arXiv preprint arXiv:2408.03326}, 2024.

\bibitem{chen2024internvl}
Z.~Chen, J.~Wu, W.~Wang, W.~Su, G.~Chen, S.~Xing, M.~Zhong, Q.~Zhang, X.~Zhu, L.~Lu \emph{et~al.}, ``Internvl: Scaling up vision foundation models and aligning for generic visual-linguistic tasks,'' in \emph{CVPR}, 2024, pp. 24\,185--24\,198.

\bibitem{bai2023qwen}
J.~Bai, S.~Bai, S.~Yang, S.~Wang, S.~Tan, P.~Wang, J.~Lin, C.~Zhou, and J.~Zhou, ``Qwen-vl: A frontier large vision-language model with versatile abilities,'' \emph{arXiv preprint arXiv:2308.12966}, 2023.

\bibitem{lin2024vila}
J.~Lin, H.~Yin, W.~Ping, P.~Molchanov, M.~Shoeybi, and S.~Han, ``Vila: On pre-training for visual language models,'' in \emph{CVPR}, 2024, pp. 26\,689--26\,699.

\bibitem{pi2024image}
R.~Pi, J.~Zhang, J.~Zhang, R.~Pan, Z.~Chen, and T.~Zhang, ``Image textualization: An automatic framework for creating accurate and detailed image descriptions,'' \emph{arXiv preprint arXiv:2406.07502}, 2024.

\bibitem{pi2024personalized}
R.~Pi, J.~Zhang, T.~Han, J.~Zhang, R.~Pan, and T.~Zhang, ``Personalized visual instruction tuning,'' \emph{arXiv preprint arXiv:2410.07113}, 2024.

\bibitem{gallegos2024bias}
I.~O. Gallegos, R.~A. Rossi, J.~Barrow, M.~M. Tanjim, S.~Kim, F.~Dernoncourt, T.~Yu, R.~Zhang, and N.~K. Ahmed, ``Bias and fairness in large language models: A survey,'' \emph{CL}, pp. 1--79, 2024.

\bibitem{shen2024anything}
X.~Shen, Z.~Chen, M.~Backes, Y.~Shen, and Y.~Zhang, ``" do anything now": Characterizing and evaluating in-the-wild jailbreak prompts on large language models,'' in \emph{ACM CCS}, 2024, pp. 1671--1685.

\bibitem{yi2024jailbreak}
S.~Yi, Y.~Liu, Z.~Sun, T.~Cong, X.~He, J.~Song, K.~Xu, and Q.~Li, ``Jailbreak attacks and defenses against large language models: A survey,'' \emph{arXiv preprint arXiv:2407.04295}, 2024.

\bibitem{guo2024vllm}
Y.~Guo, F.~Jiao, L.~Nie, and M.~Kankanhalli, ``The vllm safety paradox: Dual ease in jailbreak attack and defense,'' \emph{arXiv preprint arXiv:2411.08410}, 2024.

\bibitem{pantazopoulos2024learning}
G.~Pantazopoulos, A.~Parekh, M.~Nikandrou, and A.~Suglia, ``Learning to see but forgetting to follow: Visual instruction tuning makes llms more prone to jailbreak attacks,'' \emph{arXiv preprint arXiv:2405.04403}, 2024.

\bibitem{bachu2024unfair}
S.~Bachu, E.~Shayegani, T.~Chakraborty, R.~Lal, A.~Dutta, C.~Song, Y.~Dong, N.~Abu-Ghazaleh, and A.~K. Roy-Chowdhury, ``Unfair alignment: Examining safety alignment across vision encoder layers in vision-language models,'' \emph{arXiv preprint arXiv:2411.04291}, 2024.

\bibitem{chakraborty2021survey}
A.~Chakraborty, M.~Alam, V.~Dey, A.~Chattopadhyay, and D.~Mukhopadhyay, ``A survey on adversarial attacks and defences,'' \emph{CAAI TIT}, pp. 25--45, 2021.

\bibitem{huang2017adversarial}
S.~Huang, N.~Papernot, I.~Goodfellow, Y.~Duan, and P.~Abbeel, ``Adversarial attacks on neural network policies,'' \emph{arXiv preprint arXiv:1702.02284}, 2017.

\bibitem{chakraborty2018adversarial}
A.~Chakraborty, M.~Alam, V.~Dey, A.~Chattopadhyay, and D.~Mukhopadhyay, ``Adversarial attacks and defences: A survey,'' \emph{arXiv preprint arXiv:1810.00069}, 2018.

\bibitem{goodfellow2014explaining}
I.~J. Goodfellow, J.~Shlens, and C.~Szegedy, ``Explaining and harnessing adversarial examples,'' \emph{arXiv preprint arXiv:1412.6572}, 2014.

\bibitem{madry2017towards}
A.~Madry, ``Towards deep learning models resistant to adversarial attacks,'' \emph{arXiv preprint arXiv:1706.06083}, 2017.

\bibitem{cheng2019improving}
S.~Cheng, Y.~Dong, T.~Pang, H.~Su, and J.~Zhu, ``Improving black-box adversarial attacks with a transfer-based prior,'' \emph{NeurIPS}, vol.~32, 2019.

\bibitem{demontis2019adversarial}
A.~Demontis, M.~Melis, M.~Pintor, M.~Jagielski, B.~Biggio, A.~Oprea, C.~Nita-Rotaru, and F.~Roli, ``Why do adversarial attacks transfer? explaining transferability of evasion and poisoning attacks,'' in \emph{USENIX}, 2019, pp. 321--338.

\bibitem{qin2022boosting}
Z.~Qin, Y.~Fan, Y.~Liu, L.~Shen, Y.~Zhang, J.~Wang, and B.~Wu, ``Boosting the transferability of adversarial attacks with reverse adversarial perturbation,'' \emph{NeurIPS}, vol.~35, pp. 29\,845--29\,858, 2022.

\bibitem{perez2022ignore}
F.~Perez and I.~Ribeiro, ``Ignore previous prompt: Attack techniques for language models,'' \emph{arXiv preprint arXiv:2211.09527}, 2022.

\bibitem{yao2024fuzzllm}
D.~Yao, J.~Zhang, I.~G. Harris, and M.~Carlsson, ``Fuzzllm: A novel and universal fuzzing framework for proactively discovering jailbreak vulnerabilities in large language models,'' in \emph{ICASSP}.\hskip 1em plus 0.5em minus 0.4em\relax IEEE, 2024, pp. 4485--4489.

\bibitem{shayegani2023jailbreak}
E.~Shayegani, Y.~Dong, and N.~Abu-Ghazaleh, ``Jailbreak in pieces: Compositional adversarial attacks on multi-modal language models,'' in \emph{ICLR}, 2023.

\bibitem{tolpegin2020data}
V.~Tolpegin, S.~Truex, M.~E. Gursoy, and L.~Liu, ``Data poisoning attacks against federated learning systems,'' in \emph{ESORICs}.\hskip 1em plus 0.5em minus 0.4em\relax Springer, 2020, pp. 480--501.

\bibitem{saha2020hidden}
A.~Saha, A.~Subramanya, and H.~Pirsiavash, ``Hidden trigger backdoor attacks,'' in \emph{AAAI}, 2020, pp. 11\,957--11\,965.

\bibitem{li2020rethinking}
Y.~Li, T.~Zhai, B.~Wu, Y.~Jiang, Z.~Li, and S.~Xia, ``Rethinking the trigger of backdoor attack,'' \emph{arXiv preprint arXiv:2004.04692}, 2020.

\bibitem{zeng2021rethinking}
Y.~Zeng, W.~Park, Z.~M. Mao, and R.~Jia, ``Rethinking the backdoor attacks' triggers: A frequency perspective,'' in \emph{ICCV}, 2021, pp. 16\,473--16\,481.

\bibitem{li2021invisible}
Y.~Li, Y.~Li, B.~Wu, L.~Li, R.~He, and S.~Lyu, ``Invisible backdoor attack with sample-specific triggers,'' in \emph{ICCV}, 2021, pp. 16\,463--16\,472.

\bibitem{bagdasaryan2023ab}
E.~Bagdasaryan, T.-Y. Hsieh, B.~Nassi, and V.~Shmatikov, ``(ab) using images and sounds for indirect instruction injection in multi-modal llms,'' \emph{arXiv preprint arXiv:2307.10490}, 2023.

\bibitem{gao2024inducing}
K.~Gao, Y.~Bai, J.~Gu, S.-T. Xia, P.~Torr, Z.~Li, and W.~Liu, ``Inducing high energy-latency of large vision-language models with verbose images,'' \emph{arXiv preprint arXiv:2401.11170}, 2024.

\bibitem{lu2024test}
D.~Lu, T.~Pang, C.~Du, Q.~Liu, X.~Yang, and M.~Lin, ``Test-time backdoor attacks on multimodal large language models,'' \emph{arXiv preprint arXiv:2402.08577}, 2024.

\bibitem{wang2024stop}
Z.~Wang, Z.~Han, S.~Chen, F.~Xue, Z.~Ding, X.~Xiao, V.~Tresp, P.~Torr, and J.~Gu, ``Stop reasoning! when multimodal llms with chain-of-thought reasoning meets adversarial images,'' \emph{arXiv preprint arXiv:2402.14899}, 2024.

\bibitem{luo2024image}
H.~Luo, J.~Gu, F.~Liu, and P.~Torr, ``An image is worth 1000 lies: Adversarial transferability across prompts on vision-language models,'' \emph{arXiv preprint arXiv:2403.09766}, 2024.

\bibitem{gao2024adversarial}
K.~Gao, Y.~Bai, J.~Bai, Y.~Yang, and S.-T. Xia, ``Adversarial robustness for visual grounding of multimodal large language models,'' \emph{arXiv preprint arXiv:2405.09981}, 2024.

\bibitem{ying2024jailbreak}
Z.~Ying, A.~Liu, T.~Zhang, Z.~Yu, S.~Liang, X.~Liu, and D.~Tao, ``Jailbreak vision language models via bi-modal adversarial prompt,'' \emph{arXiv preprint arXiv:2406.04031}, 2024.

\bibitem{yang2024enhancing}
X.~Yang, X.~Tang, F.~Zhu, J.~Han, and S.~Hu, ``Enhancing cross-prompt transferability in vision-language models through contextual injection of target tokens,'' \emph{arXiv preprint arXiv:2406.13294}, 2024.

\bibitem{jang2024replace}
J.~Jang, H.~Lyu, J.~Koh, and H.~J. Yang, ``Replace-then-perturb: Targeted adversarial attacks with visual reasoning for vision-language models,'' \emph{arXiv preprint arXiv:2411.00898}, 2024.

\bibitem{dong2023robust}
Y.~Dong, H.~Chen, J.~Chen, Z.~Fang, X.~Yang, Y.~Zhang, Y.~Tian, H.~Su, and J.~Zhu, ``How robust is google's bard to adversarial image attacks?'' \emph{arXiv preprint arXiv:2309.11751}, 2023.

\bibitem{niu2024jailbreaking}
Z.~Niu, H.~Ren, X.~Gao, G.~Hua, and R.~Jin, ``Jailbreaking attack against multimodal large language model,'' \emph{arXiv preprint arXiv:2402.02309}, 2024.

\bibitem{gu2024agent}
X.~Gu, X.~Zheng, T.~Pang, C.~Du, Q.~Liu, Y.~Wang, J.~Jiang, and M.~Lin, ``Agent smith: A single image can jailbreak one million multimodal llm agents exponentially fast,'' \emph{arXiv preprint arXiv:2402.08567}, 2024.

\bibitem{tan2024wolf}
Z.~Tan, C.~Zhao, R.~Moraffah, Y.~Li, Y.~Kong, T.~Chen, and H.~Liu, ``The wolf within: Covert injection of malice into mllm societies via an mllm operative,'' \emph{arXiv preprint arXiv:2402.14859}, 2024.

\bibitem{qraitem2024vision}
M.~Qraitem, N.~Tasnim, P.~Teterwak, K.~Saenko, and B.~A. Plummer, ``Vision-llms can fool themselves with self-generated typographic attacks,'' \emph{arXiv preprint arXiv:2402.00626}, 2024.

\bibitem{ma2024visual}
S.~Ma, W.~Luo, Y.~Wang, X.~Liu, M.~Chen, B.~Li, and C.~Xiao, ``Visual-roleplay: Universal jailbreak attack on multimodal large language models via role-playing image characte,'' \emph{arXiv preprint arXiv:2405.20773}, 2024.

\bibitem{zou2024image}
X.~Zou, K.~Li, and Y.~Chen, ``Image-to-text logic jailbreak: Your imagination can help you do anything,'' \emph{arXiv preprint arXiv:2407.02534}, 2024.

\bibitem{wang2024ideator}
R.~Wang, B.~Wang, X.~Ma, and Y.-G. Jiang, ``Ideator: Jailbreaking vlms using vlms,'' \emph{arXiv preprint arXiv:2411.00827}, 2024.

\bibitem{wang2024jailbreak}
Y.~Wang, X.~Zhou, Y.~Wang, G.~Zhang, and T.~He, ``Jailbreak large visual language models through multi-modal linkage,'' \emph{arXiv preprint arXiv:2412.00473}, 2024.

\bibitem{teng2024heuristic}
M.~Teng, J.~Xiaojun, D.~Ranjie, L.~Xinfeng, H.~Yihao, C.~Zhixuan, L.~Yang, and R.~Wenqi, ``Heuristic-induced multimodal risk distribution jailbreak attack for multimodal large language models,'' \emph{arXiv preprint arXiv:2412.05934}, 2024.

\bibitem{awadalla2023openflamingo}
A.~Awadalla, I.~Gao, J.~Gardner, J.~Hessel, Y.~Hanafy, W.~Zhu, K.~Marathe, Y.~Bitton, S.~Gadre, S.~Sagawa \emph{et~al.}, ``Openflamingo: An open-source framework for training large autoregressive vision-language models,'' \emph{arXiv preprint arXiv:2308.01390}, 2023.

\bibitem{shumailov2021sponge}
I.~Shumailov, Y.~Zhao, D.~Bates, N.~Papernot, R.~Mullins, and R.~Anderson, ``Sponge examples: Energy-latency attacks on neural networks,'' in \emph{EuroS\&P}.\hskip 1em plus 0.5em minus 0.4em\relax IEEE, 2021, pp. 212--231.

\bibitem{chen2022nicgslowdown}
S.~Chen, Z.~Song, M.~Haque, C.~Liu, and W.~Yang, ``Nicgslowdown: Evaluating the efficiency robustness of neural image caption generation models,'' in \emph{CVPR}, 2022, pp. 15\,365--15\,374.

\bibitem{lu2022learn}
P.~Lu, S.~Mishra, T.~Xia, L.~Qiu, K.-W. Chang, S.-C. Zhu, O.~Tafjord, P.~Clark, and A.~Kalyan, ``Learn to explain: Multimodal reasoning via thought chains for science question answering,'' \emph{NeurIPS}, pp. 2507--2521, 2022.

\bibitem{zhang2023multimodal}
Z.~Zhang, A.~Zhang, M.~Li, H.~Zhao, G.~Karypis, and A.~Smola, ``Multimodal chain-of-thought reasoning in language models,'' \emph{arXiv preprint arXiv:2302.00923}, 2023.

\bibitem{he2024multi}
L.~He, Z.~Li, X.~Cai, and P.~Wang, ``Multi-modal latent space learning for chain-of-thought reasoning in language models,'' in \emph{AAAI}, 2024, pp. 18\,180--18\,187.

\bibitem{peng2023kosmos}
Z.~Peng, W.~Wang, L.~Dong, Y.~Hao, S.~Huang, S.~Ma, and F.~Wei, ``Kosmos-2: Grounding multimodal large language models to the world,'' \emph{arXiv preprint arXiv:2306.14824}, 2023.

\bibitem{wang2024visionllm}
W.~Wang, Z.~Chen, X.~Chen, J.~Wu, X.~Zhu, G.~Zeng, P.~Luo, T.~Lu, J.~Zhou, Y.~Qiao \emph{et~al.}, ``Visionllm: Large language model is also an open-ended decoder for vision-centric tasks,'' \emph{NeurIPS}, 2024.

\bibitem{li2021referring}
M.~Li and L.~Sigal, ``Referring transformer: A one-step approach to multi-task visual grounding,'' \emph{NeurIPS}, pp. 19\,652--19\,664, 2021.

\bibitem{wei2022chain}
J.~Wei, X.~Wang, D.~Schuurmans, M.~Bosma, F.~Xia, E.~Chi, Q.~V. Le, D.~Zhou \emph{et~al.}, ``Chain-of-thought prompting elicits reasoning in large language models,'' \emph{NeurIPS}, pp. 24\,824--24\,837, 2022.

\bibitem{chu2023survey}
Z.~Chu, J.~Chen, Q.~Chen, W.~Yu, T.~He, H.~Wang, W.~Peng, M.~Liu, B.~Qin, and T.~Liu, ``A survey of chain of thought reasoning: Advances, frontiers and future,'' \emph{arXiv preprint arXiv:2309.15402}, 2023.

\bibitem{li2022blip}
J.~Li, D.~Li, C.~Xiong, and S.~Hoi, ``Blip: Bootstrapping language-image pre-training for unified vision-language understanding and generation,'' in \emph{ICML}.\hskip 1em plus 0.5em minus 0.4em\relax PMLR, 2022, pp. 12\,888--12\,900.

\bibitem{wang2024break}
Y.~Wang, C.~Liu, Y.~Qu, H.~Cao, D.~Jiang, and L.~Xu, ``Break the visual perception: Adversarial attacks targeting encoded visual tokens of large vision-language models,'' in \emph{ACM MM}, 2024, pp. 1072--1081.

\bibitem{bird2009natural}
S.~Bird, E.~Klein, and E.~Loper, \emph{Natural language processing with Python: analyzing text with the natural language toolkit}.\hskip 1em plus 0.5em minus 0.4em\relax " O'Reilly Media, Inc.", 2009.

\bibitem{xu2023instructions}
J.~Xu, M.~D. Ma, F.~Wang, C.~Xiao, and M.~Chen, ``Instructions as backdoors: Backdoor vulnerabilities of instruction tuning for large language models,'' \emph{arXiv preprint arXiv:2305.14710}, 2023.

\bibitem{yan2024backdooring}
J.~Yan, V.~Yadav, S.~Li, L.~Chen, Z.~Tang, H.~Wang, V.~Srinivasan, X.~Ren, and H.~Jin, ``Backdooring instruction-tuned large language models with virtual prompt injection,'' in \emph{NAACL}, 2024, pp. 6065--6086.

\bibitem{tao2024imgtrojan}
X.~Tao, S.~Zhong, L.~Li, Q.~Liu, and L.~Kong, ``Imgtrojan: Jailbreaking vision-language models with one image,'' \emph{arXiv preprint arXiv:2403.02910}, 2024.

\bibitem{liang2024revisiting}
S.~Liang, J.~Liang, T.~Pang, C.~Du, A.~Liu, E.-C. Chang, and X.~Cao, ``Revisiting backdoor attacks against large vision-language models,'' \emph{arXiv preprint arXiv:2406.18844}, 2024.

\bibitem{lyu2024trojvlm}
W.~Lyu, L.~Pang, T.~Ma, H.~Ling, and C.~Chen, ``Trojvlm: Backdoor attack against vision language models,'' \emph{arXiv preprint arXiv:2409.19232}, 2024.

\bibitem{lyu2024backdooring}
W.~Lyu, J.~Yao, S.~Gupta, L.~Pang, T.~Sun, L.~Yi, L.~Hu, H.~Ling, and C.~Chen, ``Backdooring vision-language models with out-of-distribution data,'' \emph{arXiv preprint arXiv:2410.01264}, 2024.

\bibitem{sun2024safeguarding}
J.~Sun, C.~Wang, J.~Wang, Y.~Zhang, and C.~Xiao, ``Safeguarding vision-language models against patched visual prompt injectors,'' \emph{arXiv preprint arXiv:2405.10529}, 2024.

\bibitem{zhao2024bluesuffix}
Y.~Zhao, X.~Zheng, L.~Luo, Y.~Li, X.~Ma, and Y.-G. Jiang, ``Bluesuffix: Reinforced blue teaming for vision-language models against jailbreak attacks,'' \emph{arXiv preprint arXiv:2410.20971}, 2024.

\bibitem{oh2024uniguard}
S.~Oh, Y.~Jin, M.~Sharma, D.~Kim, E.~Ma, G.~Verma, and S.~Kumar, ``Uniguard: Towards universal safety guardrails for jailbreak attacks on multimodal large language models,'' \emph{arXiv preprint arXiv:2411.01703}, 2024.

\bibitem{liu2024unraveling}
Q.~Liu, C.~Shang, L.~Liu, N.~Pappas, J.~Ma, N.~A. John, S.~Doss, L.~Marquez, M.~Ballesteros, and Y.~Benajiba, ``Unraveling and mitigating safety alignment degradation of vision-language models,'' \emph{arXiv preprint arXiv:2410.09047}, 2024.

\bibitem{wang2024steering}
H.~Wang, G.~Wang, and H.~Zhang, ``Steering away from harm: An adaptive approach to defending vision language model against jailbreaks,'' \emph{arXiv preprint arXiv:2411.16721}, 2024.

\bibitem{ghosal2024immune}
S.~S. Ghosal, S.~Chakraborty, V.~Singh, T.~Guan, M.~Wang, A.~Beirami, F.~Huang, A.~Velasquez, D.~Manocha, and A.~S. Bedi, ``Immune: Improving safety against jailbreaks in multi-modal llms via inference-time alignment,'' \emph{arXiv preprint arXiv:2411.18688}, 2024.

\bibitem{gou2025eyes}
Y.~Gou, K.~Chen, Z.~Liu, L.~Hong, H.~Xu, Z.~Li, D.-Y. Yeung, J.~T. Kwok, and Y.~Zhang, ``Eyes closed, safety on: Protecting multimodal llms via image-to-text transformation,'' in \emph{ECCV}.\hskip 1em plus 0.5em minus 0.4em\relax Springer, 2024, pp. 388--404.

\bibitem{fares2024mirrorcheck}
S.~Fares, K.~Ziu, T.~Aremu, N.~Durasov, M.~Tak{\'a}{\v{c}}, P.~Fua, K.~Nandakumar, and I.~Laptev, ``Mirrorcheck: Efficient adversarial defense for vision-language models,'' \emph{arXiv preprint arXiv:2406.09250}, 2024.

\bibitem{robey2023smoothllm}
A.~Robey, E.~Wong, H.~Hassani, and G.~J. Pappas, ``Smoothllm: Defending large language models against jailbreaking attacks,'' \emph{arXiv preprint arXiv:2310.03684}, 2023.

\bibitem{xie2024gradsafe}
Y.~Xie, M.~Fang, R.~Pi, and N.~Gong, ``Gradsafe: Detecting jailbreak prompts for llms via safety-critical gradient analysis,'' in \emph{ACL}, 2024, pp. 507--518.

\bibitem{rombach2022high}
R.~Rombach, A.~Blattmann, D.~Lorenz, P.~Esser, and B.~Ommer, ``High-resolution image synthesis with latent diffusion models,'' in \emph{CVPR}, 2022, pp. 10\,684--10\,695.

\bibitem{bao2023one}
F.~Bao, S.~Nie, K.~Xue, C.~Li, S.~Pu, Y.~Wang, G.~Yue, Y.~Cao, H.~Su, and J.~Zhu, ``One transformer fits all distributions in multi-modal diffusion at scale,'' in \emph{ICML}.\hskip 1em plus 0.5em minus 0.4em\relax PMLR, 2023, pp. 1692--1717.

\bibitem{zhang2023adding}
L.~Zhang, A.~Rao, and M.~Agrawala, ``Adding conditional control to text-to-image diffusion models,'' in \emph{ICCV}, 2023, pp. 3836--3847.

\bibitem{huang2024effective}
Y.~Huang, F.~Zhu, J.~Tang, P.~Zhou, W.~Lei, J.~Lv, and T.-S. Chua, ``Effective and efficient adversarial detection for vision-language models via a single vector,'' \emph{arXiv preprint arXiv:2410.22888}, 2024.

\bibitem{lin2014microsoft}
T.-Y. Lin, M.~Maire, S.~Belongie, J.~Hays, P.~Perona, D.~Ramanan, P.~Doll{\'a}r, and C.~L. Zitnick, ``Microsoft coco: Common objects in context,'' in \emph{ECCV}.\hskip 1em plus 0.5em minus 0.4em\relax Springer, 2014, pp. 740--755.

\bibitem{chen2024dress}
Y.~Chen, K.~Sikka, M.~Cogswell, H.~Ji, and A.~Divakaran, ``Dress: Instructing large vision-language models to align and interact with humans via natural language feedback,'' in \emph{CVPR}, 2024, pp. 14\,239--14\,250.

\bibitem{helff2024llavaguard}
L.~Helff, F.~Friedrich, M.~Brack, K.~Kersting, and P.~Schramowski, ``Llavaguard: Vlm-based safeguards for vision dataset curation and safety assessment,'' \emph{arXiv preprint arXiv:2406.05113}, 2024.

\bibitem{crone2018socio}
D.~L. Crone, S.~Bode, C.~Murawski, and S.~M. Laham, ``The socio-moral image database (smid): A novel stimulus set for the study of social, moral and affective processes,'' \emph{PloS one}, p. e0190954, 2018.

\bibitem{schlarmann2024robust}
C.~Schlarmann, N.~D. Singh, F.~Croce, and M.~Hein, ``Robust clip: Unsupervised adversarial fine-tuning of vision embeddings for robust large vision-language models,'' \emph{arXiv preprint arXiv:2402.12336}, 2024.

\bibitem{li2024single}
J.~Li, Q.~Wei, C.~Zhang, G.~Qi, M.~Du, Y.~Chen, and S.~Bi, ``Single image unlearning: Efficient machine unlearning in multimodal large language models,'' \emph{arXiv preprint arXiv:2405.12523}, 2024.

\bibitem{chakraborty2024cross}
T.~Chakraborty, E.~Shayegani, Z.~Cai, N.~Abu-Ghazaleh, M.~S. Asif, Y.~Dong, A.~K. Roy-Chowdhury, and C.~Song, ``Cross-modal safety alignment: Is textual unlearning all you need?'' \emph{arXiv preprint arXiv:2406.02575}, 2024.

\bibitem{hossain2024sim}
M.~Z. Hossain and A.~Imteaj, ``Sim-clip: Unsupervised siamese adversarial fine-tuning for robust and semantically-rich vision-language models,'' \emph{arXiv preprint arXiv:2407.14971}, 2024.

\bibitem{hossain2024securing}
------, ``Securing vision-language models with a robust encoder against jailbreak and adversarial attacks,'' \emph{arXiv preprint arXiv:2409.07353}, 2024.

\bibitem{chen2024bathe}
Y.~Chen, H.~Li, Z.~Zheng, and Y.~Song, ``Bathe: Defense against the jailbreak attack in multimodal large language models by treating harmful instruction as backdoor trigger,'' \emph{arXiv preprint arXiv:2408.09093}, 2024.

\bibitem{ying2024safebench}
Z.~Ying, A.~Liu, S.~Liang, L.~Huang, J.~Guo, W.~Zhou, X.~Liu, and D.~Tao, ``Safebench: A safety evaluation framework for multimodal large language models,'' \emph{arXiv preprint arXiv:2410.18927}, 2024.

\bibitem{inan2023llama}
H.~Inan, K.~Upasani, J.~Chi, R.~Rungta, K.~Iyer, Y.~Mao, M.~Tontchev, Q.~Hu, B.~Fuller, D.~Testuggine \emph{et~al.}, ``Llama guard: Llm-based input-output safeguard for human-ai conversations,'' \emph{arXiv preprint arXiv:2312.06674}, 2023.

\bibitem{zhang2024avibench}
H.~Zhang, W.~Shao, H.~Liu, Y.~Ma, P.~Luo, Y.~Qiao, and K.~Zhang, ``Avibench: Towards evaluating the robustness of large vision-language model on adversarial visual-instructions,'' \emph{arXiv preprint arXiv:2403.09346}, 2024.

\bibitem{cheng2024typographic}
H.~Cheng, E.~Xiao, and R.~Xu, ``Typographic attacks in large multimodal models can be alleviated by more informative prompts,'' \emph{arXiv preprint arXiv:2402.19150}, 2024.

\bibitem{li2024red}
M.~Li, L.~Li, Y.~Yin, M.~Ahmed, Z.~Liu, and Q.~Liu, ``Red teaming visual language models,'' \emph{arXiv preprint arXiv:2401.12915}, 2024.

\bibitem{chen2024red}
S.~Chen, Z.~Han, B.~He, Z.~Ding, W.~Yu, P.~Torr, V.~Tresp, and J.~Gu, ``Red teaming gpt-4v: Are gpt-4v safe against uni/multi-modal jailbreak attacks?'' \emph{arXiv preprint arXiv:2404.03411}, 2024.

\bibitem{li2024mossbench}
X.~Li, H.~Zhou, R.~Wang, T.~Zhou, M.~Cheng, and C.-J. Hsieh, ``Mossbench: Is your multimodal language model oversensitive to safe queries?'' \emph{arXiv preprint arXiv:2406.17806}, 2024.

\bibitem{liu2024arondight}
Y.~Liu, C.~Cai, X.~Zhang, X.~Yuan, and C.~Wang, ``Arondight: Red teaming large vision language models with auto-generated multi-modal jailbreak prompts,'' in \emph{ACM MM}, 2024, pp. 3578--3586.

\bibitem{zhou2024multimodal}
K.~Zhou, C.~Liu, X.~Zhao, A.~Compalas, D.~Song, and X.~E. Wang, ``Multimodal situational safety,'' \emph{arXiv preprint arXiv:2410.06172}, 2024.

\bibitem{weng2024textit}
F.~Weng, Y.~Xu, C.~Fu, and W.~Wang, ``Mmj-bench: A comprehensive study on jailbreak attacks and defenses for multimodal large language models,'' \emph{arXiv preprint arXiv:2408.08464}, 2024.

\bibitem{mazeika2024harmbench}
M.~Mazeika, L.~Phan, X.~Yin, A.~Zou, Z.~Wang, N.~Mu, E.~Sakhaee, N.~Li, S.~Basart, B.~Li \emph{et~al.}, ``Harmbench: A standardized evaluation framework for automated red teaming and robust refusal,'' \emph{arXiv preprint arXiv:2402.04249}, 2024.

\bibitem{ji2024beavertails}
J.~Ji, M.~Liu, J.~Dai, X.~Pan, C.~Zhang, C.~Bian, B.~Chen, R.~Sun, Y.~Wang, and Y.~Yang, ``Beavertails: Towards improved safety alignment of llm via a human-preference dataset,'' \emph{NeurIPS}, vol.~36, 2024.

\bibitem{liu2023trustworthy}
Y.~Liu, Y.~Yao, J.-F. Ton, X.~Zhang, R.~G.~H. Cheng, Y.~Klochkov, M.~F. Taufiq, and H.~Li, ``Trustworthy llms: A survey and guideline for evaluating large language models' alignment,'' \emph{arXiv preprint arXiv:2308.05374}, 2023.

\bibitem{anwar2024foundational}
U.~Anwar, A.~Saparov, J.~Rando, D.~Paleka, M.~Turpin, P.~Hase, E.~S. Lubana, E.~Jenner, S.~Casper, O.~Sourbut \emph{et~al.}, ``Foundational challenges in assuring alignment and safety of large language models,'' \emph{arXiv preprint arXiv:2404.09932}, 2024.

\bibitem{wang2024cross}
S.~Wang, X.~Ye, Q.~Cheng, J.~Duan, S.~Li, J.~Fu, X.~Qiu, and X.~Huang, ``Cross-modality safety alignment,'' \emph{arXiv preprint arXiv:2406.15279}, 2024.

\bibitem{januspro_arxiv25}
X.~Chen, Z.~Wu, X.~Liu, Z.~Pan, W.~Liu, Z.~Xie, X.~Yu, and C.~Ruan, ``Janus-pro: Unified multimodal understanding and generation with data and model scaling,'' \emph{arXiv preprint arXiv:2501.17811}, 2025.

\bibitem{minigpt4v2_arxiv23}
J.~Chen, D.~Zhu, X.~Shen, X.~Li, Z.~Liu, P.~Zhang, R.~Krishnamoorthi, V.~Chandra, Y.~Xiong, and M.~Elhoseiny, ``Minigpt-v2: large language model as a unified interface for vision-language multi-task learning,'' \emph{arXiv preprint arXiv:2310.09478}, 2023.

\bibitem{kaufmann2023survey}
T.~Kaufmann, P.~Weng, V.~Bengs, and E.~H{\"u}llermeier, ``A survey of reinforcement learning from human feedback,'' \emph{arXiv preprint arXiv:2312.14925}, 2023.

\end{thebibliography}
}

\end{document}